\documentclass{emulateapj}
\usepackage{psfig,enumerate}
\usepackage{color}
\usepackage{atbegshi}
\usepackage{url}
\usepackage[caption=false]{subfig}
\usepackage[pdfpagemode={UseOutlines},bookmarks=true,bookmarksopen=true,
   bookmarksopenlevel=0,bookmarksnumbered=true,hypertexnames=false,
   colorlinks,linkcolor={blue},citecolor={blue},urlcolor={red},
   pdfstartview={FitV},unicode,breaklinks=true]{hyperref}
\def\hackaltaffiltext#1#2{\AtBeginShipoutNext{\footnotetext[#1]{#2}\stepcounter{footnote}}}

\newcommand{\kms}{\mbox{km s$^{-1}~$}} 
\newcommand{\kmse}{\mbox{km s$^{-1}$}}

\newcommand{\msun}{M$_{\odot}~$} 
\newcommand{\msune}{M$_{\odot}$}

\newcommand{\dgr}{$^{\circ}~$}

\newcommand{\teff}{$\rm T_{eff}$~}
\newcommand{\teffe}{$\rm T_{eff}$}
\newcommand{\logg}{$\log{g}$~}
\newcommand{\logge}{$\log{g}$}

\begin{document}

%\title{Metal-poor stars in the Magellanic Clouds as seen by the APOGEE-2 and MAPS Projects}
\title{The Lazy Giants: APOGEE Abundances Reveal Low \\
Star Formation Efficiencies in the Magellanic Clouds}
% ... in metal-poor stars of ...?

%\shorttitle{MC $\alpha$-Knee}
\shorttitle{The Lazy Giants}
\shortauthors{Nidever et al.}

%ANALYSIS GROUP
\author{
David L.\ Nidever\altaffilmark{1,2},
Sten Hasselquist\altaffilmark{3,$\dagger$},
Christian R.\ Hayes\altaffilmark{4},
Keith Hawkins\altaffilmark{5},
Joshua Povick\altaffilmark{1},
Steven R.\ Majewski\altaffilmark{4},
Verne V.\ Smith\altaffilmark{2},
Borja Anguiano\altaffilmark{4},
Guy S.\ Stringfellow\altaffilmark{6},
Jennifer S.\ Sobeck\altaffilmark{7},
Katia Cunha\altaffilmark{8,9},
%COMMENTS GROUP
Timothy C.\ Beers\altaffilmark{10},
Joachim M.\ BestenLehner\altaffilmark{11},
Roger E.\ Cohen\altaffilmark{12},
D.\ A.\ Garcia-Hernandez\altaffilmark{13,14},
Henrik J\"onsson\altaffilmark{15},
Christian Nitschelm\altaffilmark{16},
Matthew Shetrone\altaffilmark{17},
Ivan Lacerna\altaffilmark{18,19},
% ARCHITECTS GROUP
Carlos Allende Prieto\altaffilmark{13},
Rachael L.\ Beaton\altaffilmark{14,15},
Flavia Dell'Agli\altaffilmark{13,16},
Jos\'e G.\ Fern\'andez-Trincado\altaffilmark{20},
Diane Feuillet\altaffilmark{11}
Carme Gallart\altaffilmark{13},
Fred R.\ Hearty\altaffilmark{22},
Jon Holtzman\altaffilmark{23},
Arturo Manchado\altaffilmark{13},
Ricardo R.\ Mu\~noz\altaffilmark{24,25},
Robert O'Connell\altaffilmark{4},
and Margarita Rosado\altaffilmark{26}
}

\altaffiltext{1}{Department of Physics, Montana State University, P.O. Box 173840, Bozeman, MT 59717-3840 (dnidever@montana.edu)}
\altaffiltext{2}{National Optical Astronomy Observatory, 950 North Cherry Ave, Tucson, AZ 85719}
\altaffiltext{3}{Department of Physics \& Astronomy, University of Utah, Salt Lake City, UT, 84112, USA (stenhasselquist@astro.utah.edu)}
\altaffiltext{$\dagger$}{NSF Astronomy and Astrophysics Postdoctoral Fellow}
\altaffiltext{4}{Department of Astronomy, University of Virginia, Charlottesville, VA, 22904, USA}
\altaffiltext{5}{Department of Astronomy, The University of Texas at Austin, 2515 Speedway Boulevard, Austin, TX 78712, USA}
\altaffiltext{6}{Center for Astrophysics and Space Astronomy, University of Colorado, 389 UCB, Boulder, CO, 80309-0389, USA}
\altaffiltext{7}{Department of Astronomy, University of Washington, 3910 15th Ave NE, Seattle, WA 98195-0002, USA}
\altaffiltext{8}{Observatorio Nacional, Rio de Janeiro, Brazil}
\altaffiltext{9}{Steward Observatory, 933 N. Cherry St., University of Arizona, Tucson, AZ 85721, USA}
\altaffiltext{10}{Department of Physics and JINA Center for the Evolution of the Elements, University of Notre Dame, Notre Dame, IN 46556, USA}
\altaffiltext{11}{Max-Planck-Institut f{\"u}r Astronomie, K{\"o}nigstuhl 17, 69117, Heidelberg, Germany}
\altaffiltext{12}{Space Telescope Science Institute, 3700 San Martin Drive, Baltimore, MD 21218}
\altaffiltext{13}{Instituto de Astrofisica de Canarias, 38205 La Laguna, Tenerife, Spain}
\altaffiltext{14}{Department of Astrophysical Sciences, Princeton University, 4 Ivy Lane, Princeton, NJ 08544, USA}
\hackaltaffiltext{15}{The Observatories of the Carnegie Institution for Science, 813 Santa Barbara St., Pasadena, CA 91101, USA}
\hackaltaffiltext{16}{Universidad de La Laguna (ULL), Departamento de Astrofísica, E-38206, La Laguna, Tenerife, Spain}
\hackaltaffiltext{17}{Lund Observatory, Department of Astronomy and Theoretical Physics, Lund University, Box 43, SE-22100 Lund, Sweden}
\hackaltaffiltext{18}{10 Centro de Astronom{\'i}a (CITEVA), Universidad de Antofagasta, Avenida Angamos 601, Antofagasta 1270300, Chile}
\hackaltaffiltext{19}{McDonald Observatory, The University of Texas at Austin, 1 University Station, Austin, TX 78712, USA}
\hackaltaffiltext{20}{Instituto de Astronom\'ia y Ciencias Planetarias, Universidad de Atacama, Copayapu 485, Copiap\'o, Chile}
\hackaltaffiltext{21}{Instituto Milenio de Astrof\'isica, Av. Vicu\~na Mackenna 4860, Macul, Santiago, Chile}
\hackaltaffiltext{22}{Department of Astronomy and Astrophysics, The Pennsylvania State University, University Park, PA 16802, USA}
\hackaltaffiltext{23}{Department of Astronomy, New Mexico State University, Las Cruces, NM 88003, USA}
\hackaltaffiltext{24}{Departamento de Astronom\'ia, Universidad de Chile, Camino del Observatorio 1515, Las Condes, Santiago, Chile}
\hackaltaffiltext{25}{Visiting astronomer, Cerro Tololo Inter-American Observatory, National Optical Astronomy Observatory, which is operated by the Association of Universities for Research in Astronomy (AURA) under a cooperative agreement with the National Science Foundation.}
\hackaltaffiltext{26}{Instituto de Astronomía, Universidad Nacional Autonoma de Mexico (UNAM), Apdo. Postal  70-264, CP 04510, Mexico City, Mexico}

%hackaltafiltext

\begin{abstract}
We report the first APOGEE metallicities and $\alpha$-element abundances measured for 3600 red giant stars spanning a large radial range of both the Large (LMC) and Small Magellanic Clouds (SMC), the largest Milky Way dwarf galaxies. Our sample is an order of magnitude larger than that of previous studies, and extends to much larger radial distances. These are the first results presented that make use of the newly installed Southern APOGEE instrument on the du Pont telescope at Las Campanas Observatory.  Our unbiased sample of the LMC spans a large range in metallicity, from [Fe/H]=$-$0.2 to very metal-poor stars with [Fe/H]$\approx$ $-$2.5, the most metal-poor Magellanic Clouds (MCs) stars detected to date. The LMC [$\alpha$/Fe]--[Fe/H] distribution is very flat over a large metallicity range, but rises by $\sim$0.1 dex at $-$1.0~$<$~[Fe/H]~$\lesssim$~$-$0.5.  We interpret this as a sign of the known recent increase in MC star-formation activity, and are able to reproduce the pattern with a chemical evolution model that includes a recent ``starburst''. At the metal-poor end, we capture the increase of [$\alpha$/Fe] with decreasing [Fe/H], and constrain the ``$\alpha$-knee'' to [Fe/H]~$\lesssim$~$-$2.2 in both MCs, implying a low star-formation efficiency of $\sim$0.01 Gyr$^{-1}$. The MC knees are more metal poor than those of less massive Milky Way (MW) dwarf galaxies such as Fornax, Sculptor, or Sagittarius. One possible interpretation is that the MCs formed in a lower-density environment than the MW, a hypothesis that is consistent with the paradigm that the MCs fell into the MW's gravitational potential only recently.
\end{abstract}

% FIX THIS

% Outline:
% Introduction
% Observations and Data Reduction
% Results
% Discussion

% Introduction
\section{Introduction}

Dwarf galaxies are the most abundant galaxies in the Universe, and our Milky Way (MW) hosts dozens (e.g., \citealt{vandenBergh1999,Willman2010}; \citealt{Drlica-Wagner2015}),
with dozens more likely to be found in the coming decades. These galaxies span a large range in stellar mass ($\sim 10^{3}$ {\msun} to $\sim 10^{9}$ {\msun}; \citealt{Mateo1998}; \citealt{McConnachie2012}) and morphologies. They also serve as important laboratories for studying the details of galaxy formation at sub-MW scales, as well as the extent to which the MW was formed by the hierarchical buildup of such systems, as first suggested by \citet{Searle&Zinn1978}.  While a large fraction of MW dwarf galaxies do not currently contain gas, probably due to interactions with the MW in the past (\citealt{Grcevich2009}; \citealt{Pearson2016}),
they often exhibit complex and unique star-formation histories (e.g., \citealt{Weisz2014}; \citealt{Gallart2015,Bermejo-Climent2018}).

Our understanding of the formation and evolution of dwarf galaxies has grown significantly in the last two decades due to deep $HST$ and large-area, ground-based imaging and multi-object spectroscopy. Despite their apparently widely varying star-formation histories \citep[SFHs;][]{Dolphin2002,Weisz2014}, these galaxies appear to follow the well-established mass-metallicity correlation of galaxies, regardless of whether or not they are currently forming stars \citep{Kirby2013}. The underlying cause of the mass-metallicity relation is thought to be either shallower gravitational potentials, leading to lower metal retention (e.g., \citealt{Dekel1986}), a correlation between star-formation efficiency (SFE; star-formation rate divided by gas mass) and metallicities achieved during early star formation (e.g., \citealt{Matteucci1994,Calura2009}), or an initial mass function (IMF) that increases (becomes more ``top-heavy'') with increasing galaxy mass (e.g., \citealt{Koppen2007}). While there is support in the literature for each scenario, large-scale chemical-abundance studies of individual stars in dwarf galaxies are required to offer important constraints on the variation in SFE, outflows, and IMF over each dwarf galaxy's lifetime. 

The SFE, outflows, and IMF can be probed by analyzing the distribution of the $\alpha$-elements (O, Mg, Si, S, Ca, Ti) in a dwarf galaxy. These elements are primarily produced in massive stars ($M$ $>$ 8 {\msune}), and are released to the interstellar medium (ISM) in Type II supernovae (SNe) explosions. The stars that are formed are enhanced in the $\alpha$-elements relative to Fe in the first 1--2 Gyr of star formation --- i.e., before Type Ia SNe (which require the existence of white dwarf progenitors)
begin to explode. Type Ia SNe enrich the Fe abundance of the ISM without substantially enriching the $\alpha$-elements, thereby lowering the abundance of the $\alpha$-elements as compared to the Fe abundance \citep{Tinsley1979}. In the [$\alpha$/Fe]-[Fe/H] abundance plane, the point at which the mean [$\alpha$/Fe] begins to decrease is commonly referred to as the ``knee'', and the [Fe/H] where that occurs is a tracer of the early SFE of the galaxy --- i.e., prior to the system reaching that metallicity. Additionally, the level
of the high-$\alpha$ plateau can probe the IMF (e.g., \citealt{McWilliam1997}), while the slope of the knee can probe the gas outflow rate \citep[e.g.,][]{Andrews2017}. 

To date, the position of the knee has only been identified in a 
handful of dwarf galaxies. Based on these few cases, the [Fe/H] position of the knee appears to correlate with stellar mass, although variations have been observed \citep{Hendricks2014}, and the definition of such a feature is often ill-defined (e.g., \citealt{Zasowski2019}). 
Curiously, however, the position of the $\alpha$-element knee has not been measured for the two largest satellites of the Milky Way, the Large and Small Magellanic Clouds (LMC and SMC, respectively). These galaxies still contain gas, and are likely falling into the MW potential well for the first time (e.g., \citealt{Besla2007,Besla2012}). Both galaxies exhibit ongoing star formation, with a SFH
that suggests recent, strong starbursts beginning some 2--4 Gyr ago \citep{Smecker-Hane2002,Harris&Zaritsky2009,Rubele2012,Weisz2013,Meschin2014}, possibly due to a close encounter of the two galaxies with each other. 
Given a dynamical history that suggests that they have chemically evolved for the most part in near isolation, the MCs represent systems in contrast to the 
MW dwarf spheroidal satellites, which are thought to have had their evolution greatly influenced by their long association with the MW (e.g., \citealt{Wetzel2015,Fillingham2018}), which may have both incited star formation episodes through gravitational encounters, and facilitated the removal of gas and metals.

In further contrast to most MW dSphs, the MCs, due to their relative
proximity ($d$ $\sim$ 50 kpc), are more amenable
to high-resolution spectroscopy of their constituent stars. On the other hand, where the MCs present a disadvantage relative to the MW dSphs is their great angular extent,
with LMC stellar populations found to span a radial distance of at least $\sim$20\dgr \citep{Munoz2006,Majewski2009,Nidever2018,Belokurov2019}; 
thus previous high-resolution spectroscopic studies have been limited in both sample size  and spatial coverage.  Despite these limitations, previous abundance studies of the LMC have outlined its rough chemical characteristics, for example, 
that some of the $\alpha$-element abundances (Mg and O) of its stars are deficient compared to those of the MW, while other $\alpha$-elements (Si, Ca and Ti) are similarly enriched to MW stars \citep{Pompeia2008,Lapenna2012,VanderSwaelmen2013}. These studies also suggest that the $\alpha$-element knee of the LMC is at [Fe/H] $<$ $-$1.5. Further details were explored in \citet{Bekki2012}, who tried to fit the chemical-abundance tracks using models that utilized
the SFH of \citet{Harris&Zaritsky2009}, but the available chemical abundance data did not appear to be precise enough to resolve features in the [Mg/Fe]--[Fe/H] abundance plane that were predicted by their models. 
The SMC is less well studied, but the chemical-abundance results of 200 stars from \citep{Mucciarelli2014} indicate that the [$\alpha$/Fe] abundance patterns of the SMC are quite similar to the LMC, as studied by \citet{Lapenna2012}.

\begin{figure}[t]
\begin{center}
%\includegraphics[trim={0cm 3cm 0cm 3cm},clip,width=1.0\hsize,angle=0]{apogee_mcs_finalmap_2deg_smash_names2.pdf}
%  trim={<left> <lower> <right> <upper>}
\includegraphics[width=1.0\hsize,angle=0]{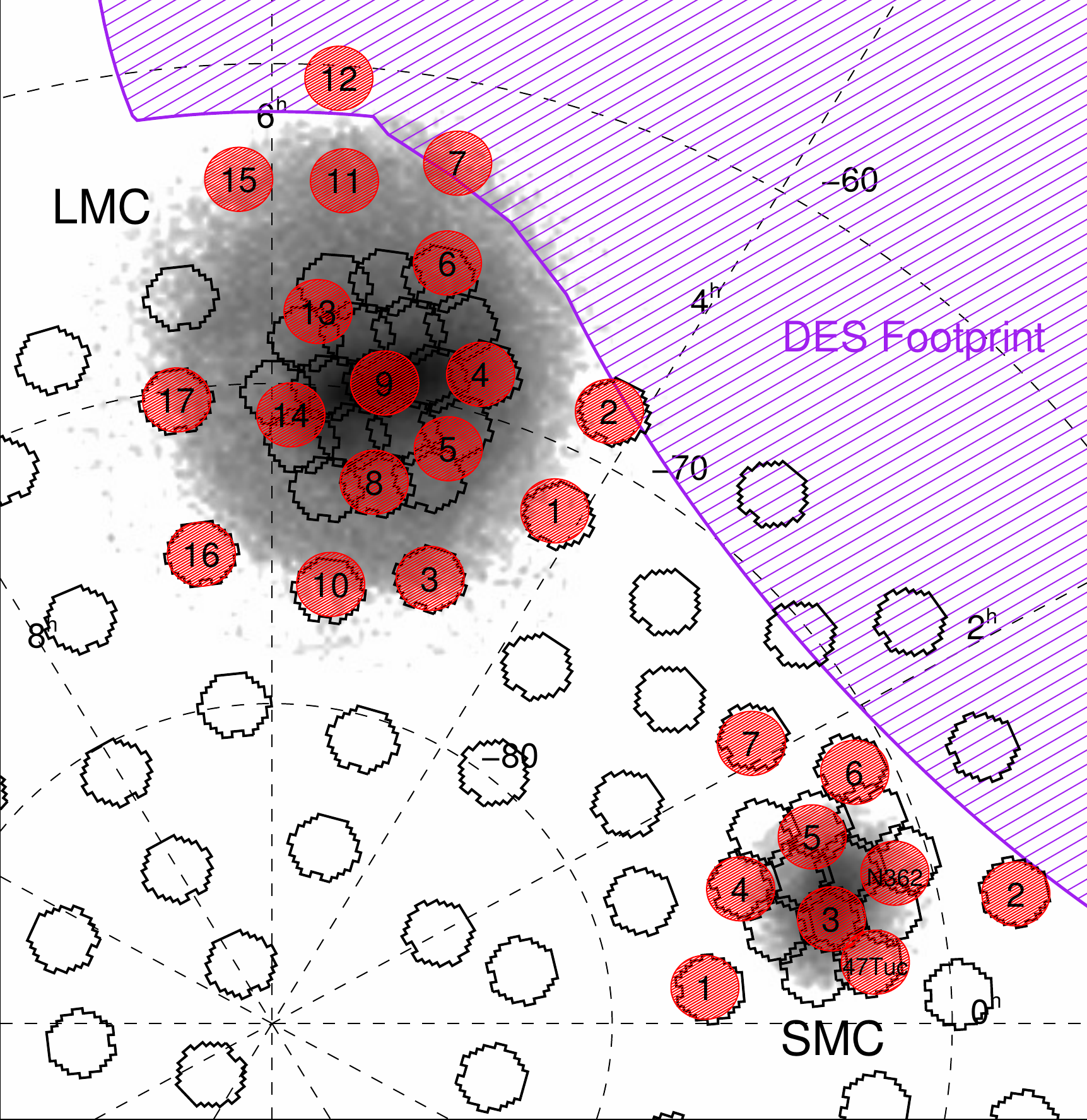}
\end{center}
\caption{Map of the APOGEE MC fields (red filled symbols) shown on top of the 2MASS red giant branch star density map. The APOGEE MC 
field names are indicated by the overplotted numbers. 
Open hexagons (2.2\dgr in diameter) with no overplotted numbers
designate other fields from the SMASH survey.  The DES footprint is shown in the
purple shaded region.  An equatorial $\alpha$/$\delta$ grid is marked by dashed lines.}
\label{fig_apogeemap}
\end{figure}

Fortunately, the Apache Point Observatory Galactic Evolution Experiment (APOGEE, \citealt{Majewski2017}), part of SDSS-III \citep{Eisenstein2011} and SDSS-IV \citep{Blanton2017}, through installation of a second APOGEE spectrograph \citep[APOGEE-2S;][]{Wilson2019} on the du Pont telescope at Las Campanas Observatory (LCO), has recently procured high-resolution, $H$-band spectra for already several thousand stars residing in the MCs. These observations cover much of the galaxies: out to 10\dgr for the LMC and 6\dgr for the SMC, and with large azimuthal coverage for both. 
An advantage of these particular data is that they can be compared directly to the vast collection of APOGEE spectra similarly obtained and analyzed for both MW stars as well as stars in a sampling of other MW satellites, which ensures that any observed differences in properties found between galactic systems is likely to be real, and not the result of systematic errors between disparate data sets.

In this paper we analyze the [$\alpha$/Fe]--[Fe/H] abundance patterns of red giant branch (RGB) stars in the Magellanic Clouds in order to understand the chemical evolution and SFHs of these galaxies. 
We probe and place upper-limits on the position of the $\alpha$-element knee in the LMC and SMC, which provides powerful constraints on the early 
SFEs of these galaxies.  In addition, we introduce a new metric, [Fe/H]$_{\alpha 0.15}$, that can be used to robustly and uniformly measure the position of a galaxy's $\alpha$-element abundance trendline.
%location of the $\alpha$-element knee in the LMC as well as an upper limit on the position of the knee for the SMC, which provides powerful constraints on the early  star formation efficiencies of these galaxies.
The layout of the paper is as follows:  Section \ref{sec:targets} provides an overview of the APOGEE MC fields and target selection, while section \ref{sec:data} describes the APOGEE observations currently in hand
and the data reduction procedures.  In Section \ref{sec:memberselection} we detail how we select bona fide MC member stars.  Checks on the veracity 
of the Southern APOGEE data are presented in Section \ref{sec:qachecks}.  Our main results are described in Section \ref{sec:results}, and the relevance and interpretation of our measurements are discussed in Section \ref{sec:discussion}.  Finally, our conclusions are summarized in Section \ref{sec:summary}.

%  alpha abundances
%  why the knee in multiple galaxies
%  independent probe of SFE
%  attempts to measure and intepret the knee, Hendricks paper
%  motivation
%   -hierarchical galaxy formation
%   -understand the formation/evolution of dwarf galaxies

% Background abundance references:
%Accurate APOGEE abundances of Sagittarius \citep{Hasselquist2017}. 
%Fornax knee \citep{Hendricks2014}.
%LMC abundances \citep{vanderswaelmen2013}.
%LMC abundances and kinematics \citep{Cole2005}.

\section{Field and Target Selection}
\label{sec:targets}

In this work we only analyze spectra of MC RGB stars, but the full APOGEE MC survey consists of many astrophysical targets, which we describe here for future APOGEE MC papers to reference.
A brief description of the APOGEE MC survey target selection is given in the SDSS-IV targeting paper \citep{Zasowski2017}, but we give more details here.
The main goal of the APOGEE MC survey is to study the galactic chemical and kinematical evolution of the Magellanic Clouds, with a particular focus on spatial variations.
RGB stars are the majority of the stars in our MC survey. However, a large number of other stellar classes were also chosen to enable a variety of astrophysical studies: massive stars, hot main-sequence stars, asymptotic giant branch (AGB) stars, and post-AGB stars. In addition, stars in the \citet{Olsen2011} metal-poor accreted stream were selected to help ascertain their properties and origin, as were some stars that were previously studied with high-resolution spectra for calibration purposes.

To explore spatial gradients in the MCs, we selected fields that spanned a wide range of radius and position angle (see Figure \ref{fig_apogeemap} and Table \ref{table_fields}). We required a field to have approximately 50 or more MC RGB candidates for which S/N=100 could be obtained with 9 $\sim$1 hour ``visits'' in the LMC and 12 visits in the SMC. This produced a maximum radius of $\sim$9.5\dgr in the LMC and $\sim$6\dgr in the SMC. To aid in the chemical analysis we picked fields that also had deep $ugriz$ photometry from
the Survey of the MAgellanic Stellar History \citep[SMASH;][]{Nidever2017} or the Dark Energy Survey
\citep[DES;][]{DES}\footnote{LMC11 and LMC15 are not currently covered by SMASH or DES because of a change in the DES footprint. Deep DECam observations of those regions will be obtained in the near future by an approved program.}.  One of the goals of SMASH is to derive accurate spatially-resolved SFHs
to old ages.  When this information is combined with the precise APOGEE abundances,
simple chemical evolution models can be robustly constrained.
At the time of this writing, the SMASH SFHs are still being generated; therefore, we did not use them in our current analysis.

\begin{figure}[t]
\begin{center}
\includegraphics[width=1.0\hsize,angle=0]{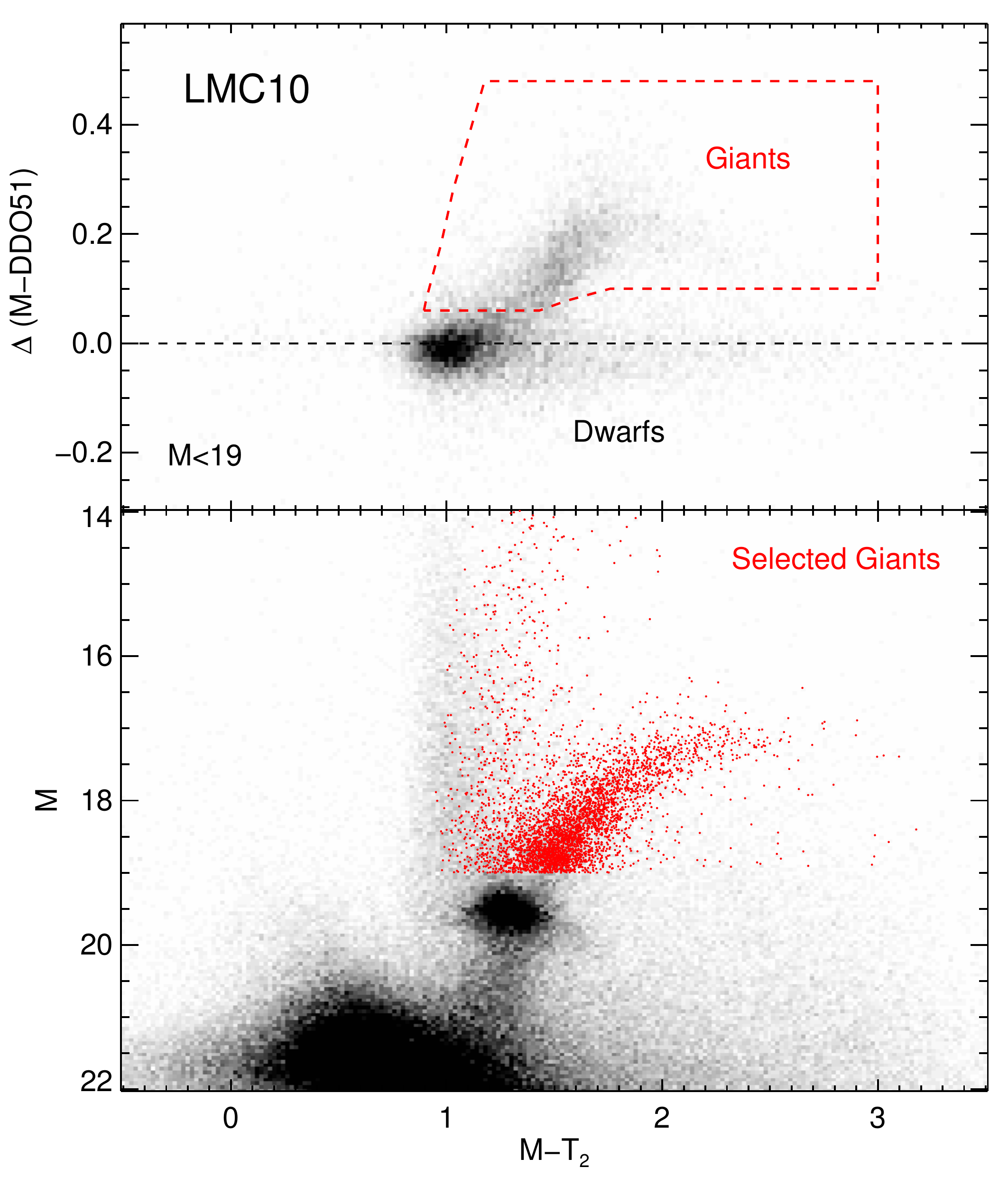}
\end{center}
\caption{
An illustration of how the Washington $M$, $T_2$ and $DDO$51 photometry is used to help select MC RGB targets for field LMC10.
(Top) A modified color-color diagram (with the shape of the dwarf locus subtracted from M-$DDO$51) showing the dwarf/giant separation (with $M$ $<$19 mag).  The giant selection box is indicated by the red dashed polygon.  (Bottom) The color-magnitude diagram with stars falling in the giant selection box shown as red filled circles. The selected stars are dominated by LMC red giant branch and foreground MW halo giants.
}
\label{fig_washphot}
\end{figure}

To help target likely Magellanic giant stars, especially blue metal-poor giants that can be heavily contaminated by MW foreground stars, we obtained Washington $M$, $T_2$, and DDO51 imaging with the WFI camera on the MPG/ESO-2.2m telescope at La Silla.  The data were reduced with the {\em IRAF} CCDRED package, and photometric parameters derived with the PHOTRED photometry pipeline \citep{Nidever2017}.
Figure \ref{fig_washphot} shows an example of the ($M$-$T_2$, $\Delta$[$M$-$DDO$51]) color-color diagram \citep[top panel;][]{Majewski2000} that is used to select giant star candidates (red polygon), and shown in the ($M$-$T_2$, $M$) CMD as red filled circles (bottom panel).  The photometry and giant candidates are used later on during the selection of MC RGB targets (see below).

\begin{figure}[t]
\begin{center}
\includegraphics[width=1.0\hsize,angle=0]{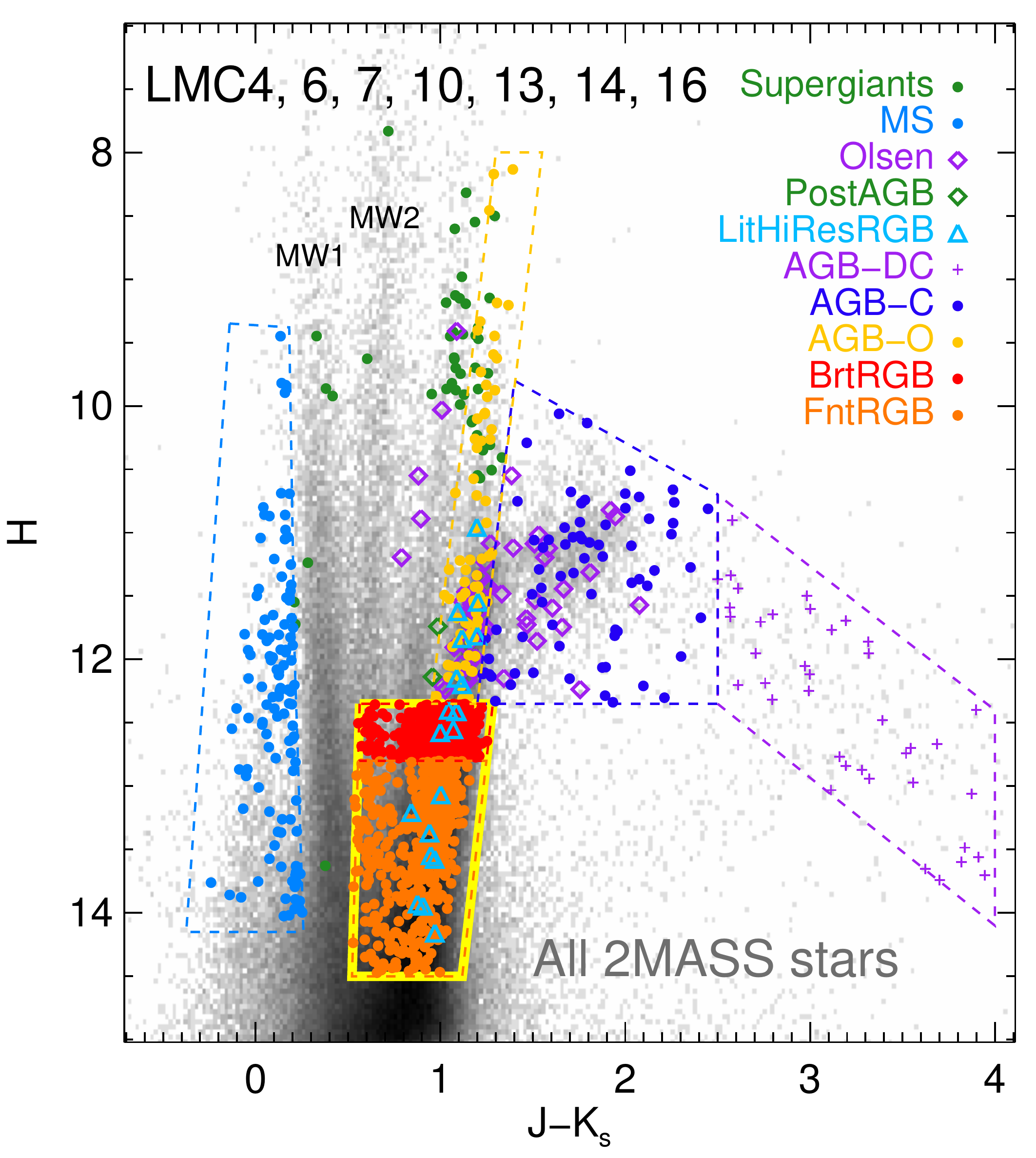}
\end{center}
\caption{The 2MASS near-infrared color-magnitude diagram of seven representative LMC fields with the density of all stars shown in grayscale and the APOGEE targets identified by their targeting class.  The RGB stars that are the main focus of this analysis are highlighted with a thick yellow polygon.  The blue sequence of 2MASS stars at $J-K_{\rm s}$ $\sim$0.3 (MW1) is dominated by foreground Milky Way disk F-K dwarfs and the middle sequence at $J-K_{\rm s}$ $\sim$0.7 (MW2) is dominated by foreground Milky Way disk K dwarfs and halo red clump stars \citep{Nikolaev2000}.}
\label{fig_targets}
\end{figure}

While the Washington+$DDO51$ photometry was used to prune out any potential MW foreground dwarfs, the selection of RGB targets otherwise used a wide ($J$-$K_{\rm s}$) range to avoid producing a bias against metal-poor stars as seen in Figure \ref{fig_targets}.  This figure shows the 2MASS photometry of all stars (grayscale) in seven representative LMC fields (with foreground MW sequences identified) with our APOGEE targets in color points to illustrate our targeting strategy. The RGB stars which are used in this analysis are shown as red (bright RGB) and orange (fainter RGB) filled dots, and were selected using the Washington and $DDO$51 photometry.

Initial estimates indicate that there is $\sim$20--40\% vignetting in the outer portion of the field of view of the Southern APOGEE fiber plugplates
\citep[0.8\degr$<$$R$$<$0.95\degr;][]{Wilson2019}.  For this
reason, most APOGEE-2S fields restrict targets to a field radius of 0.8\degr.  This same restriction was used in the MC fields when there were enough bright targets to fill the fibers allotted to each target class for that field.  Exceptions were made for bright but rare targets (e.g., supergiants, main-sequence stars) and RGB targets in outer fields, where the density is low. Note that an initial set of MC targets with a 0.8\dgr outer radius for all target types and fields was created and observed in the first APOGEE-2S MC observing run in October 2017.  Unfortunately, with the smaller radius restriction a large fraction of intended MC RGB targets were ``lost'' in the lower density outer fields.  For later observing runs the plates were redesigned to include the original targets in the outer radial zones, and observed starting in December 2017.  Therefore, some stars will have more visits than the originally planned 9/12 for the LMC/SMC.

\begin{figure}[ht]
\begin{center}
\includegraphics[width=1.0\hsize,angle=0]{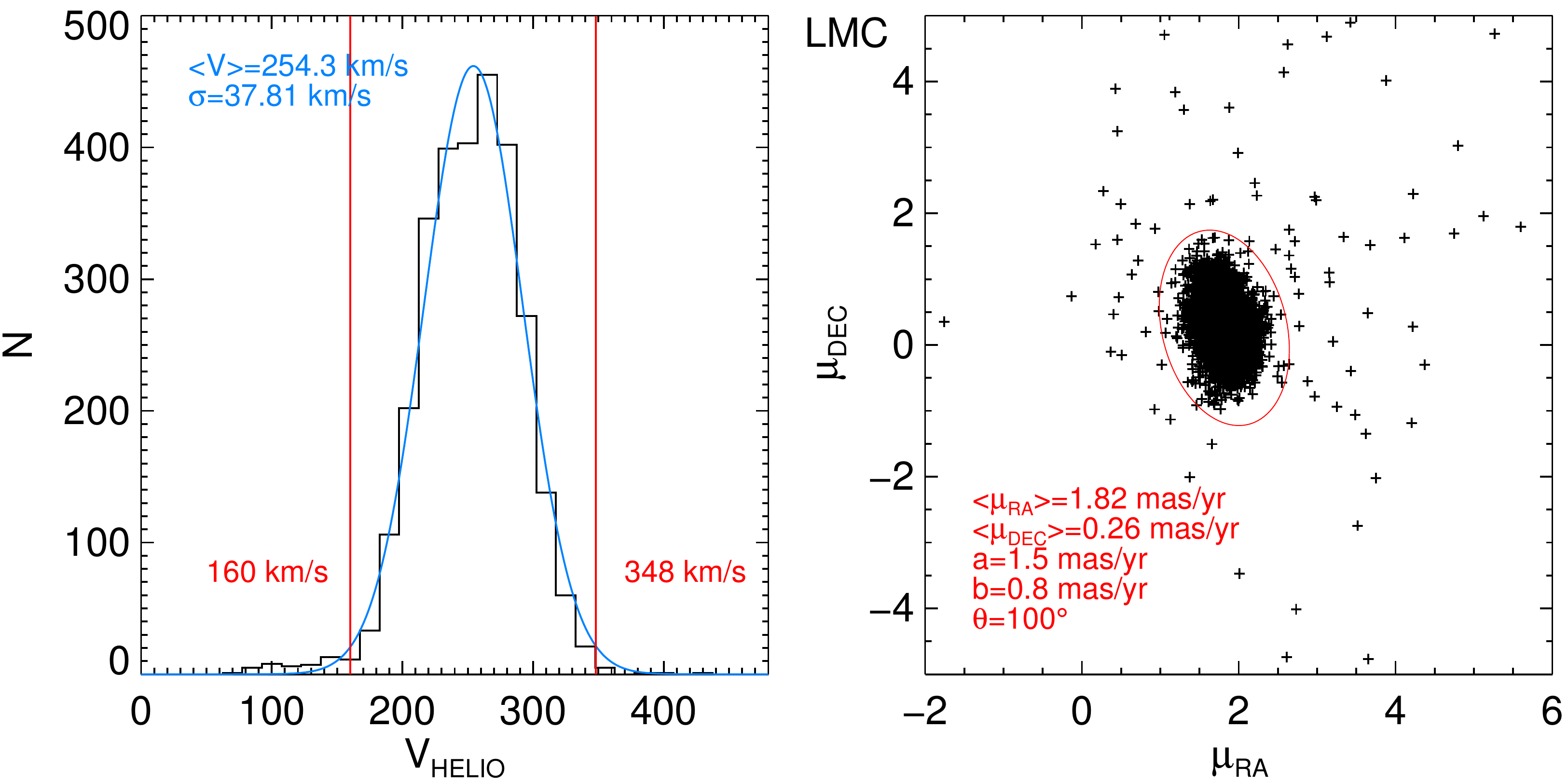} \\
\includegraphics[width=1.0\hsize,angle=0]{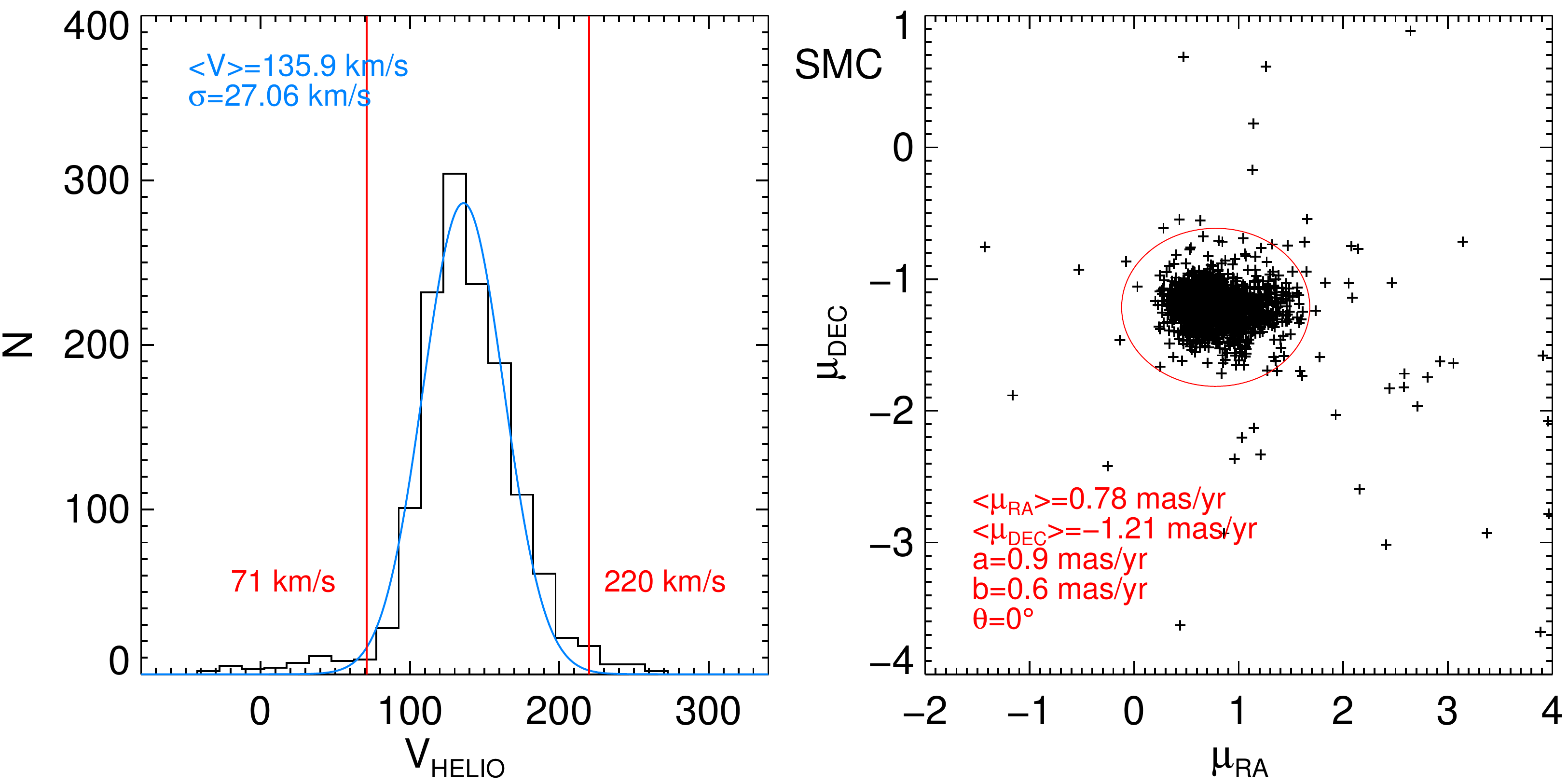}
\end{center}
\caption{The distribution of heliocentric radial velocities (left) and Gaia DR2 proper (right) motions of APOGEE RGB targets in the LMC (top) and SMC (bottom) fields. 
The applied velocity cuts to remove MW contamination} are shown in the red lines (for radial velocity) and ellipses (proper motions).  The cuts remove 4\% and 8\% of the LMC and SMC RGB sample, respectively.
\label{fig_velcuts}
\end{figure}

The target selection for each MC field proceeds as follows.  All targets are selected for each class according to the criteria described in the list below.  An absolute priority is given for all targets in the full list in a two-step process using first the ``group priority'' (the priority of one group of targets to each other; in the order listed below), and then by the priority within the group using a sorting algorithm (none, random, magnitude, or CMD-uniform). 
Next, a limit on the number of targets for each class that can be assigned fibers is imposed, with a 50\% excess as a buffer for dealing with fiber collisions and other contingencies.
Targets with bright neighbors (brighter than 2 mag and within 6\arcsec) and targets that collide with a higher priority targets within 56\arcsec~(the minimum allowed fiber-to-fiber spacing; \citealt{Zasowski2013}) are removed.  Any extra targets above the limit for that class are removed.  In addition, the total number of targets for that entire field (250 for MC fields and 205 for the NGC362 and 47 Tucanae calibration-cluster fields) is imposed and any excess targets are removed
from the low priority end of the list. Finally, the list of targets and their information, as well as the selection function fraction (the number of targets selected divided by the total number in that class) are saved.  
Unlike the standard APOGEE observing for deep fields, where a ``cohorting" scheme is utilized to allow swapping of brighter targets from field visit to field visit to increase the number of targets observed \citep{Zasowski2013},
for the MC observing the same targets are observed for the full number of visits each field is visited.  
The MC target selection software is available at \url{https://svn.sdss.org/public/repo/apogee/apogeetarget/trunk/pro/mcs/}

% The maximum alloted fibers with no buffer is imposed on each class as well as the total number of fibers for this field (260 for MC fields and 205 for the  N362 and 47Tuc calibration cluster fields).

%Below we detail how the 10 target classes were selected (in priority order):
Details on how the ten target classes were selected are given below (in target-class priority order): 

\begin{enumerate}

\item {\bf Supergiants:}  Massive Magellanic stars were selected from \citet{Neugent2012} and \citet{Bonanos2009}, and combined in one catalog.  Blue stars that overlap the ``main-sequence" target class (see below) were removed.  Generally a maximum of 20 fibers per field (but increased to $\sim$35 in some inner fields) were allotted for the supergiants, and a limiting radius of 0.95\dgr was used to improve the yield.
%To date 144 MC supergiants have been observed.
%SRM: Not illustrated in Fig. 3 so not clear where those would lie in CMD?

\item {\bf Main-sequence:} Young, blue main-sequence targets were selected using the cyan box in the ($J$-$K_{\rm s}$, $H$) diagram, and are shown as filled light blue circles in Figure \ref{fig_targets}.  Roughly 20 fibers per field were allotted for the main-sequence stars, and the stars were selected with magnitude priority sorting and a limiting field radius of 0.95\degr.
%To date 403 main-sequence stars have been observed.

\item {\bf Olsen stream:} \citet{Olsen2011} discovered a stream of metal-poor stars in the LMC that might be accreted SMC stars.  A maximum of 20 fibers per field
were allotted for the Olsen stream stars and these were selected with magnitude priority sorting and a limiting field radius of 0.95\degr. 
%To date 71 Olsen stream stars have been observed.

\begin{figure}[t]
\begin{center}
\includegraphics[width=1.0\hsize,angle=0]{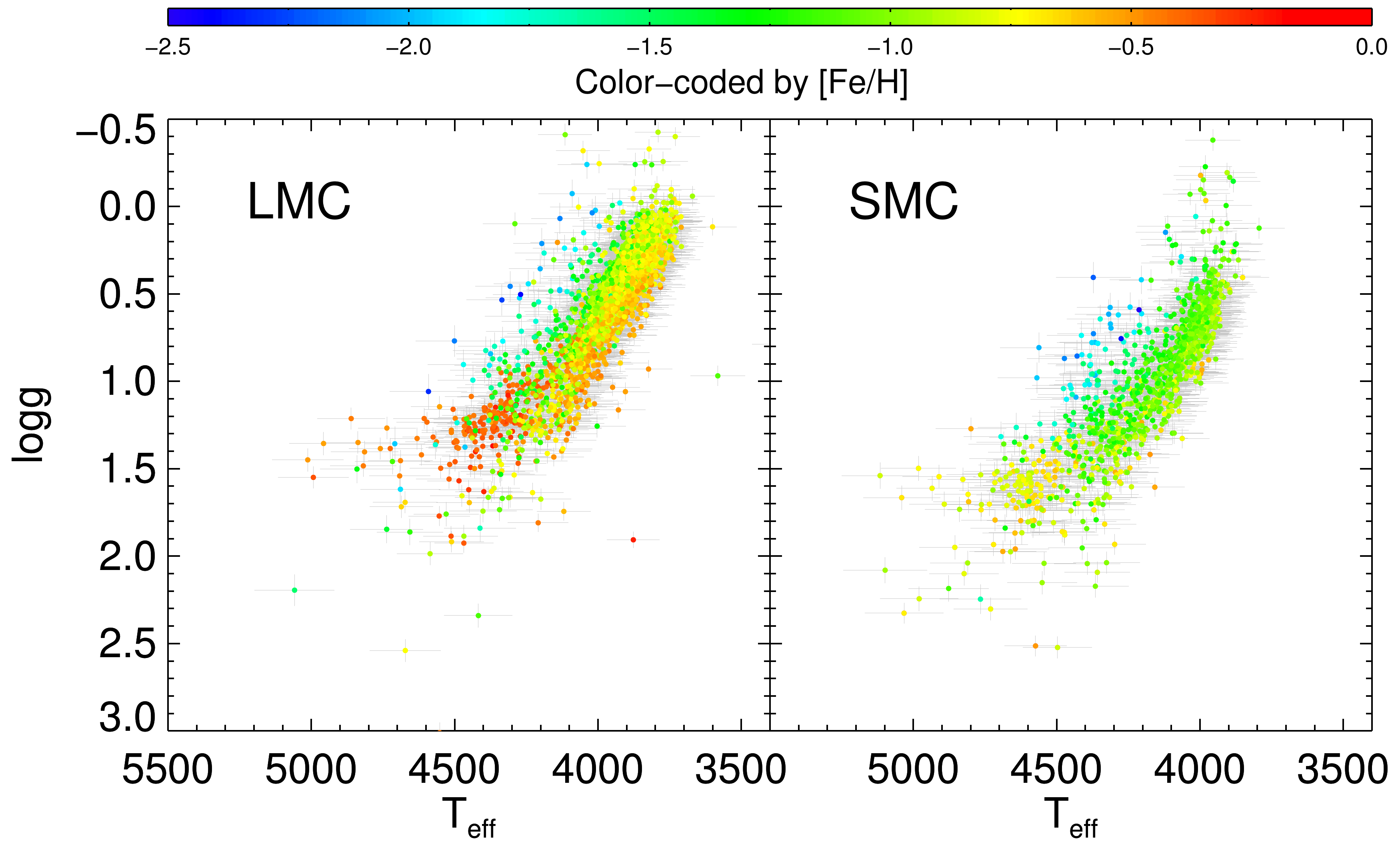}
\end{center}
\caption{The distribution of the selected MC RGB stars in \teff-- \logg space, and color-coded by [Fe/H] for stars with no ASPCAP {\tt STARBAD} flag set.}
\label{fig_tefflogg}
\end{figure}

\item {\bf Post-AGB:} Cool post-AGB stars were selected from \citet{Kamath2014,Kamath2015}. A maximum of 10 fibers per field were allotted for the post-AGB stars and these were selected with random priority sorting and a limiting field radius of 0.80\degr.
%To date 20 Post-AGB stars have been observed.

\item {\bf Literature High-resolution RGB:} To check our results, we observed RGB stars that other groups have also studied with high-resolution spectra (\citealt{VanderSwaelmen2013} and Carrera et al., in prep).  A maximum of 10 fibers per field were allotted for the literature RGB stars, and these were selected with magnitude priority sorting and a limiting radius of 0.80\dgr.  Some of the literature targets are shown as blue open diamonds in Figure \ref{fig_targets}.  
To date, 24 literature RGB stars have been observed, but only 1 star has sufficient S/N for reliable abundances.  A meaningful comparison cannot be performed at this time.

\item {\bf AGB-DC:} The AGB targeting strategy was guided by the AGB selections in \citet{Nikolaev2000} using 2MASS \citep{Skrutskie2006}, and \citet{DellAgli2015a,DellAgli2015b} using $Spitzer$.
A maximum of 15 fibers per field were allotted to dusty carbon-rich AGB stars.  These targets were selected using the magenta box in the ($J$-$K_{\rm s}$, $H$) diagram, using a ``CMD-uniform'' sampling designed to select stars uniformly across the 2-D CMD space.  These targets were also selected with a limiting field radius of 0.80\degr. 
%The 102 AGB-DC targets observed to date 
Some of the AGB-DC targets observed to date are shown as filled magenta crosses in Figure \ref{fig_targets}.  
%To date 102 AGB-DC targets have been observed.

\item {\bf AGB-C:} 
%Carbon-rich AGB stars.  
A maximum of 15 fibers per field were allotted to carbon-rich AGB stars, and the
targets were selected using the blue box in the ($J$-$K_{\rm s}$, $H$) diagram with
CMD-uniform sampling and a limiting field radius of 0.80\degr.  The AGB-C targets are shown as filled blue circles in Figure \ref{fig_targets}. 
%To date 241 AGB-C targets have been observed.
  
\item {\bf AGB-O:} Oxygen-rich AGB stars were assigned to a maximum of 15 fibers per field (but this was increased to $\sim$40 in some inner regions).  These targets were selected using the yellow box in the ($J$-$K_{\rm s}$, $H$) diagram with CMD-uniform sampling and a limiting field radius of 0.80\dgr (except for the field SMC7 which used 0.95\degr).  Some of the AGB-O targets are shown as filled yellow circles in Figure \ref{fig_targets}.
%, with 284 observed to date.

\item {\bf Bright RGB:} The largest group of targets are the red giant branch stars, selected in two groups -- ``bright" and ``faint" (see below). The higher priority group are those for which S/N=100 spectra will be obtained in 9/12 visits for the LMC/SMC.  Targets were selected from among those that passed the Washington photometry giant star selection (see above and Figure \ref{fig_washphot}) and that lie within the red-bounded polygonal area in the ($J$-$K_{\rm s}$, $H$) diagram. The RGB CMD selection box extends to fairly blue colors ($\sim$0.55) to ensure we capture metal-poor MC stars 
at the risk of some contamination from metal-poor MW halo stars \citep{Nikolaev2000}.
These stars span a magnitude range of 12.35$<$$H$$<$12.8 for the LMC and 12.9$<$$H$$<$13.2 for the SMC. Random sampling and a limiting field radius of 0.80\dgr (except for some outer fields where 0.95\degr was allowed) were used for the selection.  All of the unassigned fibers remaining after attempting to fill them with stars in any of the above target classes were allotted to bright RGB stars, up to a maximum of 250.  The bright RGB targets are shown as filled red circles in Figure \ref{fig_targets}.  Thus far, 2505 bright RGB targets have been observed, with 2374 having S/N$>$40.

\item {\bf Faint RGB:} Faint RGB stars were targeted if not enough bright RGB targets were available.  Targets were selected from among those that passed the Washington photometry giant cut and that lie within the orange polygon in the ($J$-$K_{\rm s}$, $H$) diagram.  Magnitude priority sorting and a limiting field radius of 0.80\dgr (except for some outer fields that used 0.95\degr) were adopted. All of the remaining fibers were allotted up to a maximum of 250.  The faint RGB targets are shown as filled orange circles in Figure \ref{fig_targets}.  Thus far, 2061 faint RGB targets have been observed, with 1490 having S/N$>$40.
\end{enumerate}

The CMDs were not dereddened for the target selection, because it is challenging to calculate accurate star-by-star extinctions in the inner MCs (but see \citealt{Choi2018a}). However, our CMD selection polygons were made large enough to capture the relevant stellar populations, even with the moderate reddening seen in the NIR.
For the SMC fields, the CMD selection polygons were shifted in magnitude by +0.4 mag due to the larger SMC distance compared to that of the LMC.

The selection of targets in the 30 Dor ``APOGEE-2S first light'' plate was focused on getting spectra for massive stars, and so used different criteria than used for the normal MC fields.  A number of massive star types (e.g., Luminous Blue Variables, Wolf-Rayet, Blue Supergiant, main-sequence stars) were selected manually from existing catalogs, and will be discussed in full in a subsequent paper (Stringfellow et al., in preparation).  In addition, 72 RGB stars were selected from the CMD in a manner similar to that shown in Figure \ref{fig_targets}, but with a magnitude range of 12.30$<$$H$$<$13.0, and a less expansive blue limit of $J-K_{\rm s}$=0.9.

\section{Observations and Reductions}
\label{sec:data}

% FIGURES
% - some demonstration that the southern abundances are okay
%     compare north and south MW disk abundances, just alpha
%     side-by-side alpha vs. feh of north and south
%     only use LMC Teff/logg range, 4600<Teff<3600, -1<logg<3.0
%     density with trend lines, density are different b/c sample different parts of the galaxy
% - target selection, include DDO51 2CD selection
% - teff vs. logg
%     red, warm, metal-rich stars are ~3 Msun young stars, blue loop
% - Maybe lead with current Fig 4. LMC and MW alpha vs. feh
% - 3 panel LMC/SMC abundances, discuss that vdS and APOGEE sample different regions
%      don't see a strong knee in their data
% - GCE figure, show 1 and 5 as well
%    show discuss what we find for SFE for the MW

For the last seven years, the original APOGEE instrument \citep{Wilson2019}
has been taking data of the Northern sky using the SDSS-2.5m telescope \citep{Gunn2006} at Apache Point Observatory.   
However, to allow the APOGEE-2 project to access the entire Milky Way, as well as the Magellanic Clouds, a Southern copy of the APOGEE instrument \citep{Wilson2019}, was installed at the du Pont-2.5m telescope at LCO in January 2017, with first light on February 16, 2017, using the 30 Dor plate.  The Southern instrument is nearly identical to the Northern one, with some small modifications, such as a revised LN$_2$ tank suspension system and protections against seismic events \citep[APOGEE-2S;][]{Wilson2019}.  APOGEE-2S uses 300 fibers and plugplates with a 2\dgr diameter field-of-view.

Since the inaugural ``first light" observations of the 30 Doradus plate, the first full season featuring Magellanic Cloud targeting began in September 2017 and are continuing. 
We present data from the 16th SDSS data release (DR16), which includes data taken through August 31, 2018, which will be released in December 2019 and described in detail by J{\"o}nsson et al. (in prep.).
These data were reduced through the standard APOGEE reduction pipeline \citep{Nidever2015}, and stellar parameters and abundances were obtained via the ASPCAP pipeline \citep{Holtzman2015,Garcia2016}, which uses a library of synthetic spectra \citep{Zamora2015}. 
ASPCAP first derives stellar parameters by fitting to a 7-D synthetic spectral grid (\teff, \logg, [M/H], [$\alpha$/M] (O, Mg, Si, S, Ca, Ti), [C/M], [N/M], and
$v_{\rm micro}$), and then derives individual element abundances (C, C I, N, O, Na, Mg, Al, Si, P, S, K, Ca, Ti, Ti II, V, Cr, Mn, Fe, Co, Ni, Cu, Ge, Rb, Yb, Ce, and Nd) by fitting spectral ``windows'' that are sensitive to variations in those individual element abundances.
Changes since DR14 \citep{dr142018} include improvements to the linelist, the use of MARCS atmospheres \citep{Gustafsson2008} for all grids, no internal temperature-dependent abundance calibrations were applied, and improved derivation of uncertainties.
Constant abundance offsets were determined from the mean offset of solar metallicity stars in the solar neighborhood from solar abundance ratios.  The additive offsets for the elements relevant to this study are 0.009, 0.038, 0.002, 0.000, and 0.033 dex for Mg, Si, Ca, Fe and $\alpha$, respectively.
%O        0.022
%Mg       0.009
%Si       0.038
%S        0.040
%Ca       0.002
%Fe       0.000
%M        0.000
%alpha    0.033
For our analysis, only stars that had reliable stellar parameters (no bad quality flags set) were used. 
Statistical uncertainties are derived by looking at repeat observations of stars in different overlapping fields or with different visits and then fitting the logarithm of the scatter as a linear function of \teffe, [M/H], and S/N.  It is possible that the uncertainties are somewhat underestimated.  We conservatively increase our uncertainties to match the intrinsic abundance scatter of first generation stars in globular clusters which effectively acts as an upper limit to our actual uncertainties. The factors are 1.5$\times$, 1.7$\times$ and 2.7$\times$ for [Mg/Fe], [Si/Fe] and [Fe/H], respectively, while [$\alpha$/Fe] and [Ca/Fe] do not need to be adjusted.
Because this is the first paper to utilize data from the Southern spectrograph, in Section \S \ref{sec:qachecks} we perform quality checks of the stellar parameters and abundances to verify the reliability of their results.

%{\bf The ASPCAP abundances use \citet{Grevesse2007} solar abundance scale in two ways.  First, \log{gf} values in the APOGEE linelist (Shetrone) were adjusted to the solar and the Arcturus FTS spectra using the \citet{Grevesse2007} solar abundance and the XX abundance for Arcturus. Second, the \citet{Grevesse2007} solar abundance scale is used when creating our Turbospectrum (REF) synthetic stellar spectra and sets the scale for [X/Y]=0.  The DR16 individual element abundance zero-points were slightly adjusted to set [X/Fe]=0 for the solar neighborhood stars (HOW WAS THAT SELECTION MADE).  The adjustments are small for most elements.  For the $\alpha$-elements and Fe the additive adjustments are $+$0.022, $+$0.008, $+$0.037, $+$0.002, $+$0.009, $-$0.010 for O, Mg, Si, Ca, Ti, and Fe.}
%O     +0.022
%Na    +0.022
%Mg    +0.008
%Si    +0.037
%Ca    +0.002
%Ti    +0.009
%Fe    -0.010
%alpha +0.034}

\begin{figure*}[t]
\begin{center}
%\includegraphics[trim={0cm 3cm 0cm 3cm},clip,width=1.0\hsize,angle=0]{apogee_mcs_finalmap_2deg_smash_names2.pdf}
%  trim={<left> <lower> <right> <upper>}
\includegraphics[trim={2.5cm 13.0cm 2.5cm 2.5cm},clip,width=1.0\hsize,angle=0]{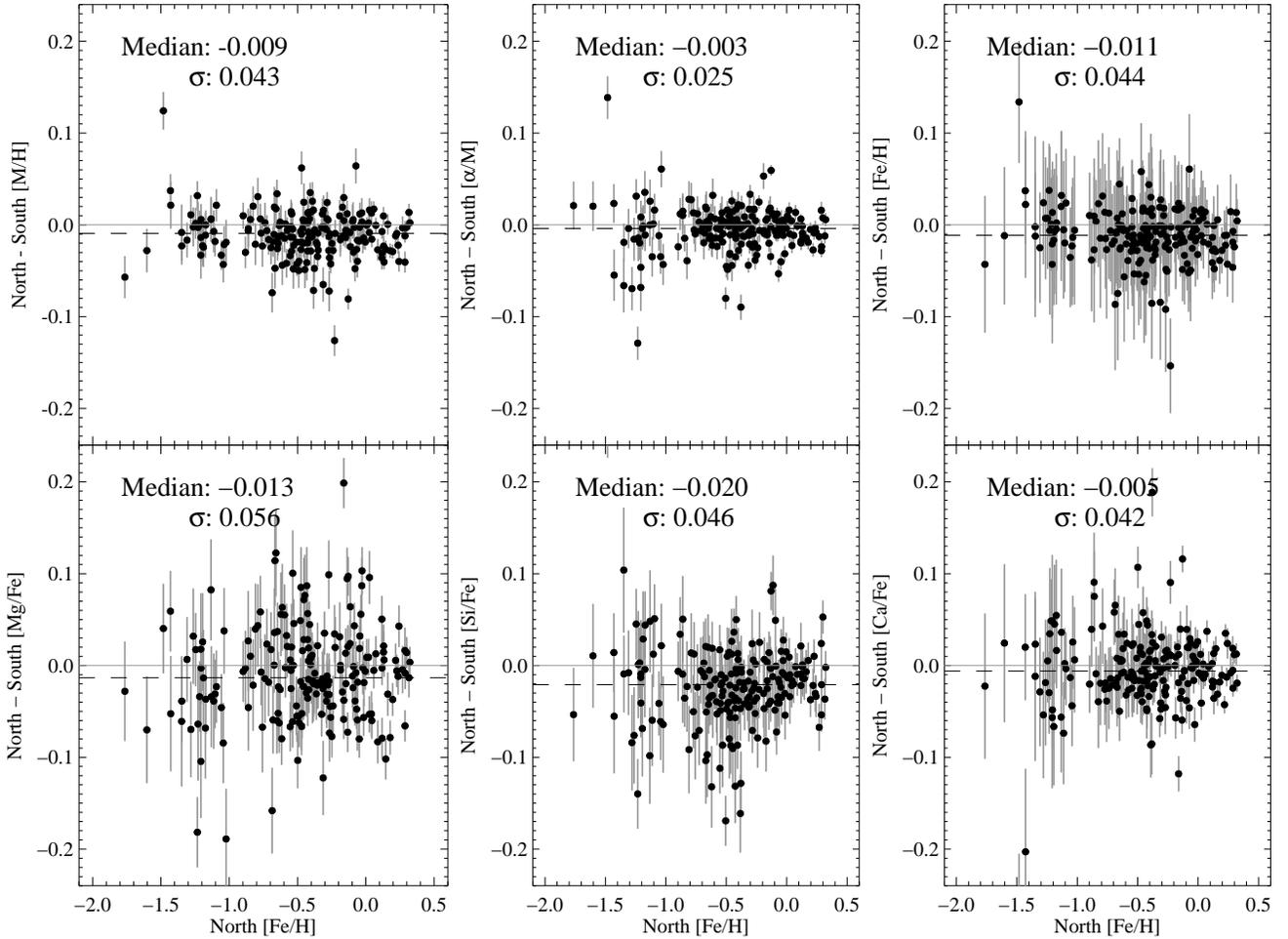}
\end{center}
\caption{Comparisons of the derived abundances for 203 stars observed from both the Northern and Southern telescopes. Medians and standard deviations of the abundance differences are indicated in each panel.}
\label{fig_northsouth}
\end{figure*}

% Separate target selection section
\section{Selection of Magellanic Cloud Member Stars}
\label{sec:memberselection}

While steps were taken to try to select as many MC RGB stars as possible (e.g., selection using the Washington photometry), it is likely that we have some MW contamination in the APOGEE MC fields, especially from MW halo stars. To refine our analysis sample from those stars that were observed, we use ASPCAP stellar parameters, radial velocity, and \emph{Gaia} DR2 \citep{GaiaDR2} proper motions (unavailable when the target selection was made) to select bona fide MC RGB stars. To this end, we required reliable stellar parameters for all the stars from the ASPCAP pipeline and \teff$<$5200 K and \logg $<$ 3.4.  The RV and proper motion selections are shown in Figure \ref{fig_velcuts}.  A Gaussian fit to the LMC RV distribution yields a mean of 254.3 \kms with $\sigma_{\rm V}$=37.8 \kmse, and we use lower/upper thresholds of 160/348 \kmse, which are roughly $\pm$2.5$\sigma$.  The same values for the SMC are mean/$\sigma_{\rm V}$=135.9/27.1 \kms with lower/upper thresholds of 71/220 \kmse.  Ellipses with the parameters, determined by eye, and shown in Figure \ref{fig_velcuts}, are used for the proper motion cuts.  Of the 2897/1363 LMC/SMC stars with S/N$\geq$20, 122/121 are pruned with the velocity cuts, or 4\%/9\% of the samples.
Figure \ref{fig_tefflogg} shows the \teffe--\logg distribution of the final LMC and SMC RGB samples color-coded by metallicity.
Our selection is higher up the giant branch than \citet{VanderSwaelmen2013} and \citet{Lapenna2012} studies, but includes large portions of the giant branch.
Table \ref{table_fields} gives information on all our fields and MC targets.

%We used measurements of metal-poor MW globular clusters to check the reliability of the metallicities and $\alpha$ abundances for the metal-poor stars.  To date, roughly %$\sim$900 northern APOGEE stars (DR14) have been reobserved at LCO.  The metallicities and $\alpha$ abundances of the two data sets compare very well with no 
%M4
%-[Fe/H]=-1.5, same as lit
%-[Si/Fe] st.dev ~ 0.1
%M10
%-[Fe/H]=-1.5, same as lit
%-[Si/Fe] std.dev = 0.06
%M55
%-[Fe/H]=-1.9 lit, -1.8 apg
%-[Si/Fe] std.dev = 0.08 for high S/N

\section{Quality Checks of APOGEE-S data}
\label{sec:qachecks}
This work is among the first to utilize data from the Southern Hemisphere component of APOGEE-2. Therefore, we perform a number of quality checks to verify the reliability of the spectroscopic parameters:

\begin{itemize}
    \item Comparison of abundances between stars observed from both the Northern and Southern APOGEE spectrographs.
    \item Comparison of the APOGEE cluster abundances to the optical cluster abundances of \citet{Carretta2009c,Carretta2009b}.
    \item Comparison of the APOGEE abundances to the abundances from GALAH DR2 \citep{Buder2018}.
    \item Comparison of APOGEE ASPCAP abundances to abundances derived from APOGEE spectra in a semi-independent way using BACCHUS \citep{Masseron2016}.
\end{itemize}

%[[Remove this part and just put in the north vs. south overlap? That's what the referee suggests.]]
%Because this work focuses on $\alpha$ abundance patterns, we first ensure that the conclusions one would draw about the MW [$\alpha$/Fe]-[Fe/H] abundance patterns are the same whether or not northern or southern data are used. In Figure \ref{fig_mw_comparison}, we show the [$\alpha$/Fe] vs. [Fe/H] distribution for a subsample of MW stars that have similar \teff and \logg distributions as the LMC stars (3700 $<$ \teff $<$ 4500 and \logg $<$ 1.8) for data taken from the north (left) and south (right). Because the southern survey is still largely incomplete, we also only compare similar regions of the MW disk. We only select stars that have $|b|$ $<$ 15$\degr$, and for the south we select 220$\degr$ $<$ $l$ $<$ 330$\degr$, and for the north we select 30$\degr$ $<$ $l$ $<$ 140$\degr$. We find a very slight ($-$0.035 dex) systematic difference in [$\alpha$/Fe] between the two hemispheres, but the overall trends are identical, indicating that the southern and northern APOGEE data are largely measuring $\alpha$-element abundances in the same way across the Galaxy. Still, when analyzing the LMC data, we only compare the LMC to other southern data to reduce the effect this small offset may have on our interpretations.

\subsection{Comparison of APOGEE-2N and APOGEE-2S}

First, we compare the derived abundances for APOGEE stars observed from both the Northern and Southern telescopes (none from the MC fields). We select overlap stars that have similar effective temperatures to the LMC (3750--4500 K) and S/N $>$ 70 from both spectrographs. The 203 stars that satisfy these criteria are shown in Figure \ref{fig_northsouth}. We find that the abundances agree within the reported uncertainties with no obvious trends in [Fe/H], but find that there is a small offset in [Si/Fe] of $\sim$ 0.02 dex.

\begin{figure*}[t]
\begin{center}
\includegraphics[trim={10cm 4cm 8cm 1cm},clip,width=1.0\hsize,angle=0]{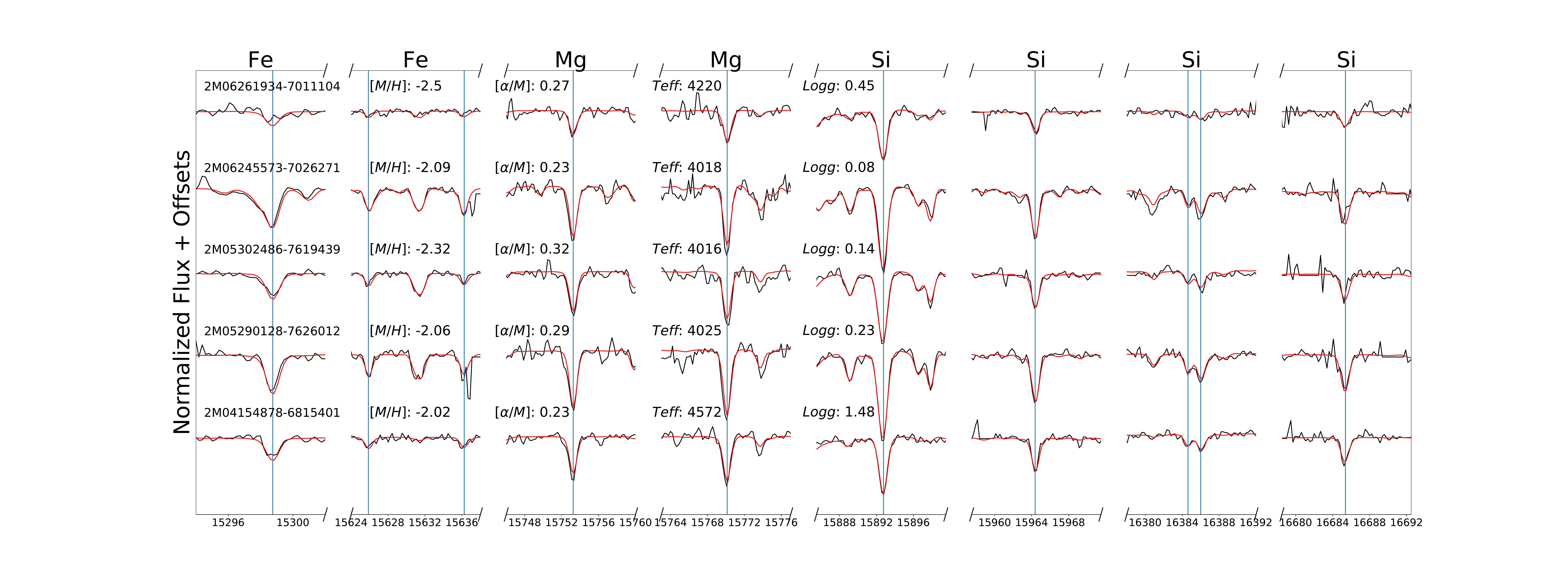}
\includegraphics[trim={0cm 0cm 17cm 0cm},clip,width=1.0\hsize,angle=0]{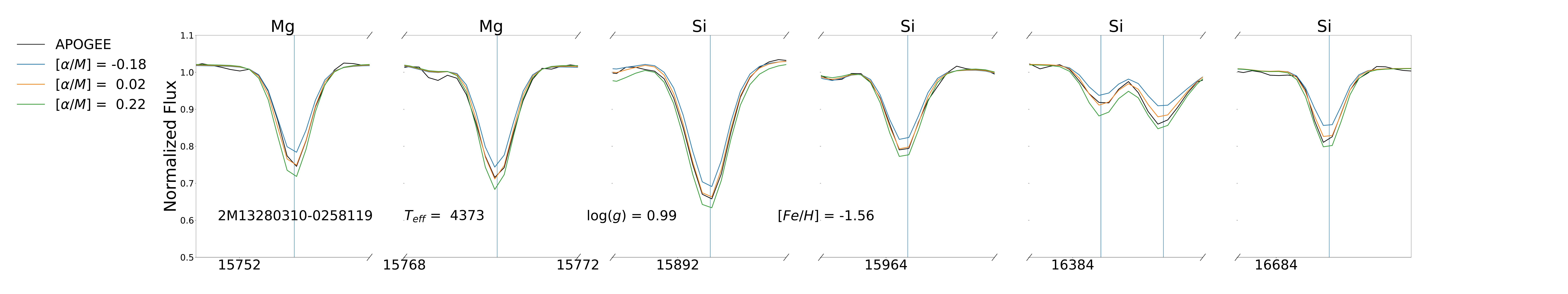}
%\includegraphics[trim={0cm 0cm 0cm 0cm},clip,width=1.0\hsize,angle=0]{overtop_features.pdf}
%\includegraphics[trim={0cm 0cm 0cm 0cm},clip,width=1.0\hsize,angle=0]{specfeatures_w_mgsi.pdf}
%  trim={<left> <lower> <right> <upper>}
\end{center}
\caption{({\em Top}) The most prominent abundance features of Fe, Mg and Si in {\bf 5} example metal-poor LMC stars (black) and the best-fitting ASPCAP synthetic spectrum (red). The central wavelengths of the lines are indicated by vertical blue lines. ({\em Bottom}) An example of a metal poor APOGEE DR16 spectrum focused on the most prominent Si and Mg lines indicating how they vary in strength with $\alpha$-abundance.  The APOGEE spectrum is shown in black, the best-fitting ASPCAP synthetic spectrum in orange, and synthetic spectra enhanced and depleted by 0.2 dex in [$\alpha$/M] (relative to the best-fitting parameters) shown in green and blue, respectively.}
\label{fig_metalpoorfeatures}
\end{figure*}

%While the bulk abundance patterns are the same for stars with $-$1.0 $<$ [Fe/H] $<$ +0.5, 
While the abundance comparisons across telescopes appear reasonable, the overlap sample has only a few stars with [Fe/H] $<$ $-$1.5. Therefore, it is also important to verify the quality of the spectra of the most metal-poor stars of our sample, as they drive our ability to measure the $\alpha$-poor knee. In the top panel of Figure \ref{fig_metalpoorfeatures}, we show the spectral features of Fe, Mg, and Si for 5 LMC stars with $-$2.5 $<$ [Fe/H] $<$ $-$2.0. This illustrates that there are plenty of measurable features in these southern spectra that are fit well by ASPCAP, and brings confidence that these are really metal-poor, $\alpha$-enhanced stars.  In addition, the bottom panel of Figure \ref{fig_metalpoorfeatures} shows how the strengths of the Mg and Si lines vary with $\alpha$-abundance for one metal-poor APOGEE star. Therefore, this indicates that APOGEE spectra are sensitive to $\alpha$ abundance even for these metal-poor stars.

%\begin{figure*}[t]
%\begin{center}
%\includegraphics[trim={8cm 5cm 8cm 4cm},clip,width=1.0\hsize,angle=0]{metalpoor_features_dr16beta.pdf}
%%  trim={<left> <lower> <right> <upper>}
%\end{center}
%\caption{The most prominent abundance features of Fe, Mg and Si in 10 example metal-poor LMC stars (black) and the best-fitting ASPCAP synthetic spectrum (red). The %central wavelengths of the lines are indicated by vertical blue lines. The stellar parameters for each star are listed above each spectrum. 
%%NEED TO INCREASE FONT SIZE.}
%}
%\label{fig_metalpoorfeatures}
%\end{figure*}
%For a complete analysis of the accuracy of APOGEE abundance results for metal-poor stars, see \citet{Meszaros2013}. [[Remember, however, that this paper makes it seem like the APOGEE metallicities at low metallicity are bad news bears. We need to mention something about the OH lines affecting the spectra? Here is what Szabolcs said: "at low metallicities may be related to the decreasing number of metal lines and increasing number of α-element lines (mostly OH), which leads to strong correlations between the two quantities (see Section 3.5)"]]

%\subsection{Cluster Abundance Comparison to Carretta et al.}
\subsection{Cluster Abundances}

\begin{figure}[ht]
\begin{center}
\includegraphics[width=1.0\hsize,angle=0]{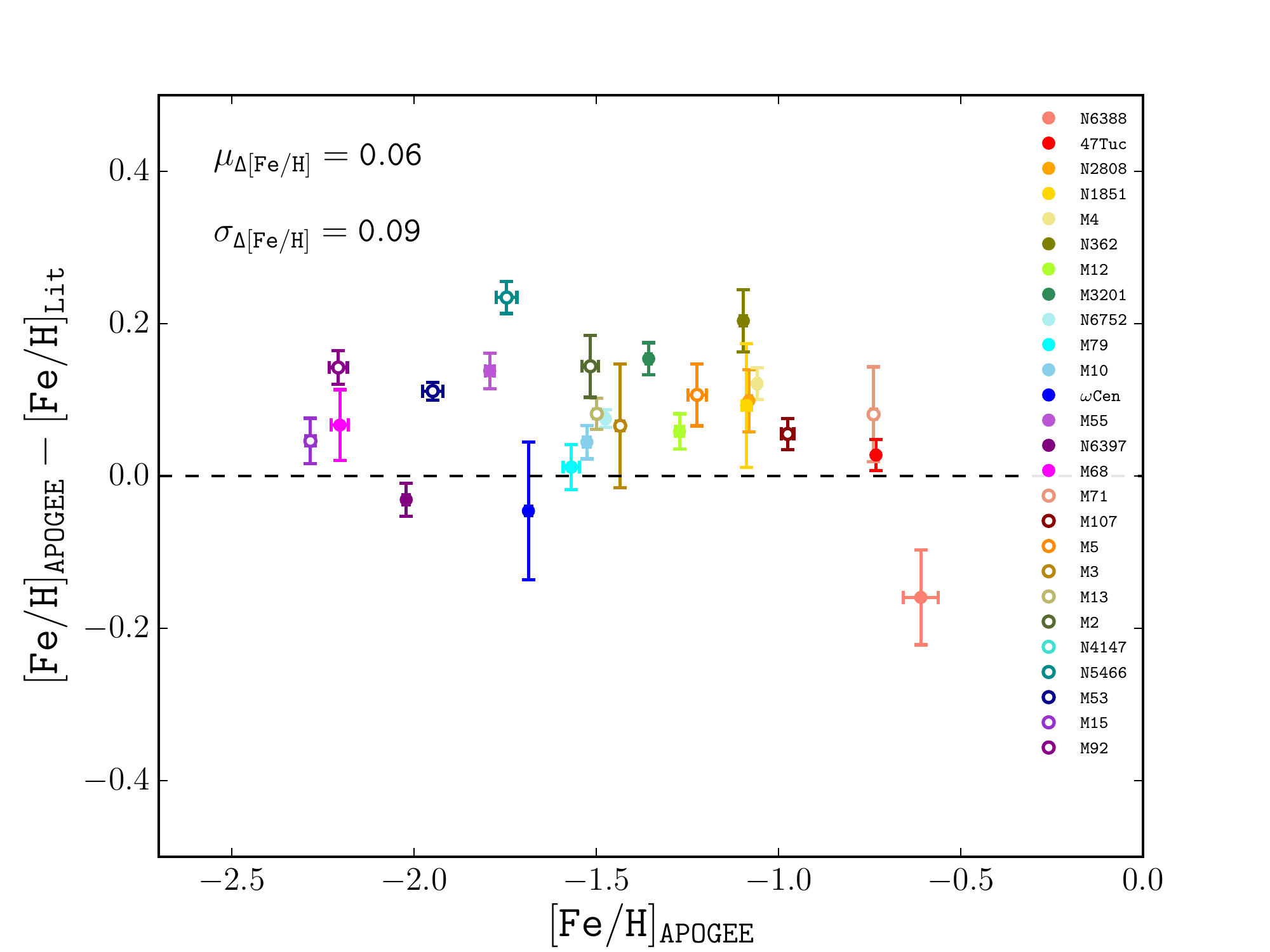}
\end{center}
\caption{A comparison of the mean [Fe/H] of the APOGEE-2 globular clusters observations (using first-generation globular cluster stars as described in the text) with Southern clusters (filled points) from \citet{Carretta2009c} and Northern clusters (empty points) compiled by \citet{Meszaros2013}.  While there are some deviations (e.g., NGC6388 and NGC6397) the measurements show an offset of 0.06 dex, with a scatter of 0.09 dex.}
\label{fig_fehcomparison}
\end{figure}

In addition, we compare the APOGEE metallicities of 26 globular clusters (15 from the South and 11 from the North) with measurements from \citet{Carretta2009c} and \citet{Meszaros2013} external globular cluster data. The differences in the mean metallicities are shown in Figure \ref{fig_fehcomparison}.  For each of these clusters, we select members using spatial and radial velocity cuts, and compute the reported mean [Fe/H] for each cluster and standard error of the mean from first-generation globular cluster members (defined as [Al/Fe] $<$ $+$0.4) with relatively high S/N ($> 70$), $3700 < T_{\rm eff} < 5500$K, and that are not flagged with poor spectra or stellar parameters.  The metallicities agree fairly well even to low metallicities, with a slight 0.06 dex offset between the two sets of measurements (APOGEE is more metal-rich than \citeauthor{Carretta2009c}).  This agrees largely with \citet[][using DR10 data]{Meszaros2013}, and \citet[][using DR14 data]{Masseron2019} although we find that the DR16 metallicities of the most metal-poor globular clusters (M92 and M15) agree better with optical studies than the previous APOGEE metallicities.

\begin{figure*}[t]
\begin{center}
\includegraphics[trim={0.5cm 1.5cm 0.5cm 3cm},clip,width=1.0\hsize,angle=0]{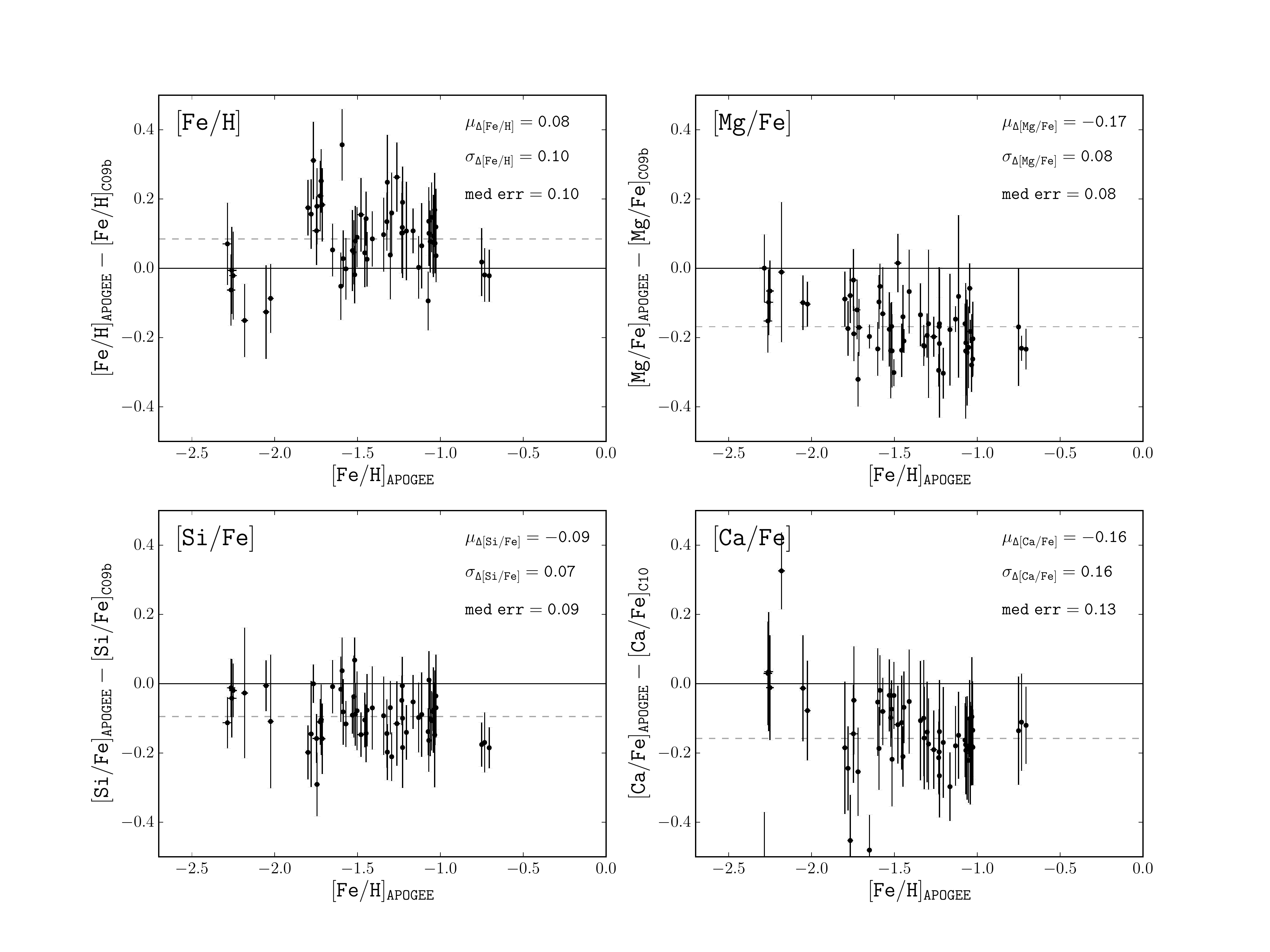}
%clusterstars_carretta09b_elemcomp_snr70_dr16_err_1g_ca.pdf}
%\includegraphics[trim={0.5cm 3cm 0.5cm 3cm},clip,width=1.0\hsize,angle=0]{clusterstars_carretta_no_o.pdf}
%\includegraphics[trim={0.5cm 3cm 0.5cm 3cm},clip,width=1.0\hsize,angle=0]{clusterstars_carretta_no_o_x_h.pdf}
%  trim={<left> <lower> <right> <upper>}
\end{center}
\caption{A comparison of the abundances ([Fe/H], [Mg/Fe], [Si/Fe] and [Ca/Fe]) from \citep{Carretta2009b} for first-generation globular cluster stars in 11 southern globular clusters. The scatter in the abundance measurements are consistent with the uncertainties but we see offsets of $+$0.08 dex in [Fe/H], $-$0.17 dex in [Mg/Fe], $-$0.09 dex in [Si/Fe], and $-$0.16 dex in [Ca/Fe].}
\label{fig_carrettacomparison}
\end{figure*}

\begin{figure*}[ht]
\begin{center}
\includegraphics[width=1.0\hsize,angle=0]{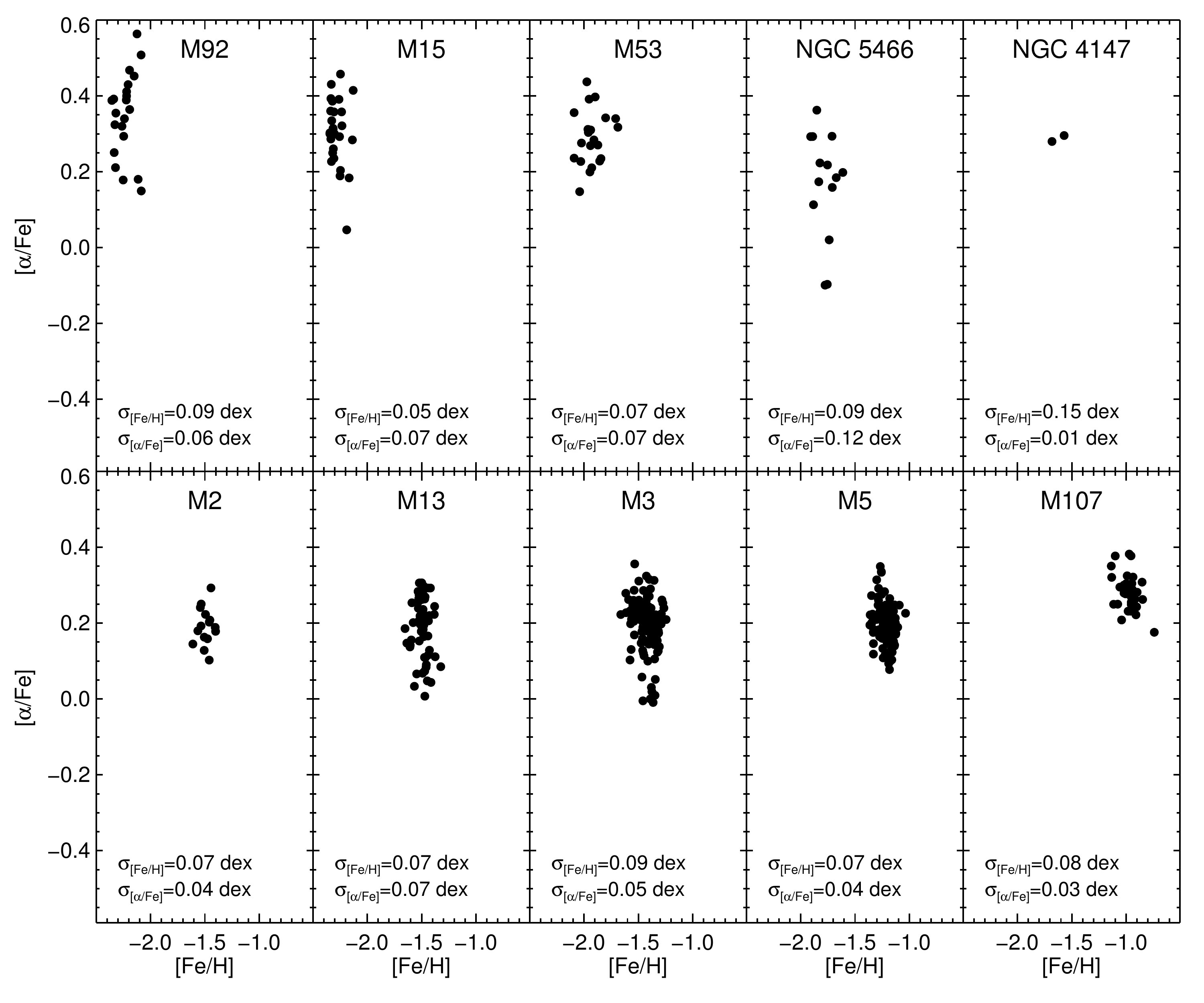}
\end{center}
\caption{The APOGEE DR16 [$\alpha$/Fe] versus [Fe/H] distributions of first generation stars ([Al/Fe]$<$0.4) of ten globular clusters.  While these cluster stars showed some anti-correlation of [$\alpha$/M] with [M/H] in \citet{Meszaros2013} for DR14, the removal of the second generation stars and using abundances relative to Fe essentially removes any correlation and gives small scatters.}
\label{fig_gc_alphafe}
\end{figure*}

Figure \ref{fig_carrettacomparison} compares individual [Fe/H], [Mg/Fe],  [Si/Fe], and [Ca/Fe] of 58 individual first-generation giants ([Al/Fe] $< 0.4$) in 11 Southern globular clusters observed by APOGEE to those measured by \citet{Carretta2009b}, using the same cuts on S/N, effective temperature, and flags employed to compare mean globular cluster metallicity above.  The scatter in the abundance differences are consistent with the measurement uncertainties, however there are offsets (across a wide range of metallicity) of $+$0.08 dex in [Fe/H] (consistent with the mean metallicity comparison), $-$0.17 dex in [Mg/Fe], $-$0.09 dex in [Si/Fe], and $-$0.16 in [Ca/Fe].

\citet{Meszaros2013} showed the APOGEE DR10 [$\alpha$/M] versus [M/H] distributions of stars from ten globular cluster and found there to be a correlation between [$\alpha$/M] and [M/H] (see their Figure 12).  This correlation still exists in DR16 but at a lower level.  The correlation is likely due to the fact that at low metallicities the [M/H] of second generation stars (which have high [Al/Fe]) is dominated by strong Al lines which pushes the [M/H] to higher values than those of their first generation (and low [Al/Fe]) counterparts.  As the second generation stars also have lower $\alpha$-abundances then the first generation stars, this effect causes a correlation between [$\alpha$/M] and [M/H].  The anti-correlation is mostly removed by using abundances relative to Fe rather than M, but not entirely.  The removal of the ``problematic'' second generation stars essentially eliminates the anti-correlation issue.  Figure \ref{fig_gc_alphafe} shows the APOGEE DR16 abundances of first generation stars of the same ten Meszaros et al.\ globular clusters (relative to Fe).  No anticorrelation is apparent and the scatters are small.  We note that the LMC stars have low [Al/Fe] and do not suffer from the globular cluster second generation star antic-correlation effect.
% The [alpha/M] vs. [M/H] anticorrelation is already not as bad in DR16 but definitely still there.

\subsection{Comparison to GALAH DR2}

\begin{figure*}[t]
\begin{center}
\includegraphics[width=1.0\hsize,angle=0]{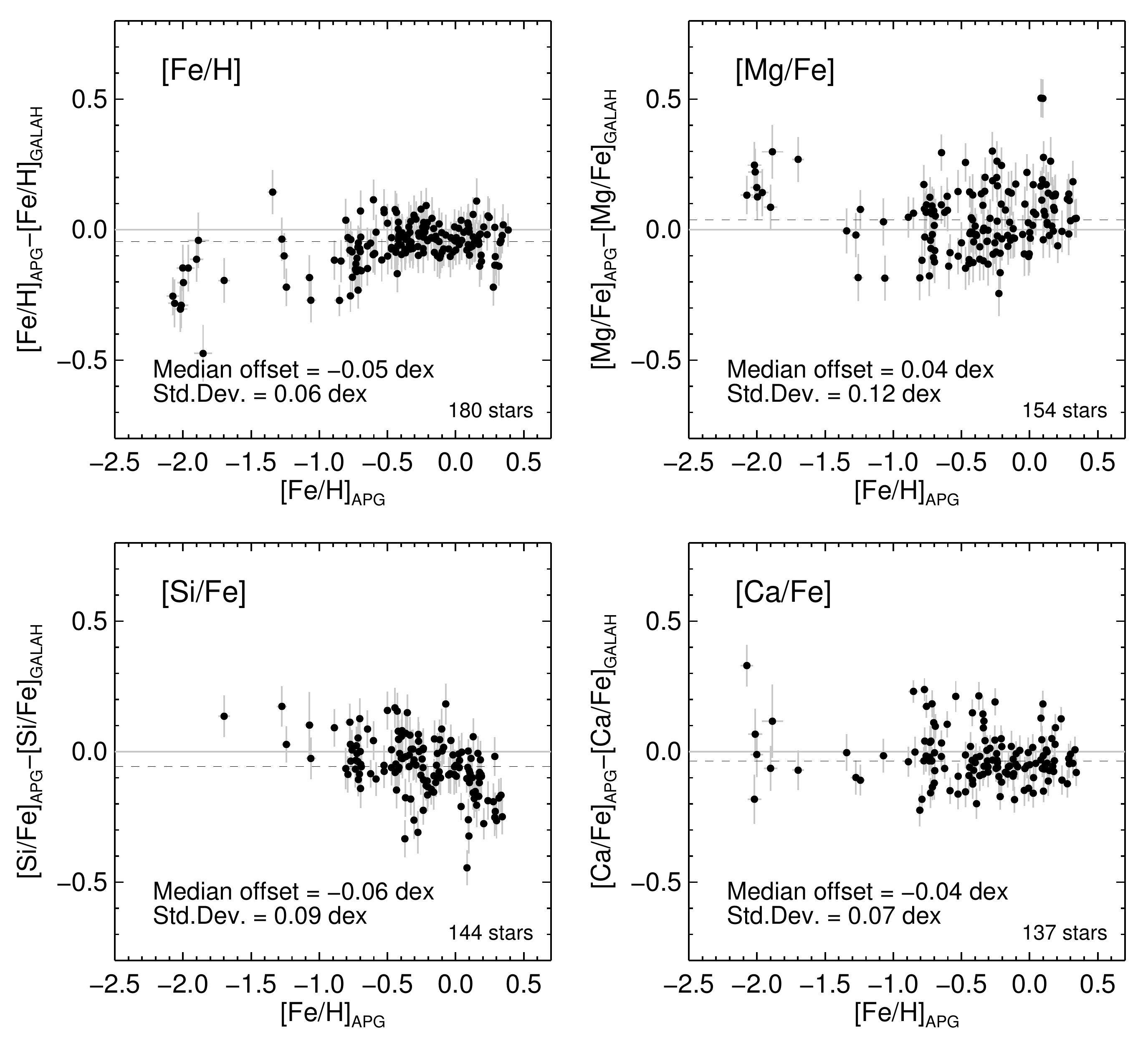}
\end{center}
\caption{A comparison of the abundances from the GALAH survey and the APOGEE-2S stars for [Fe/H], [Mg/Fe], [Si/Fe] and [Ca/Fe].  We select stars with S/N$>$60 in each survey, \teff $<$ 5000, no \texttt{STARBAD} flag set in APOGEE, and \texttt{cannon\_flag}=0 and \texttt{flag\_X\_fe}=0 for GALAH as recommended by \citet{Buder2018}.
Offsets and standard deviations are indicated in each panel.
}
\label{fig_galahcomparison}
\end{figure*}

Figure \ref{fig_galahcomparison} shows a similar comparison of the APOGEE-2S data with MW giant stars also observed by GALAH DR2 \citep{Buder2018}.  The selected stars have S/N$>$60 in both surveys, \teff$<$5000K, \logg$<$3.9, no \texttt{STARBAD} flag set in APOGEE, and \texttt{cannon\_flag}=0 and \texttt{flag\_X\_fe}=0 for GALAH, as recommended by \citeauthor{Buder2018}.  The stellar parameters span a range of 3900$<$\teffe$<$5000K, 0.8$<$\logge$<$3.9, and $-$2.1$<$[Fe/H]$<$ $+$0.3. The largest difference is the offset of approximately $-$0.2 dex in [Fe/H] at the metal-poor end ([Fe/H] $<$ $-$1.0).  This is opposite in direction of the smaller offset seen between APOGEE and \citet{Carretta2009c} in Figures \ref{fig_fehcomparison} and \ref{fig_carrettacomparison}.
GALAH silicon and calcium show slight offsets at the $\sim$0.05 dex level, but the scatter is consistent with the measurement uncertainties.  The [Si/Fe] offset is consistent with the one seen when comparing APOGEE and \citeauthor{Carretta2009b}, but smaller in magnitude.  The magnesium abundance offset is small, but in the opposite sense of the \citeauthor{Carretta2009b} comparison.

Based on these comparisons, the largest discrepancies are the [Mg/Fe] offset with \citeauthor{Carretta2009b} and the low-metallicity offset in [Fe/H] with GALAH.  However, we find that these offsets are opposite in sign across the two optical samples. Furthermore, \citet{Jonsson2018} found that magnesium was the most reliable APOGEE abundance element in DR14, with no offset and a scatter of 0.09 dex, when comparing to an optical study and constrained to [Fe/H] $>$ $-$1.  There might be small offsets in the APOGEE [Si/Fe] and [Ca/Fe] abundances, on the order of $\sim$0.05 dex, which would push the APOGEE $\alpha$-element abundance trends higher.

% Show figure of 
% check logg/teff overlap between APOGEE and vdS samples
% individual-element comparison 

%\subsection{Manual Analysis}
%\label{subsec:manualanalysis}
%
%{\bf Verne's manual analysis of a few stars with MOOG.}
%
%[[Do we still need this comparison?]]
\subsection{BACCHUS Analysis}
\label{subsec:BACCHUS}

We further check the ASPCAP $\alpha$-element abundances by completing a semi-independent stellar-abundance analysis of $\sim$90 stars randomly selected from the LMC sample, but ensuring good coverage of the metal-poor stars, and a comparable number of stars randomly selected from Milky Way field population observed with APOGEE. The abundance analysis was carried out using the Brussels Automatic Code for Characterizing High accUracy Spectra \citep[BACCHUS,][]{Masseron2016}. The BACCHUS analysis is ``semi-independent'' in that some stellar parameters (listed below) are fixed to the ASPCAP values. However, BACCHUS uses a more select list of lines to derive Fe and $\alpha$-element abundances, and generates new, self-consistent (in the abundances) synthetic spectra during the fitting process, in contrast to ASPCAP, which interpolates in a large-dimensional grid.
%utilizes a different method than what is used in ASPCAP for fitting the line abundances.

The current version of BACCHUS makes use of the the MARCS model atmosphere grid \citep{Gustafsson2008}, and the radiative transfer code TURBOSPECTRUM \citep{Alvarez1998, Plez2012} to generate synthetic spectra. The MARCS model atmosphere grid includes both carbon and $\alpha$-element enhancements.  Similar to \cite{Hawkins2016b}, we fix the \teffe\ and \logg\ stellar atmospheric parameters to the best-fit values from the ASPCAP pipeline (i.e., the uncalibrated values of \teffe and \logg, which can be found in the FPARAM column in the ASPCAP results table). In addition to \teffe\ and \logg, we also fix the C and N values to those found in the best fit from the ASCAP pipeline. The [Fe/H] and microturbulent velocity were then determined using the same procedure as in \cite{Hawkins2016b}. In short, microturblence was determined by ensuring no correlation between derived Fe abundance and reduced equivalent width (i.e., equivalent width divided by wavelength). The line selection for the various elements is the same as in \cite{Hawkins2016b}.

\begin{figure}[t]
\begin{center}
\includegraphics[width=1.0\hsize,angle=0]{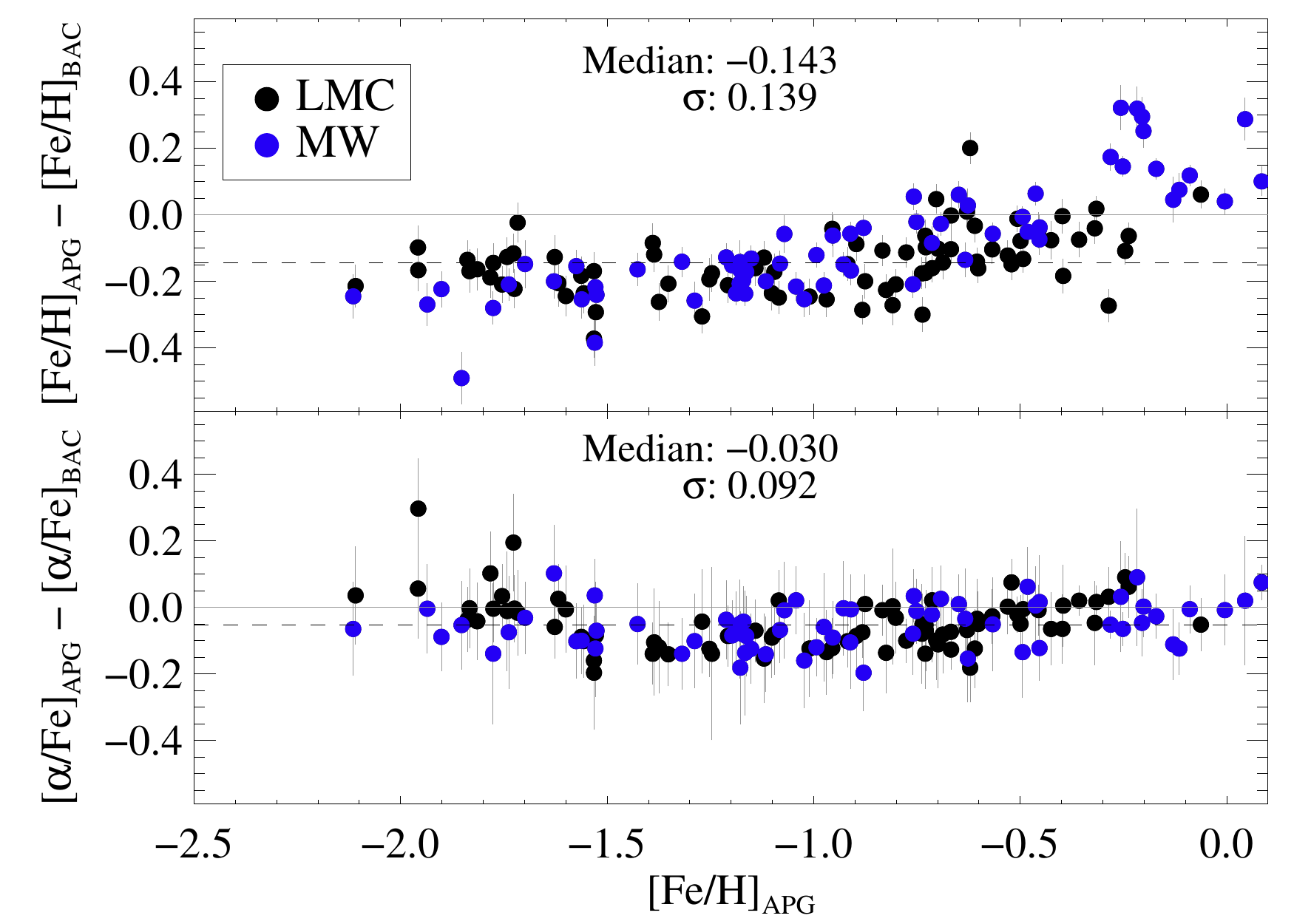}
\end{center}
\caption{(Top) The difference of ASPCAP minus BACCHUS-derived [Fe/H] as a function of ASPCAP [Fe/H] for $\sim$90 LMC (black circles), and MW (blue square) stars both observed from APOGEE-2 south.  The median offset is $-$0.143 dex with a scatter of 0.139 dex. (Bottom) Same as top panel but for the average [$\alpha$/Fe] ([Mg/Fe], [Ca/Fe], and [Si/Fe]) as a function of metallicity.  The median offset is $-$0.030 dex with a scatter of 0.092 dex.}
\label{fig:alphaBACCHUS}
\end{figure}

After fixing the  \teffe, \logg, [C/H], and [N/H], and deriving [Fe/H] and microturbulent velocity, the individual abundances for C, N, O, Mg, Si, Ca, and Ti were derived by using $\chi^2$ minimization between the observed spectra and the synthetic spectra. 
In the top panel of Figure~\ref{fig:alphaBACCHUS}, we show the difference of ASPCAP and BACCHUS-derived [Fe/H], as a function of ASPCAP [Fe/H],
%BACCHUS-derived [Mg+Si+Ca/Fe] as a function of [Fe/H]
for the LMC (black circles) as well as a Milky Way comparison sample also observed in the APOGEE-2S survey (blue squares).  In the bottom panel of Figure~\ref{fig:alphaBACCHUS}, we show the difference of ASPCAP and BACCHUS-derived [Mg+Si+Ca/Fe], as a function of ASPCAP [Fe/H], for the same stars as the top panel.  The median offset in [Fe/H] is $-$0.143 dex and $-$0.03 dex in [$\alpha$/Fe] for both the LMC and MW sample. We find that the results from ASPCAP appear to be robust and confirmed with a semi-independent BACCHUS analysis.
%
%{\bf Keith's BACCHUS analysis of $\sim$100 LMC and MW stars.}
%

% Talk about the data
% -put plot of alphas from N and S to demonstrate they are okay
% -
%Discuss the reliability of the southern data and abundances.
% south twin of north

%Show some of the metal-poor spectra.  We can see the Fe and alpha lines.

%Put in Verne's by-hand analysis of metal-poor stars.

\section{Results}
\label{sec:results}

%[[What's the best way to tell this story? I think we should have two subsections: one that describes the data, and then one that describes the chemical evolution modeling.]]

\subsection{$\alpha$-element Abundance Patterns of RGB stars in the Magellanic Clouds}
%Lead with the 6-panel (which I would argue should be only 4-panel) plot. We should probably also remove the Sgr stars from the MW sample as they stick out as a dense cloud below the LMC. We will introduce them later in the bottom panels. I will write this sections as if the 6-panel plot is a 4-panel plot, where the lower two panels are as is, but the upper panels are condensed together.

% alpha-knee
%Figure \ref{fig_trendline_mw} shows the $\alpha$ abundance distribution of our LMC stars (filled black circles) with the
%density of Milky Way stars in the background (orange) from the same southern APOGEE instrument.  Comparing the two populations
%qualitatively, they have very different distributions.  The LMC is more metal-poor reaching up to solar metallicity but having
%a low $\alpha$ abundance not unlike the low-$\alpha$ MW thin disk population.  In addition, a significant number of metal-poor ([Fe/H]$\lesssim$$-$1.5) LMC stars and the presence of a high-$\alpha$ plateau, i.e. the ``$\alpha$-knee",
%as revealed by the red trendline.  We do not find a similar knee in the SMC, as seen in the top right of Figure \ref{fig_alphafeh}, likely
%due to the lack of enough metal-poor stars. This implies that the knee in the SMC is at [Fe/H] $<$ 

\begin{figure*}[ht]
\begin{center}
\includegraphics[width=1.0\hsize,angle=0]{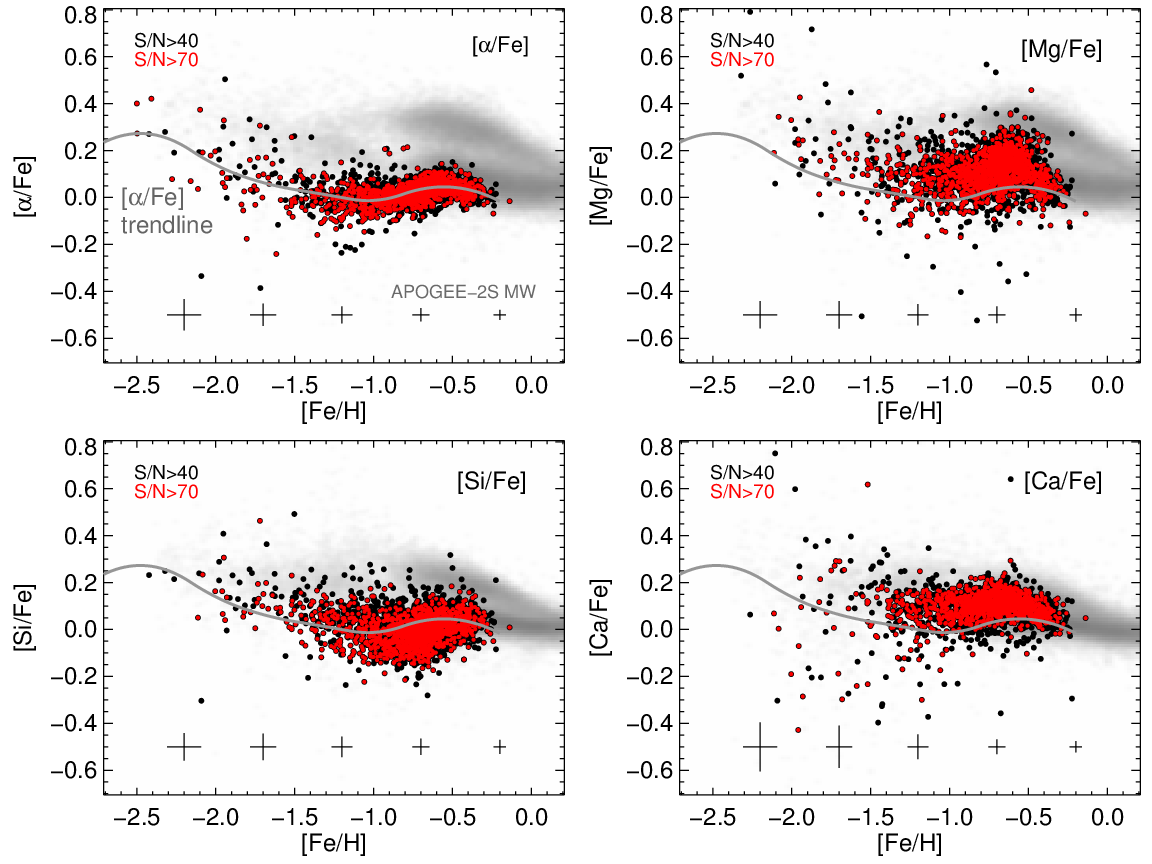}
\end{center}
\caption{The $\alpha$ abundances for the APOGEE LMC stars for S/N$>$40 (filled black circles) and S/N$>$70 (red filled circles).  The density of APOGEE-2S Milky Way disk field stars with similar \teffe, \logg and S/N as the LMC stars is shown in grayscale for reference.  The trendline of the parameter-level [$\alpha$/Fe] (upper-left panel) is shown
in each panel as a fiducial.  At the bottom of each panel are median ASPCAP errors bars for the S/N$>$40 sample}.
\label{fig_apogee_alphaelem}
\end{figure*}

Figure \ref{fig_apogee_alphaelem} shows the APOGEE ASPCAP parameter-level
%{\bf (from the 7-D synthetic grid parameter fit)}
$\alpha$-element abundance (parameter-level [$\alpha$/M] $+$ parameter-level [M/H] $-$ windowed [Fe/H]) distribution of the MC RGB stars.  The parameter-level [$\alpha$/Fe] (upper left), which fits all of the $\alpha$-elements (O, Mg, Si, S, Ca, and  Ti) simultaneously (keeping their relative abundances identical) in ASPCAP's 7-D synthetic grid parameter fitting, is more accurate than the individual $\alpha$-elements, but somewhat more challenging to interpret because the relative ``weight'' of different $\alpha$-elements changes with metallicity and \teffe.  Therefore, we compare the parameter-level [$\alpha$/Fe] to the windowed [Mg/Fe], [Fe/Si], and [Ca/Fe] abundances, shown in the other panels.  These $\alpha$-elements exhibit qualitatively similar abundance patterns to the parameter-level [$\alpha$/Fe], although with some small differences and offsets.  The general pattern seen in all four panels is that the LMC $\alpha$-element abundances are quite flat in metallicity from [Fe/H]=$-$1.7 to $-$0.3.  There is a slight increase in $\alpha$ at [Fe/H] $>$ $-$1.0, most clearly seen in the parameter-level [$\alpha$/Fe], [Mg/Fe], and [Si/Fe].  At the metal-poor end the $\alpha$-element abundances increase but do not reach their maximum until the most metal-poor stars at [Fe/H]$\approx$ $-$2.2.  Figure \ref{fig_apogee_alphaelem} also shows both the S/N$>$40 stars (black) and S/N$>$70 stars separately.  The two sets of stars exhibit nearly identical abundance patterns; this illustrates the reliability of the individual element $\alpha$-element abundances down to S/N=40, although perhaps slightly less so for Ca at the lowest metallicities.
% [[SH to DLN: can we use the erorr bars to comment on the "reliability"? Do those metal-poor stars at low-S/N have much larger error bars?]]

\begin{figure*}[ht]
\begin{center}
\includegraphics[width=0.45\hsize,angle=0]{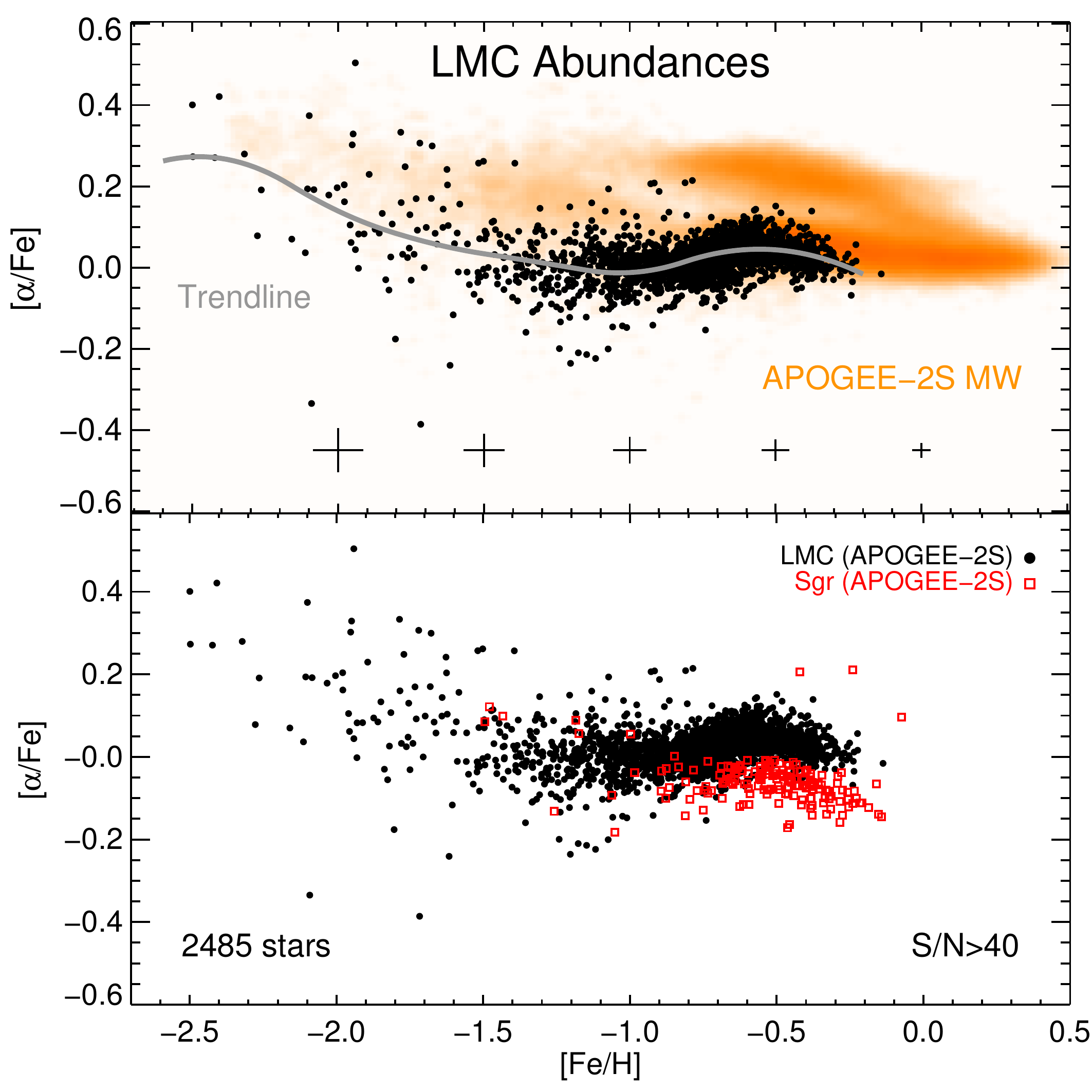}
\includegraphics[width=0.45\hsize,angle=0]{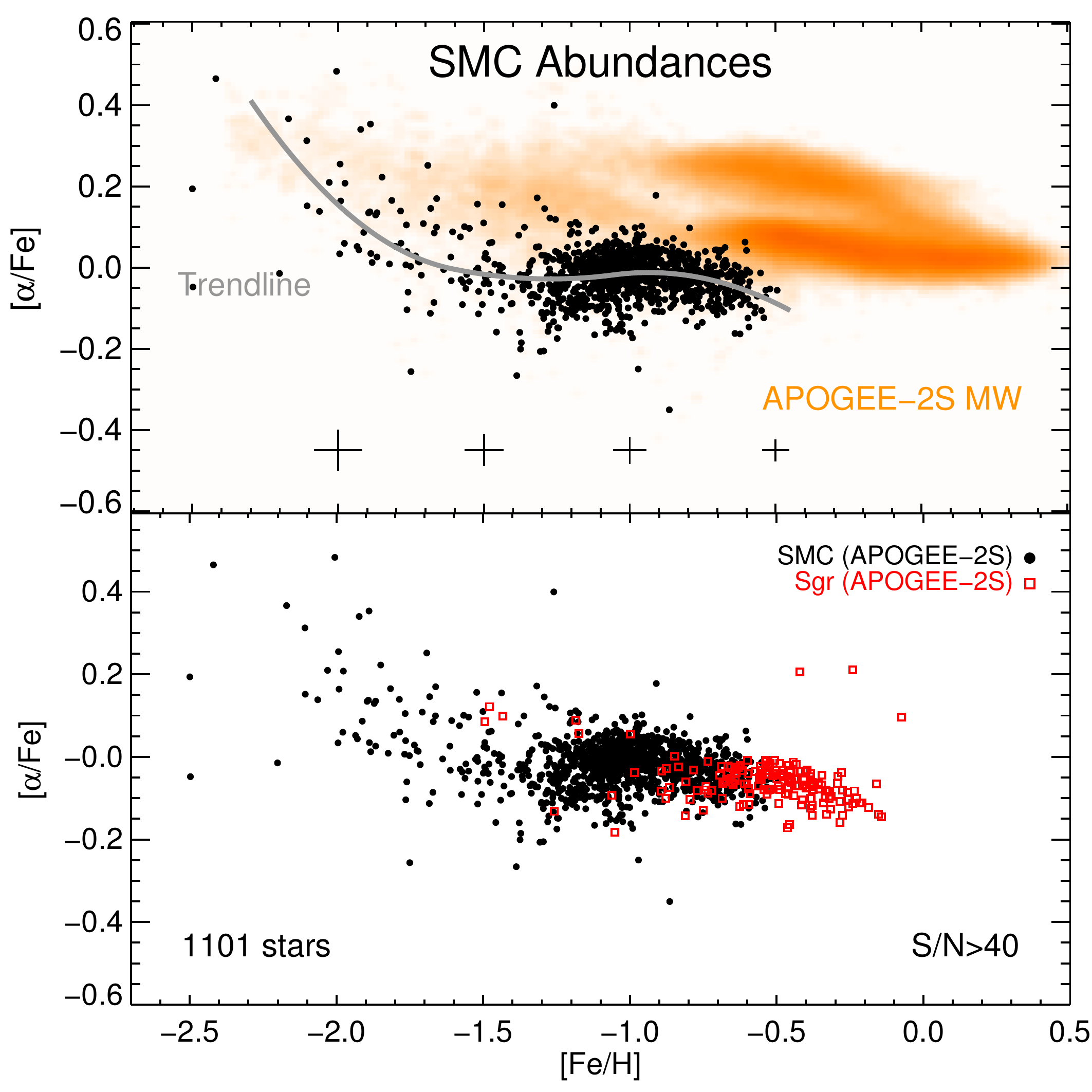}
\end{center}
\caption{The $\alpha$ abundances of the Magellanic Clouds ([$\alpha$/Fe] vs.\ [Fe/H]). ({\em Left}) LMC and ({\em Right}) SMC with the APOGEE MC as black filled circles. ({\em Top}) The density of APOGEE-2S MW stars (orange), and a B-spline trendline for the MC stars (gray). ({\em Bottom}) comparison of the LMC/SMC to Sagittarius (APOGEE-2S, open red squares). Median uncertainties from ASPCAP are shown in the top panels.}
%\caption{The $\alpha$ abundances of the Magellanic Clouds ([$\alpha$/Fe] vs.\ [Fe/H]). ({\em Top}) LMC with APOGEE data in black and %\citep{vanderswaelmen2013} in purple. ({\em Bottom}) SMC APOGEE data.}
\label{fig_alphafeh}
\end{figure*}

Now that we have verified that the parameter-level [$\alpha$/Fe] values yield abundance patterns consistent with those of the individual $\alpha$-elements and are reliable to low S/N, henceforth we use the more precise parameter-level [$\alpha$/Fe].  In the top two panels of Figure \ref{fig_alphafeh} we show the $\alpha$-element abundance distribution of both the LMC and SMC in comparison to those for MW stars, and in the bottom two panels we compare our MC data to other dwarf galaxies abundance data from the literature. % We slightly reduce the minimum S/N to 40 as this enables us to pick up more metal-poor stars and allows us to determine the $\alpha$-plateau better at the metal-poor end.
%In the top two panels of Figure \ref{fig_alphafeh} we show the $\alpha$-element abundance distribution of the MC stars (filled black circles) as compared to the density of MW stars in the background (orange).

Both the APOGEE MC and MW stars come from the Southern APOGEE instrument. While both MCs differ from the MW, in that they are generally more $\alpha$-poor at fixed [Fe/H], the MCs also differ from each other. We find that the LMC reaches metallicities as high as [Fe/H]$\approx$ $-$0.2, whereas the SMC is only as metal-rich as [Fe/H] $\approx$ $-$0.5. We also appear to be lacking the most metal-poor stars in our SMC sample, likely due to the (current) lower S/N in the SMC sample compared to the LMC, because the stars are $\sim$0.4 mag fainter. However, we are able to measure the increase in [$\alpha$/Fe] (with decreasing [Fe/H]) and constrain the $\alpha$-knee in both the LMC and SMC to [Fe/H] $\sim$ $-$2.2.  Although there is no clear evidence yet of a high-$\alpha$ plateau in either galaxies.

%[[I think we want to remove this figure, but then change the 6 panel figure. It should be just 4 panels, where the top two panels are combined.]]
%\begin{figure*}[ht]
%\begin{center}
%\includegraphics[width=1.0\hsize,angle=0]{lmc_alphafefeh_mw.eps}
%\end{center}
%\caption{The $\alpha$ abundances for the LMC with trendline and a comparison to the MW stars.}
%\label{fig_trendline_mw}
%\end{figure*}

%\begin{figure}[ht]
%\begin{center}
%\includegraphics[width=1.0\hsize,angle=0]{highalpha_dr16beta.pdf}
%\end{center}
%\caption{The kinematical distribution of the 53 metal-rich, high-$\alpha$ LMC stars (red filled circles) compared to that of the full LMC sample.  The top panel shows the chemical selection while the bottom panels show the Gaia DR2 proper motion and APOGEE radial velocities.  The kinematics of the high-$\alpha$ stars (2.7\% of the sample) are consistent with those of the bulk of the LMC stars.}
%\label{fig_highalpha}
%\end{figure}

In the bottom two panels of Figure \ref{fig_alphafeh} we show the $\alpha$-element abundance distributions of the MCs as compared to other MC spectroscopic samples: individual stars from \citet[][purple diamonds]{VanderSwaelmen2013} and individual cluster stars as compiled in \citet[][blue crosses]{Sakari2017}. They agree reasonably well with the APOGEE results at higher metallicities, but the \citeauthor{VanderSwaelmen2013} sample does not cover the low-metallicity region as our data. The \citeauthor{VanderSwaelmen2013} stars do seem to contain a larger fraction of $\alpha$-element enhanced LMC stars than the APOGEE sample. We find that the APOGEE stars largely follow a tight sequence in [$\alpha$/Fe]-[Fe/H] space, with only a few stars scattering to super-solar $\alpha$-element abundance. This discrepancy could be due to the fact that the \citeauthor{VanderSwaelmen2013} stars are at the center of the LMC, with half of their stars belonging to the bar. In our APOGEE sample, we have only a small fraction of stars that overlap spatially with the \citeauthor{VanderSwaelmen2013} sample, and likely have (relatively) very few bar stars.  We further discuss the comparison of our APOGEE abundances of the LMC with literature values below in \S \ref{subsec:lmclitcomparison}.

\begin{figure*}[ht]
\begin{center}
$\begin{array}{cc}
\includegraphics[width=0.5\hsize,angle=0]{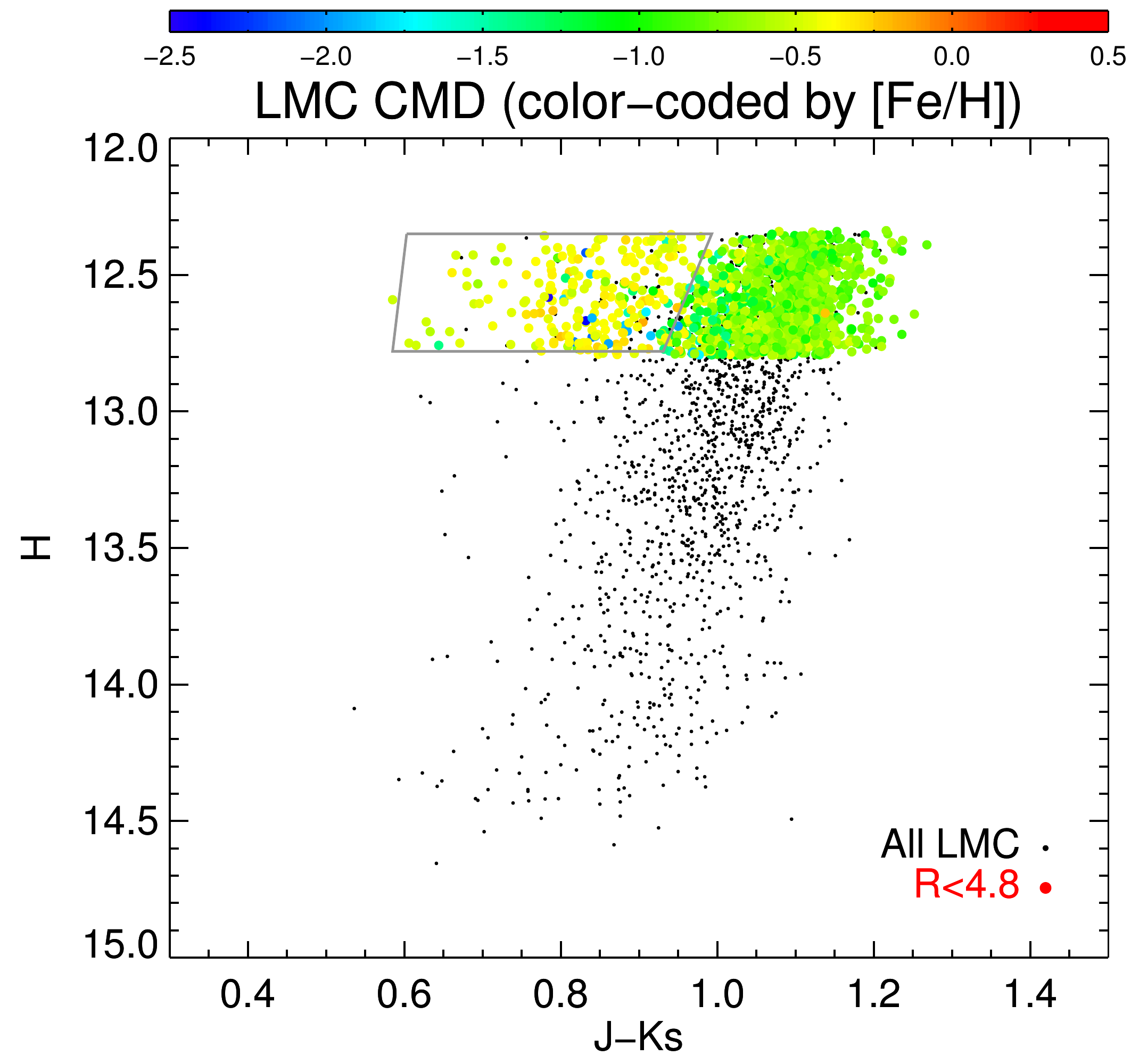} &
\includegraphics[width=0.5\hsize,angle=0]{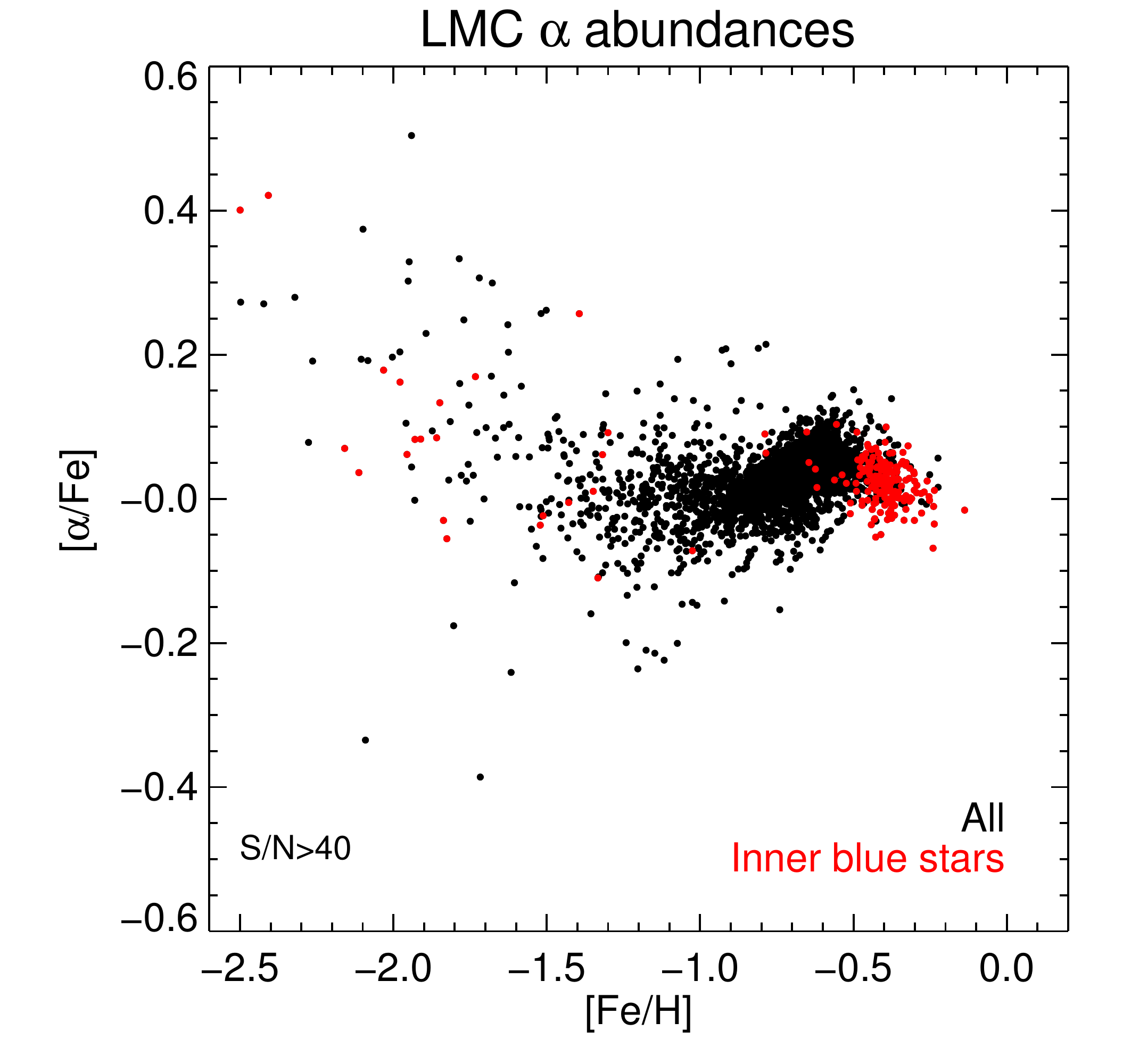}
\end{array}$
\end{center}
\caption{The young metal-rich, blue LMC population. (Left) The color-magnitude diagram of all the APOGEE-selected LMC stars.  The inner fields ($R$ $<$4.8\degr) are shown as larger symbols and are color-coded by [Fe/H].  The gray polygon more or less selects the inner, metal-rich blue stars.  (Right) The [$\alpha$/Fe] vs.\ [Fe/H] abundances for the high-S/N (greater than 50) LMC stars, with the stars selected in the left panel marked as red filled dots.  These stars are predominantly metal-rich, and include the most metal-rich stars in the LMC sample.  Of the 125 LMC stars with [Fe/H]$>$ $-$0.30, 90 of them are in this young population (72\%), and a significant fraction of the rest are just redward of the selection box.  This supports the notion that the most metal-rich population are young (blue-loop) stars in the inner region of the LMC.}
\label{fig_youngmetalrich}
\end{figure*}

The abundances for Fornax from \citet[][shown as green crosses in the bottom two panels of Figure \ref{fig_alphafeh}]{Hendricks2014} have a plateau at slightly
higher metallicity than that for the LMC. The APOGEE-2S Sagittarius (Sgr) sample (shown as red open squares in Figure \ref{fig_alphafeh}) exhibits sub-solar $\alpha$-element abundances that have been interpreted as being due to a top-light IMF \citep{Hasselquist2017}. We do not measure the position of the metal-poor knee in the Sgr data, but literature works find a knee in the range of $-$1.5 $\lesssim$ [Fe/H] $\lesssim$ $-$1.2 \citep{deBoer2014,Carlin2018}, all more metal-rich than the LMC, even though these galaxies were thought to once have similar stellar mass and have both enriched themselves to [Fe/H]$\approx$0.0.

\begin{figure*}[t]
\begin{center}
\includegraphics[width=1.0\hsize,angle=0]{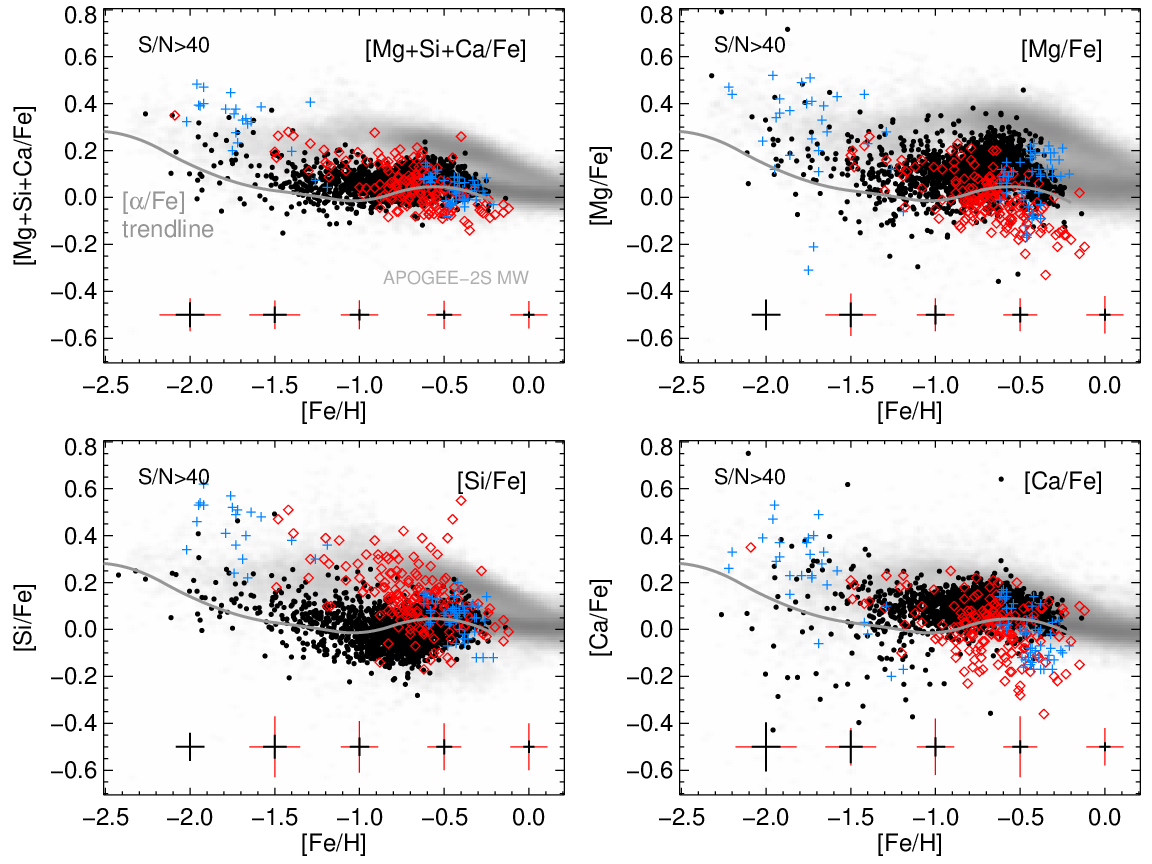}
\end{center}
\caption{A comparison of the $\alpha$-element abundances for the APOGEE LMC stars (filled black circles) to those of \citet{VanderSwaelmen2013}
(open red diamonds) and LMC clusters \citep[blue crosses; compiled by][]{Sakari2017}.  The APOGEE-2S MW distribution for each element is shown in the
grayscale in the background for reference.  At the bottom of each panel the APOGEE abundance uncertainties are shown in black and the \citeauthor{VanderSwaelmen2013} uncertainties in red.}
\label{fig_apogee_alphaelem_sakarivds}
\end{figure*}

\begin{figure}[t]
\begin{center}
\includegraphics[width=1.0\hsize,angle=0]{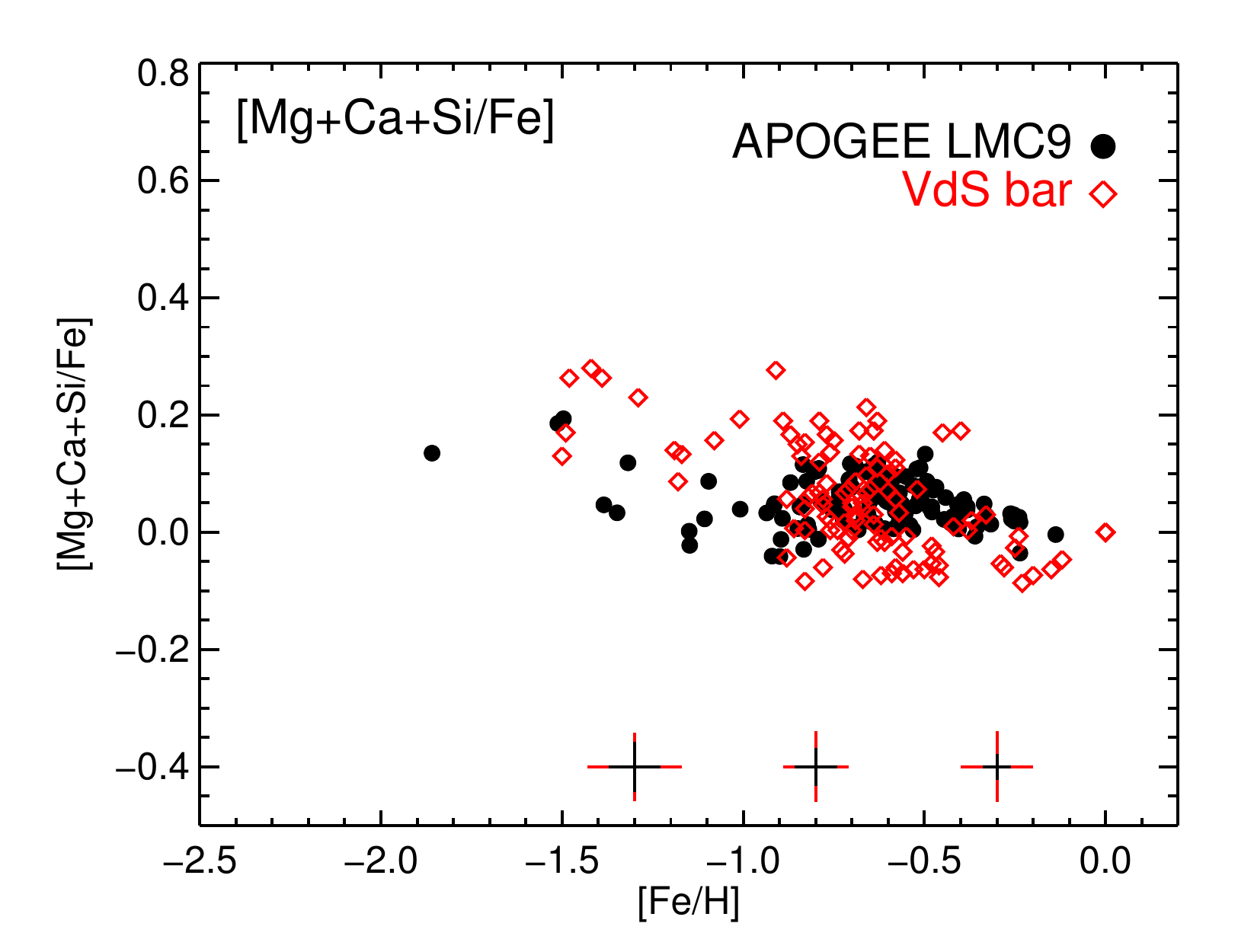}
\end{center}
\caption{A comparison of the $\alpha$-element abundances ([Mg+Ca+Si/Fe]) of 129 APOGEE LMC stars with S/N$>$40 in the central field LMC9 (filled black circles) to the 113 ``bar'' field stars from \citet{VanderSwaelmen2013} (open red diamonds). The distributions of the two samples are very similar with only a slight offset in $\alpha$-abundance of $\sim$0.1 dex at the metal-poor end.}
\label{fig_vds_lmc9_comparison}
\end{figure}

%[[What about high-alpha SMC stars??]]
% 2 panel figure, all LMC and metal-rich high-alpha stars
% (left) PMDEC vs. PMRA
% (right) Vhelio vs. LMCRAD

% rise in alpha
Additionally, we see a rise in $\alpha$-element abundance with metallicity in the LMC, beginning at [Fe/H] = $-$1.0. This implies that after the turn on of Type Ia SNe in the LMC, which lowered the [$\alpha$/Fe] abundance to near-solar at [Fe/H] = $-$1.5, the LMC may have experienced an increase in the number of Type II SNe that were able to enrich the ISM with $\alpha$-elements. An increase in the Type II SNe can be due to a starburst, where the LMC quickly begins to form stars at a much faster rate. We also observe a slight turnover in [$\alpha$/Fe] at [Fe/H] $\sim$ $-$0.5, which could be a sign of the Type Ia SNe exploding from the stars formed at the beginning of the starburst, adding Fe to the ISM of the LMC at the exclusion of the $\alpha$-elements. We further explore this starburst scenario in \S \ref{subsec:cheme}.

%SH: Added Hendricks et al. 2014 reference to end of this paragraph. Any 
%other references?
The SMC does not exhibit an increase of [$\alpha$/Fe] with increasing [Fe/H] beyond the apparent knee position, but it does exhibit a flat pattern starting at [Fe/H] $\sim$ $-$1.5, with perhaps a slight decrease beginning at [Fe/H] $\sim$ $-$0.7. Flat $\alpha$-element abundance patterns can also be indicative of a starburst, or series of starbursts, but these starbursts are not sufficiently powerful to substantially enrich the gas already present with $\alpha$-elements (e.g., \citealt{Hendricks2014}).

Because of our blue color-cut, applied in order to prevent a metallicity bias in our LMC RGB sample, we have also serendipitously observed young blue-loop, metal-rich stars, as identified in the CMD in Figure \ref{fig_youngmetalrich} (left panel). While a detailed analysis of the reliability of the blue-loop abundances is beyond the scope of this paper, these stars appear to be among the most metal-rich stars in the sample.  This fits the scenario where the LMC is currently forming stars (in the inner regions where we observe these blue-loop stars) at a metallicity of [Fe/H] = $-$0.3.

%[[Rephrase this? Goes something like, "It turns out some of our RGB target candidates are blue loop stars."]]
%Finally, we note that our most metal-rich LMC stars are likely very young, blue loop stars.  Figure \ref{fig_youngmetalrich} shows the CMD of all APOGEE-observed LMC stars (left panel).  The inner fields (R$<$4.8\degr) are color-coded by [Fe/H] and shows that the most metal-rich stars are the bluest stars.  This is counter-intuitive because metal-rich RGB stars are generally redder.  The right panel shows [$\alpha$/Fe]--[Fe/H] distribution of these stars (red) compared to that of the rest of the stars, and makes clear that they are the most metal-rich stars in the entire LMC sample.  The properties of these stars are consistent with young and metal-rich blue loop stars that recently evolved off of the main-sequence and that are known to be present in the central regions of the LMC where active star formation is currently ongoing.

\subsection{Comparison to Previous LMC Studies}
\label{subsec:lmclitcomparison}

%[[SH to DLN: In this section we need to be careful about what agrees "Well" and what doesn't. I think that having the error bars on the optical studies will help out, but I think we are way too hopeful in this paragraph when discussing how things agree.]
Previous $\alpha$-element abundance studies of the LMC \citep[e.g.,][]{Smith2000,Pompeia2008,Lapenna2012,VanderSwaelmen2013,Sakari2017} did not find the relatively flat $\alpha$-element distribution at [Fe/H] $>$ $-$1.0, or the metal-poor $\alpha$-knee that we constrain with our APOGEE dataset.  To further investigate this discrepancy, Figure \ref{fig_apogee_alphaelem_sakarivds} shows a comparison of the APOGEE LMC $\alpha$-element abundances with those of \citet{VanderSwaelmen2013}, in red open diamonds, and the LMC cluster star abundances compiled by \citet{Sakari2017} in blue crosses.  The abundances of all $\alpha$ elements are fairly consistent at the metal-rich end.  In addition, the [Ca/Fe] and [Mg/Fe] abundances agree well between the three datasets. 
%DLN: I don't think we need these two sentences below now.
%The APOGEE and \citeauthor{VanderSwaelmen2013} [Mg/Fe] abundances agree within the scatter of each dataset, although the average trends are slightly different.  
%The latter sample has no stars below [Fe/H]=$-$1.5 but there are metal-poor clusters that extend to [Fe/H]=$-$2.2 and they show an increase in $\alpha$-abundance very similar to what is seen in the APOGEE data, although shifted slightly more metal-rich. 
The [Si/Fe] abundances are the most disparate with the literature values being higher at the metal-poor end than for APOGEE.  As we previously showed in Section \ref{sec:qachecks}, our [Si/Fe] abundances compare well to \citet{Carretta2009b}, with a constant offset but no trends with metallicity (Figure \ref{fig_carrettacomparison}).
% but with the comparison to GALAH DR2 suggesting (Figure \ref{fig_galahcomparison}) that our [Si/Fe] values are too {\em high} at the metal-poor end. 
%This is inconsistent with the offsets suggested by the \citeauthor{VanderSwaelmen2013} and cluster results (i.e., APOGEE [Si/Fe] too low by $\sim$0.3 dex at the lowest metallicities).
It is challenging to say which [Si/Fe] abundances are correct in this instance, however we do note that the abundance patterns in APOGEE are more similar to the $\alpha$-element abundance patterns in other elements, whereas the \citet{VanderSwaelmen2013} Si abundances are slightly different from their other $\alpha$-elements.  Finally, the abundance trends of the [Mg+Si+Ca/Fe] average values (upper-left panel of Figure \ref{fig_apogee_alphaelem_sakarivds} are qualitatively quite similar between the three datasets (although the APOGEE metallicities are a bit more metal poor), suggesting that they are reliable.

To make a clearer comparison between our data and that of Van der Swaelmen et al.\ we plot the [Mg+Ca+Si/Fe] abundances of our central field (LMC9; 129 stars) and the 113 ``bar'' stars from Van der Swaelmen et al.\ in Figure \ref{fig_vds_lmc9_comparison}.  The distributions compare very well in both their mean values and ranges in both [Fe/H] and the $\alpha$-abundances.  There is a slight offset in the mean $\alpha$-abundance of $\sim$0.1 dex at the metal-poor end.

%NLTE effects not large.

Although APOGEE data sample more luminous (lower \logg) giants than the \citeauthor{VanderSwaelmen2013} sample, the \teffe--\logg distributions overlap considerably for \logg$>$0.7.  We investigated whether the abundances of the most luminous giants could be systematically off by splitting the APOGEE sample at \logg=0.7, but found only minor differences in the $\alpha$-element abundances of the two subsets at the metal-poor end. 
%If anything, the less luminous giants (\logg$>$0.7) were shifted slightly to the metal-poor end.
%DLN: I think the above sentence was a mistake on my part.
Therefore, we suspect that the APOGEE abundances are not affected by our sampling higher up the giant branch.
%[[SH to DLN: Again, are we saying that the lower gravity stars are more metal-poor? Extend to lower metallicities?]] DLN: done

Besides the effects described above, it's important to keep in mind some differences between the APOGEE and the \citeauthor{VanderSwaelmen2013} samples.
The APOGEE dataset is significantly larger (by $\sim$12$\times$) and it has much more spatial coverage to larger radii, which allows us to detect more metal-poor stars. In addition, our parameter-level [$\alpha$/Fe] abundances are quite precise, even down to S/N=40.  The median precision at S/N$>$70 and [Fe/H]=$-$1.5/$-$0.7 is 0.033/0.021 dex and at 40$<$S/N$<$60 and [Fe/H]=$-$1.5/$-$0.7 is 0.038/0.024 dex.
The combination of these factors likely allows us to discern abundance features not previously seen by previous studies ($\alpha$-knee at [Fe/H] $<$ $-$2.0 and $\alpha$-element increase at [Fe/H] $>$ $-$1.0), especially at the metal-poor end.

%\begin{figure}[ht]
%\begin{center}
%\includegraphics[width=1.0\hsize,angle=0]{mc_mdf_snr10.png}
%\end{center}
%\caption{The metallicity distribution function of the LMC and SMC.}
%\label{fig_mcmdf}
%\end{figure}

%Figure X shows the metallicity distribution function (MDF) of $\sim$3800 APOGEE-2S RGB stars (with S/N$>$10) in the
%Magellanic Clouds.  The LMC distribution peaks at [Fe/H]$\approx$$-$0.5 which compares well to previous measurements
%\citep[e.g.,][]{Pagel1998,Cole2005,vanderswaelmen2013}.  The skew-negative LMC distribution with a long, metal-poor tail is similar
%expected distribution from a classical one-cell galactic chemical evolution model \citep[e.g.,][]{Andrews2017} indicative of the galaxy's
%long period of past and on-going star formation.  The SMC distribution peaks at a much lower metallicity of [Fe/H]$\approx$$-$1.1 which
%is also consistent with literature values.  However, it's distribution is more complex than the LMC's with both metal-poor and metal-rich
%tails, suggesting a more complex chemical evolution history.

%Figure X shows the metallicity distribution function (MDF) of the Magellanic Clouds.

\subsection{Chemical Evolution Model}
\label{subsec:cheme}

%Outline
%- Reason for chemical evolution modeling
%	- Begin to unravel the chemical evolution history and SFH of the clouds? (LMC definitely)
%- Models we use
%	- flexCE models
%	- important parameters
%- Modeling the metal-poor end
%- modifying flexCE
%	- past studies suggest a recent increase in star formation in the LMC
%    - chemical abundances consistent with an increase in Type II/core collapse supernovae
%    - try to model a burst or increase in SFR by turning up the SFE (``suddenly easier to form stars''), perhaps through interaction
%    - mention that the models assume Kennicutt-Schmidt SF, but this may not quite be right for dwarfs (interactions may bump this up)
%    - mention that this is just a model, and many other factors play in (inflow from SMC during a possible interaction, strongly changing potential and maybe outflows, etc.)
%- reproducing the metal-rich end
%	- mention degeneracies (abundances depend on SFR, but varying peak location, duration and intensity can still produce similar SFH), so the best fit is not unique (require precise SFHs to start to get an idea of what the specific conditions were)
%- results (or the meaning of the results)
%	- discuss final SFH and MDF (MDF of APOGEE has not been accounted for selection function (maybe this is something we could do?))
%- Comparison with lit

Using this large sample of MC stars with precise chemical abundances we can begin to detect signatures of their detailed chemical evolution and start to unravel their SFHs.  To do so, we use chemical evolution models to assist in interpreting the LMC $\alpha$-element abundance pattern.  To produce the chemical evolution models used here, we use the flexCE\footnote{flexCE is available for public download at \url{http://bretthandrews.github.io/flexce}} code \citep{Andrews2017}, which is a one-zone, open-box, chemical evolution modeling program, utilizing nucleosynthetic yields for core-collapse SNe from \citet{Chieffi2004} and \citet{Limongi2006}, Type Ia SNe from \citet{Iwamoto1999}, and AGB stars from \citet{Karakas2010a}. To match the observed chemical-abundance patterns in the MCs the flexible chemical evolution modeling from flexCE allows us to vary several parameters pertaining to star formation, such as the initial gas mass, inflow rate, and time dependence, the outflow mass-loading parameter ($\eta$), and SFE.  Because the present SMC sample is limited to relatively lower S/N, we focus here on the chemical evolution modeling of the LMC, and leave the SMC for future work when the continuing APOGEE-2S observations improve the S/N of the SMC stars.

Fiducial chemical evolution models with parameters like SFE that are constant over time typically have difficulty in fitting the flat or increasing $\alpha$-element abundance patterns over $\sim$1 dex in metallicity, as seen in the LMC at metallicities [Fe/H] $\gtrsim$ $-$1.5.  So we first attempt to generate models that can match the LMC [$\alpha$/Fe]--[Fe/H] abundance pattern at lower metallicities, in the range $-$2.5 $<$ [Fe/H] $<$ $-$1.2. The parameters we adopt for the chemical evolution models that are held constant are:
\begin{itemize}
\item Initial gas mass, $M_{gas, \, i} = 2 \times 10^{9} \, M_{\odot}$;
\item Inflow rate according to a delayed tau model ($t\,e^{-t/\tau}$), with a mass normalization of $M_{1} = 6 \times 10^{10} \, M_{\odot}$, and an inflow time scale of $\tau_{1} = 2$ Gyr;
\item Outflow mass-loading factor $\eta = 10$;
\item Fiducial IMF \citep[Kroupa;][]{Kroupa2001} and Type Ia SNe delay-time distribution from \citet{Andrews2017}.
\end{itemize}
This initial mass and mass normalization is about one-tenth (modified slightly to alter the ratio of initial gas mass to the inflow mass) of the fiducial model parameters used by \citet{Andrews2017} to model the MW.  Motivated by cosmological hydrodynamical simulations
\citep[][]{Simha2014}\footnote{However, \citet{Simha2014} found, and we discuss below, that the tau model inflow rate with a constant SFE cannot alone reproduce the necessary SFH of late star-forming galaxies like the MCs.}, we use a delayed tau model for its exponential decline in inflow rate at late times (with a delayed start to the inflow) and reaches a maximum later.
%We use a delayed tau model for its exponential decline in inflow rate at late times, but does not turn on inflow immediately and reaches a maximum later and is motivated by cosmological hydrodynamical simulations \citep[][although as found by this study and discussed below, this inflow rate with a constant SFE cannot alone reproduce the necessary SFH of late star forming galaxies like the MCs]{Simha2014}.
A shorter inflow timescale helps produce a slightly steeper [$\alpha$/Fe] slope, but we choose an inflow timescale long enough such that the peak inflow of gas happens after a few billion years rather than immediately.  We use a outflow mass loading factor of 10, which also steepens the [$\alpha$/Fe] slope and is roughly consistent with values found in simulations showing that MW mass galaxies have $\eta$ of order unity, whereas SMC-mass galaxies have $\eta \sim 10-20$ \citep[e.g.,][]{Hopkins2013}.
%SH: I am pretty happy with our choice for eta, but Barger et al. 2016 have 
%measured the outflow rate of the LMC, and her value implies an eta
%of something like 2-4. Not to say it couldn't be higher in the past,
%but maybe we should address this?

\begin{figure}[t]
\begin{center}
\includegraphics[width=1.0\hsize,angle=0]{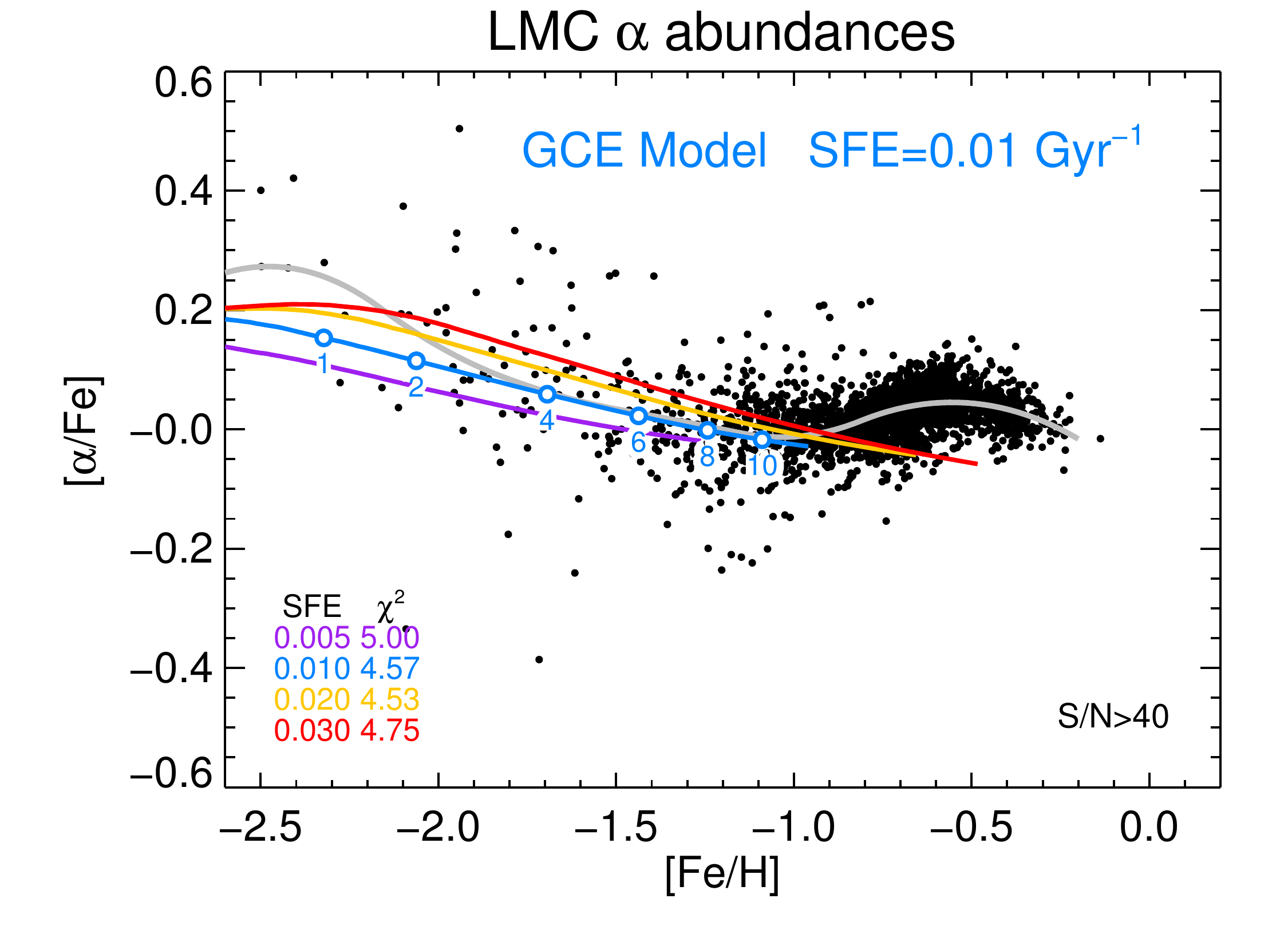}
\end{center}
\caption{The $\alpha$ abundances for the LMC with various galactic chemical evolution model sampling a range of SFE.  The model $\alpha$ abundance is the mean of [Mg/Fe], [O/Fe], [Ca/Fe] and [Si/Fe].  The best-fitting model to the metal-poor stars ([Fe/H]$<$$-$1.2) has SFE=0.01 Gyr$^{-1}$. Certain times are indicated by open circles.  None of these models match the flat (and rising) $\alpha$ distribution at the metal-rich end.}
\label{fig_gcemodel}
\end{figure}

We then alter the remaining parameter, the SFE, to generate a series of models whose SFE varies from 0.005 Gyr$^{-1}$ to 0.03 Gyr$^{-1}$, as shown in Figure \ref{fig_gcemodel}.  Comparing to the observed abundance trend in the LMC, we find that the model with SFE = 0.01 Gyr$^{-1}$ returns the lowest $\chi^{2}$ for the data in the range $-$2.5 $<$ [Fe/H] $<$ $-$1.2.

\begin{figure}[ht]
\begin{center}
\includegraphics[width=1.0\hsize,angle=0]{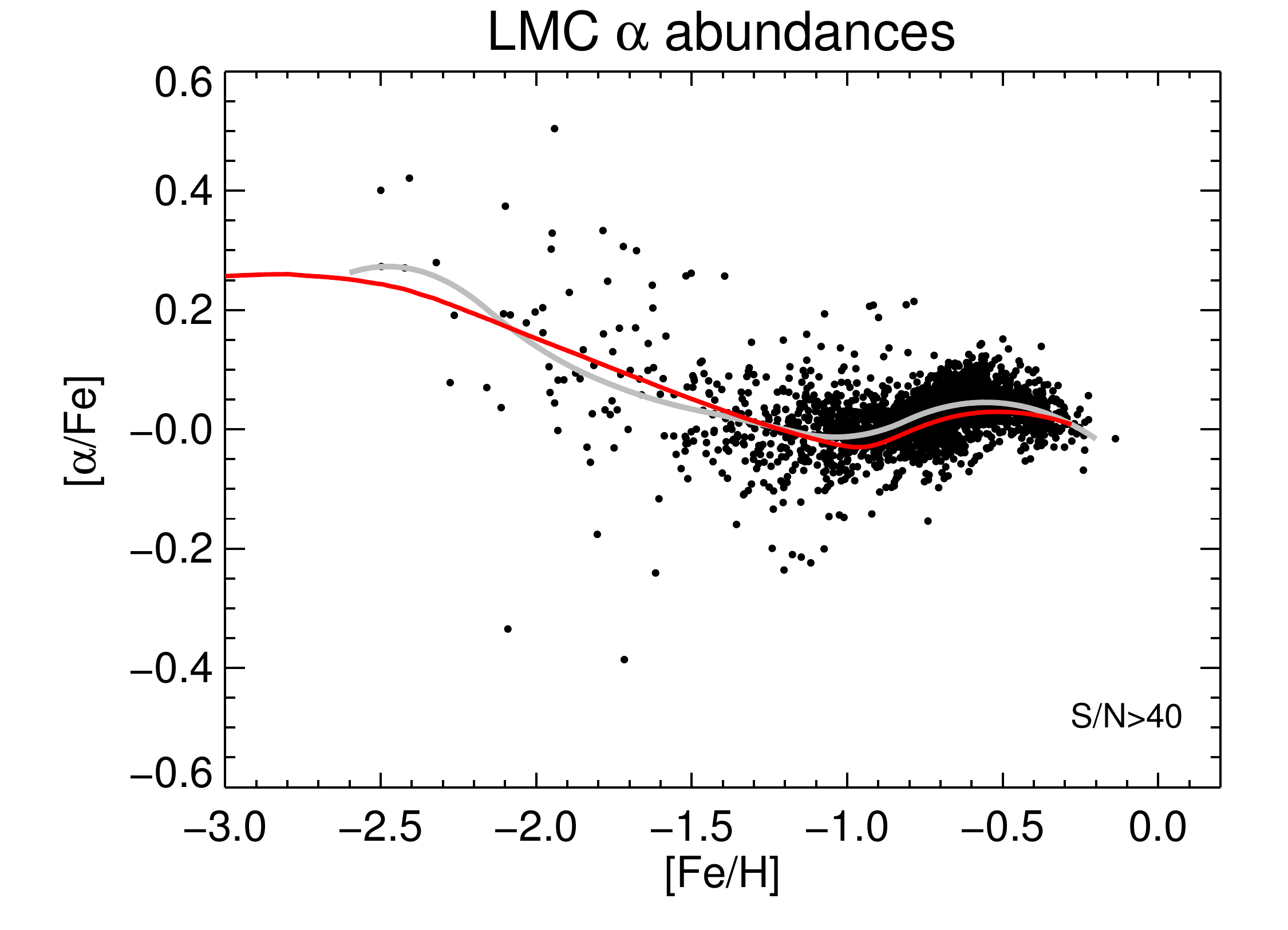}
\includegraphics[width=1.0\hsize,angle=0]{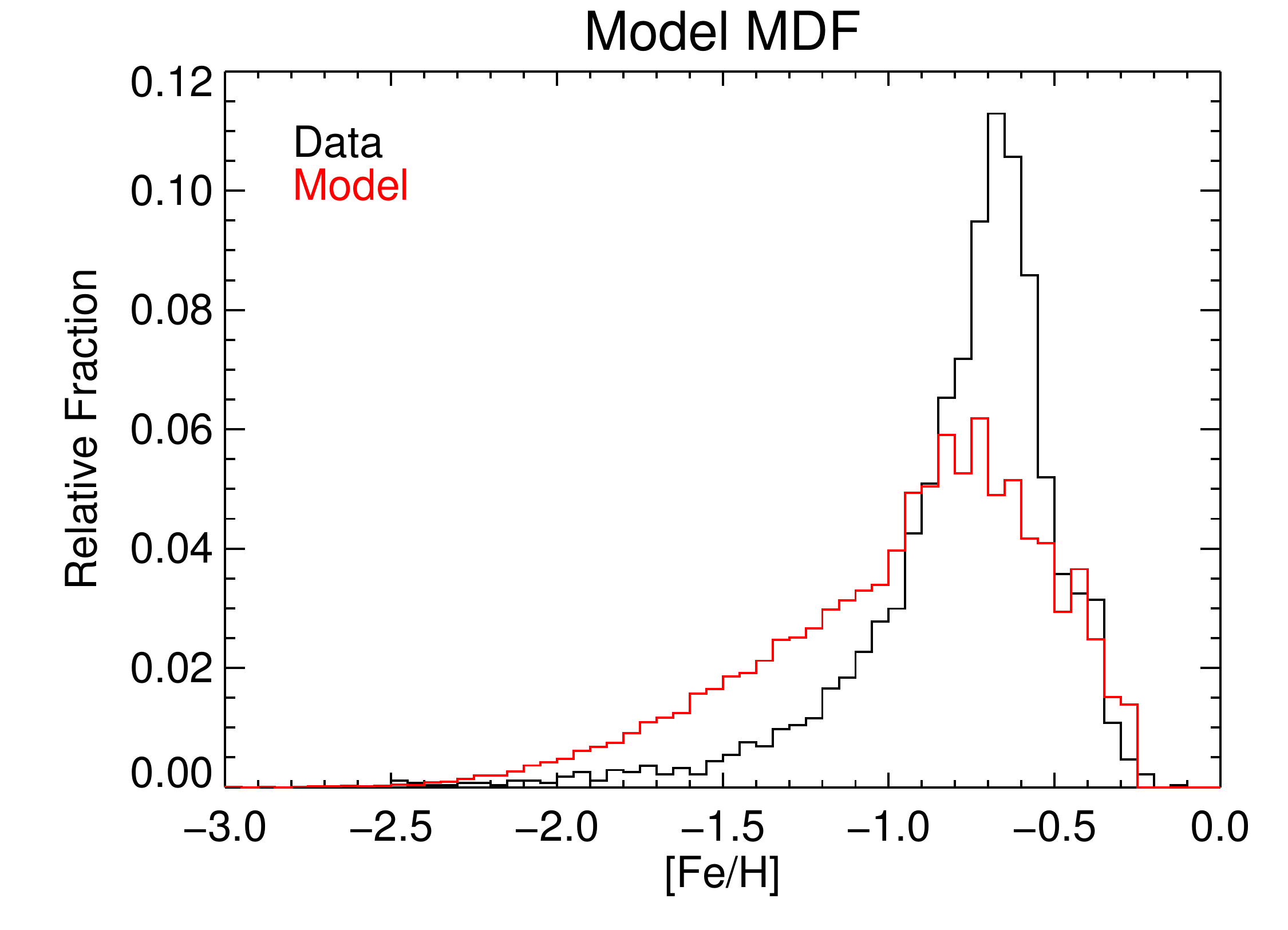}
\includegraphics[width=1.0\hsize,angle=0]{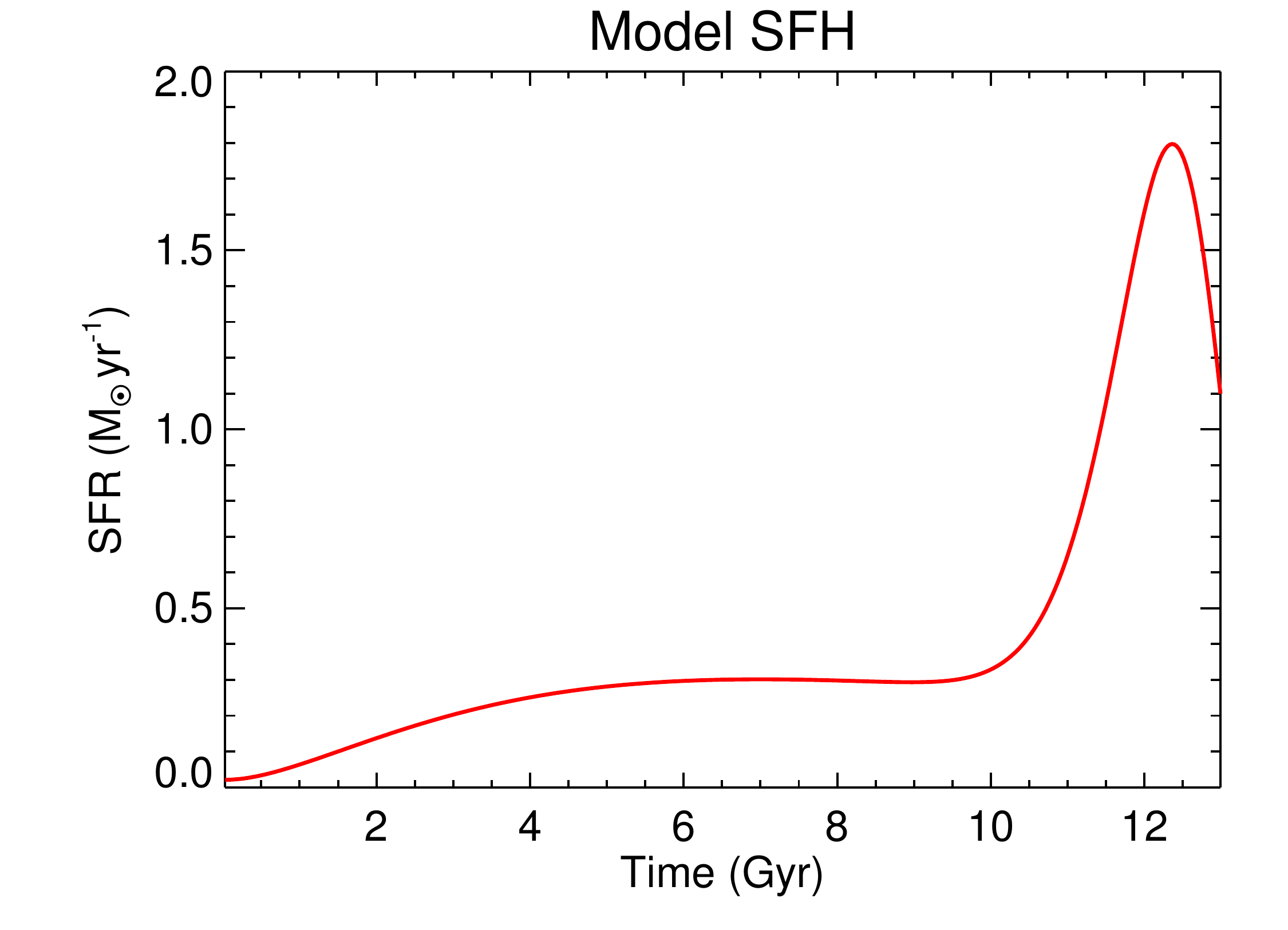}
\end{center}
\caption{Comparison of our best-fitting galactic one-zone chemical ``starburst'' evolution model with a recent starburst.
(Top) The $\alpha$ abundances for the LMC (black dots) with a trendline (gray) and chemical evolution model (red).
(Middle) The metallicity distribution function of the data (black) and model (red).  The model matches the shape
of the data well but is slightly too metal-poor by$-$0.15 dex.
(Bottom) The star formation history of the chemical evolution model indicating the rapid increase in the star formation
rate from $\sim$0.24$\times$10$^{-10}$ \msun yr$^{-1}$ pc$^{-2}$ for the first $\sim$10 Gyr to 2.1$\times$10$^{-10}$ \msun yr$^{-1}$ pc$^{-2}$
(an increase of $\sim$8.75$\times$) at the peak of the ``starburst" over the last couple Gyrs.}
\label{fig_gcemodel_burst}
\end{figure}

Despite the fit to the metal-poor stars, most of these models barely enrich to [Fe/H] $\sim$ $-$1.0, and none reproduce the flat or increasing $\alpha$-element abundance pattern seen in the LMC at higher metallicities.
%This is due largely to the fact the adopted SFE implies a gas consumption timescale of 100 Gyr. Clearly, this initially low SFE was kickstarted at some point in the LMC's history.
%Previous studies have attempted to model the SFH of the LMC and explain the $\alpha$-element abundance pattern, and using Bekki \& Tsujimoto 2012, modified by the SFH of the LMC measured by (reference et al. 2009), invoked a starburst model in which the LMC undergoes a flurry of star formation beginning some 2 Gyr ago. In their models, they noticed that their starburst models predicted an increase in [$\alpha$/Fe] from -1.0 $<$ [Fe/H] $<$ -0.5 followed by a slight decrease from -0.5 $<$ [Fe/H] $<$ 0.0. This is qualitatively similar behavior to what we see in the LMC $\alpha$-element abundance patterns.
An increasing $\alpha$-element abundance pattern at these metallicities is, however, not entirely unexpected.  \citet{Bekki2012} modeled the LMC $\alpha$-element abundance patterns using a variety of SFHs, including models where the LMC underwent a burst of SFH starting about 2 Gyr ago, which produces a bump in [Mg/Fe] ratios as a function of metallicity owing to an increased contribution of core-collapse SNe for a relatively short duration that gives way once again to a dominating Fe contribution from Type Ia SNe.  While the model comparisons to the data in \citet{Bekki2012} were not entirely conclusive that this was the best SFH model, they were consistent with producing a modest increase in [Mg/Fe] beginning at [Fe/H] $\sim$ $-$1.0 followed by a decrease in [Mg/Fe] beginning at [Fe/H] $\sim$ $-$0.5.  With a larger and more precise sample, we are now able to see this predicted ``bump'' in the [Mg/Fe] ratios (and other $\alpha$-elements), which motivates trying to model a burst or increase in SFR to reproduce this feature following the lead of \citet{Bekki2012}.

% Might be worth more comparison with Bekki, since they suggested the need for winds and found no plateau in Ca which they concluded meant that Ca was preferentially lost, but I think might hint at the fact that the ``knee'' was at lower metallicity for all elements

To modify our best-fit model in flexCE to simulate a starburst in the LMC's chemical evolution history, we alter the SFE to be time dependent, allowing for bursts of star formation that could be triggered by events such as interactions with the SMC, without requiring an infall of gas (which could also be triggered by an interaction with the SMC).  We model a burst or increase in star formation by adding a Gaussian shaped increase in SFE on top of the constant SFE already in the models (see bottom panel of Figure \ref{fig_gcemodel_burst}).  By varying the time of the burst, its duration, and its strength, we find that we can not only reach higher metallicities than constant SFE models, but can also produce the increase and subsequent decrease in [$\alpha$/Fe] seen in the abundance pattern of the LMC at metallicities [Fe/H] $\gtrsim$ $-$1.0.

%Figure \ref{burst_param} shows the our best burst model for the LMC as well as the results of changing the parameters of the burst, i.e., the time of the burst, its duration, and its strength. 
Our best burst model, shown in Figure \ref{fig_gcemodel_burst}, uses essentially the
%[[{\bf FIX THIS DEPENDING ON THE CHANGES MADE TO THE ABOVE MODELS:  BEST BURST HAS 0.0125 GYR-1 AND TIMESCALE OF 2 GYR}]]
%(a SFE of 0.0125 Gyr$^{-1}$ instead of 0.01 Gyr$^{-1}$)
same parameters as the best-fit model to the metal-poor end of the LMC abundance distribution given above, except with a slightly modified base SFE$_{i}$ = 0.0125 Gyr$^{-1}$ that later undergoes an increase following a Gaussian form with a duration of $\sigma_{\tau} = 1.5$ Gyr, peaking at $\mu_{\tau} = 15.25$ Gyr, with a 35$\times$ increase in SFE (i.e., SFE$_{\rm peak}$ = 0.35 Gyr$^{-1}$).  Because the peak of the SFE is outside the model run time of 13 Gyr, the actual peak SFR occurs around 12.25 Gyr, because the gas available for star formation depletes more rapidly during the starburst, and produces a factor of 6 increase in the SFR.
%SH: In above paragraph, what's the increase in SFE? 6x?

%CHRIS FIX TEXT WHEN FINAL VERSIONS OF MODELS:  OLD burst model SFE 0.01 Gyr-1 and 4 Gyr inflow.  Best model uses 0.0125 and 2 Gyr.  Also for burst param the old model is in the text now, but the better model has duration 1.5 Gyr time 15.25 Gyr and intensity of a factor of 90 increase which produces a similar shape

Altogether, this best model produces a SFH with a low and slow SFR for the majority of LMC's history, which produces the low-metallicity end of the LMC's chemical-abundance patterns, which is kickstarted within the past 2--3 Gyr with a recent burst of star formation producing the high-metallicity population of the LMC.  Looking at the $\alpha$-element abundance pattern of our burst model, we can see that the model passes through the low-metallicity end of the LMC abundance distribution, although there is a large scatter that is likely due to a combination of higher uncertainties in metal-poor stars and their intrinsic scatter.  This model also reproduces the rise and slight turnover seen in the [$\alpha$/Fe] ratio at metallicities [Fe/H] $\gtrsim -1$ that were difficult to reproduce with constant SFE models.  

By binning the model in metallicity, and normalizing by the surviving population of stars at each time step, we can also compare the expected metallicity distribution function (MDF) of the chemical evolution model to the observed MDF (normalized by the number of stars in our LMC sample with S/N $> 50$).  In general, the shape of the model MDF is similar to the observed MDF, although it is shifted to lower metallicities by $\sim$0.15 dex.
%This could be due to the APOGEE's selection of LMC stars being more heavily weighted to the (on average) brighter metal-rich RGs. 
This shift could suggest that a chemical evolution model with a slightly higher base SFE may be more appropriate, although higher SFEs were slightly less preferred when fitting to only the metal-poor LMC abundance profiles.

%\begin{figure}[t]
%\begin{center}
%\includegraphics[width=1.0\hsize,angle=0]{lmc_alphafefeh_lit_trendline.eps}
%\end{center}
%\caption{A comparison of the $\alpha$-element abundance patterns of various Local Group galaxies.  The MW, LMC, SMC and Sgr trendlines come from APOGEE-2S data while that for Fornax is from \citet{Hendricks2014} and that for Sculptor is from \citep{Kirby2010}.}
%\label{fig_trendlit}
%\end{figure}
%% get better SGR trendline from Chris

\begin{figure*}[t]
\begin{center}
\includegraphics[width=0.50\hsize,angle=0]{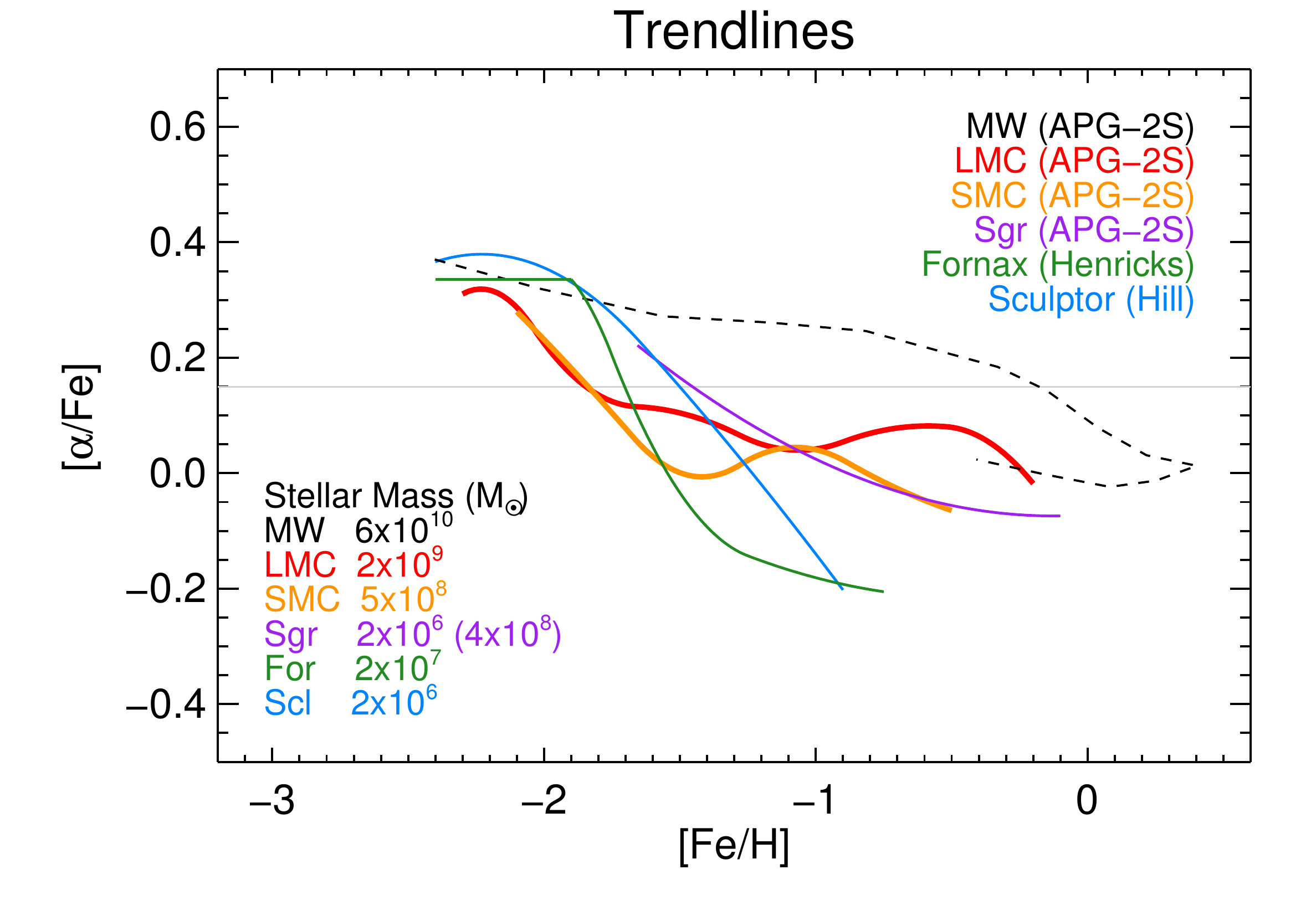}
\includegraphics[width=0.49\hsize,angle=0]{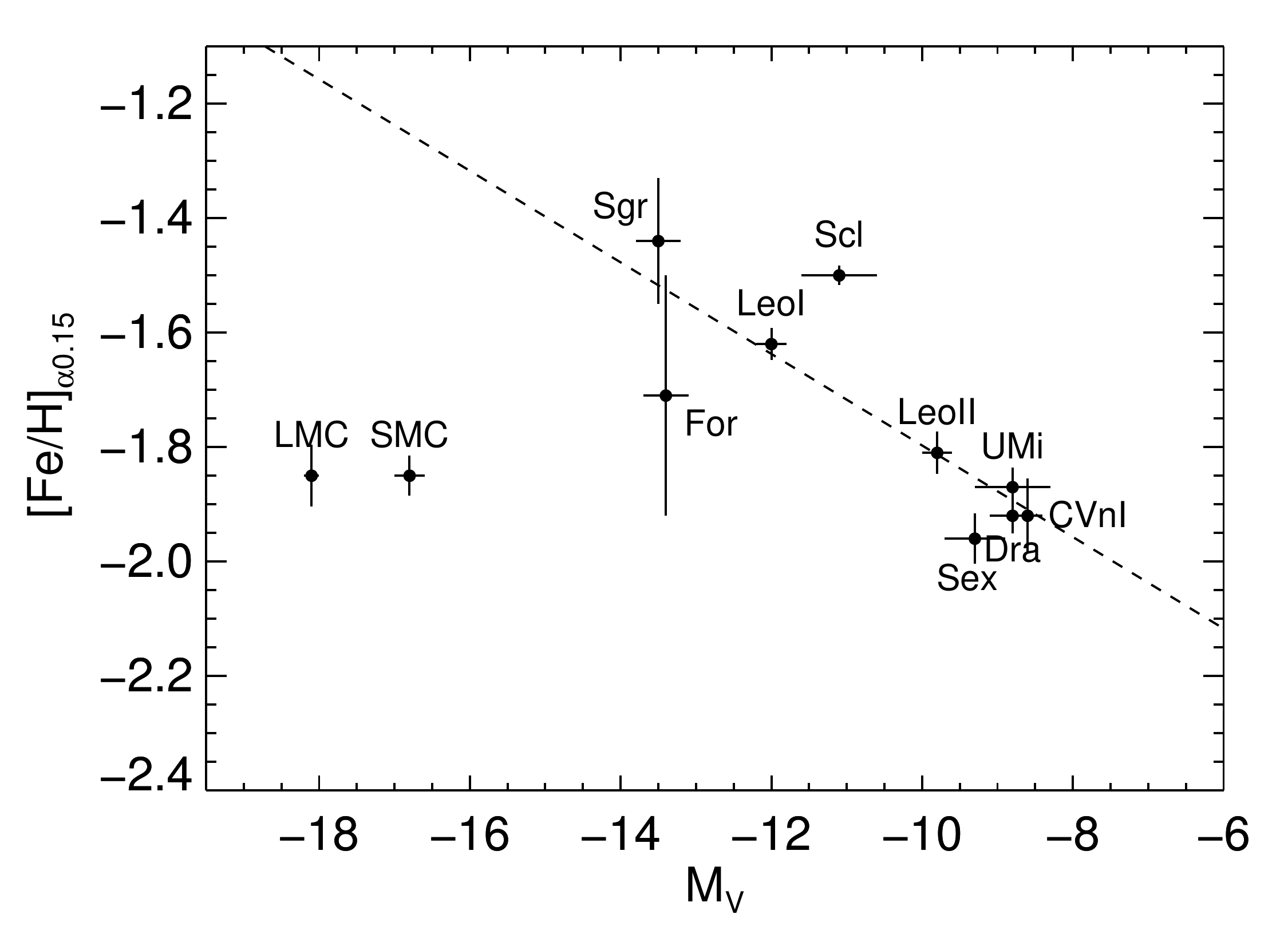}
\end{center}
\caption{(Left) A comparison of the $\alpha$-element abundance patterns of various Local Group galaxies.  The MW, LMC, SMC, and Sgr trendlines come from APOGEE-2S data (using [Mg+Ca+Si/Fe]) while that for Fornax (using [Mg/Fe]) is from \citet{Hendricks2014} and that for Sculptor (using [Mg+Ca+Si/Fe]) is from \citep{Hill2018}.  The [$\alpha$/Fe]=$+$0.15 line used to measure [Fe/H]$_{\alpha0.15}$ is shown in gray.
(Right) The [Fe/H] position where the $\alpha$-element abundance trendline crosses [$\alpha$/Fe]=$+$0.15 ([Fe/H]$_{\alpha0.15}$) versus $M_V$ for various dwarf galaxies with bootstrap uncertainties. The APOGEE-2 abundances were used for the LMC, SMC, and Sgr, while the \citet{Kirby2010} abundances were used for the rest of the dwarf galaxies.  The best linear fit to the low-luminosity galaxies is shown as a black-dashed line.  The Magellanic Clouds are clearly outliers, with significantly more metal-poor $\alpha$-element abundance trendlines than expected for their absolute magnitudes or mass.}
\label{fig_alphaknees}
\end{figure*}

Qualitatively, this old and slow SFH with a recent burst in the past few Gyr is similar to the SFHs found by photometric studies \citep[e.g.,][]{Meschin2014}.  However, the presence of stars with low $\alpha$-element abundances, even at metallicities of [Fe/H] $\sim -1.5$, and an increasing [$\alpha$/Fe] abundance pattern with decreasing metallicity seem to imply a relatively slower early SFH than found by the photometric studies, which are more limited in their metallicity discrimination.  However, this discrepancy should be revisited when more metal-poor stars are included in LMC chemical evolution analyses.  Regardless of the specific SFE at low metallicities, the fact that metallicities above [Fe/H] $\sim -1$ in the model are produced within the burst of star formation within the last 2--3 Gyr of the model, implies that the majority of stars we have observed in the LMC using APOGEE have been formed within the past 2--3 Gyr.  While an age analysis is beyond the scope of this paper, we do find that the most metal-rich stars in our sample are likely blue-loop stars, as shown in Figure \ref{fig_youngmetalrich}. If so, this would imply that their masses are $\sim$ 3 $M_{\odot}$, which supports that the metal-rich stars are rather young.

\section{Discussion}
\label{sec:discussion}

% lead with mass vs. environment on knee position
% henricks says there's a relation between mass and alpha-knee
% but other paper (jablonka) suggests that there is no change in alpha-knee
% with  mass
% it's a hard thing to measure, defined in different ways, some might not
% show a clear knee
% we have defined our own metric using the "shin" to compare SFE across dwarfs
% we do see some kind of correlation, but the MCs are outliers suggesting
% it's an environmental effect.  need a mass-environment relation
% falling into MW-galaxy can start SF, isolated galaxies evolve slower
% 

% OLD TEXT
Compared to many other MW dwarf galaxies, the $\alpha$-element abundance trendlines of the MCs (as measured by APOGEE) are more metal poor.  Trendlines using B-splines are shown in the left panel of Figure \ref{fig_alphaknees}, where we compare the [$\alpha$/Fe]--[Fe/H] trendlines of other galaxies in the literature to our MC trendlines.
%We do not actually measure the metal-poor knee in Sgr
%%[[We actually may have a better idea with latest data!]],
%but other studies have measured the knee at [Fe/H]$\approx$ $-$1.3 %\citep[e.g.,][]{deBoer2014}.
%which is as metal poor as the APOGEE Sgr sample goes.
We find that the MCs $\alpha$-element abundance trendlines (and their knees) are more metal poor than Fornax, Sculptor, and Sgr,
even though the MCs are a factor of $\sim$100--1000 more massive in stellar mass. 

The MW dwarf galaxies show a range in $\alpha$-knees, as shown by the compilation in \citet{Hendricks2014}.  The $\alpha$-knee is sensitive to the early SFR and \citeauthor{Hendricks2014} suggest that there is a relationship between the $\alpha$-knee of the galaxy and its absolute magnitude (and possibly stellar mass).  
However, the $\alpha$-knee is challenging to measure because it requires having reliable $\alpha$-element abundances for a statistically significant number of metal-poor stars, and because some dwarf galaxies only show a linear increase at low metallicity without any clear sign of reaching a plateau.  Therefore, to reliably compare the $\alpha$-element abundances of the dwarf galaxies, we establish a new metric that measures the position of the steeply falling portion of the $\alpha$-element abundances at low metallicity or ``$\alpha$-shin''.  We define [Fe/H]$_{\alpha0.15}$ as the metallicity where the $\alpha$-element trendline crosses [$\alpha$/Fe]=$+$0.15.  The right panel of Figure \ref{fig_alphaknees} shows [Fe/H]$_{\alpha0.15}$ versus $M_{\rm V}$ for eleven MW dwarf galaxies using APOGEE-2 abundances for LMC, SMC, and Sgr, while the \citet{Kirby2010} abundances for the rest of the dwarf galaxies.  For the APOGEE data the trendlines were fit using a B-spline to the [$\alpha$/Fe] abundances, while for the \citeauthor{Kirby2010} data trendlines were fit for each $\alpha$-element separately and then averaged.  Uncertainties in [Fe/H]$_{\alpha0.15}$ were determined using a bootstrap technique using 100 mocks.
For the lower luminosity dwarfs ($M_{\rm V}$ $>$ $-$14) an anti-correlation is apparent, similar to the anti-correlation presented by \citet{Hendricks2014}.  The best-fit linear trend is indicated by the dashed line, and has a Pearson correlation coefficient of $-$0.705, which indicates a strong anti-correlation.  The MCs clearly fall off this correlation, and their $\alpha$-element abundance trends are significantly more metal-poor than expected for their higher luminosities.

%There is some discussion in the literature that the position of the knee correlates with the mass of the galaxy. This was summarized in \citet{Hendricks2014}, and they found that Fornax potentially had a knee position that was too metal-poor for its mass. In Figure \ref{fig_alphaknees} we replicate the knee [Fe/H]--$M_{v}$ plot from Hendricks et al., adding to it our measurements for the MCs. We find that the MCs fall way off this trend, having knee positions that are uncharacteristically metal-poor for their luminosities.

So why are the MCs so ``lazy'' early-on in their evolution? One major difference between the MCs and the rest of the dwarf galaxies in Figure \ref{fig_alphaknees} is the environment in which they formed and evolved. The MCs are likely falling into the MW gravitational potential well for the first time \citep[e.g.,][]{Besla2007,Besla2012}, and have presumably been forming stars in isolation for roughly 10 Gyr. If the other galaxies with more metal-rich $\alpha$-element knees fell into the MW potential well early on in their formation, it is possible they have artificially high SFE from tidal interactions with the MW, and/or experienced enhanced star formation during ram pressure stripping.
\citet{Gallart2015} derived precise star formation histories from deep $HST$ imaging of LG dwarf galaxies, and found that the early star formation rate of the dwarf galaxies depended on the density of their environment, with ``slow'' dwarfs forming in relatively isolated regions.  In this context, the slow early star formation of the MCs adds even more support to the hypothesis that the MCs fell into the MW potential only recently.  If the low SFE of the MCs is an environmental effect, then we would expect that Magellanic satellites, such as the ones recently discovered \citep[Hor1, Car2, Car3, and Hyi1;][]{Bechtol2015,Torrealba2018,Li2018,Koposov2018,Kallivayalil2018}, should also have low SFE and metal-poor $\alpha$-element abundance trendlines.  Future abundance studies will help resolve whether this is indeed the case.

%(reference list). [[List some isolated galaxies that warrant follow-up spectroscopy using 8m class telescopes to find alpha knee.]]

%SH: In following paragraph, we need to add in references that suggest
%that the starbursts in the LMC and SMC are temporally correlated.
However, as discussed in \S \ref{subsec:cheme}, it is not possible for the LMC to enrich its gas to the level seen today with the SFE responsible for the metal-poor knee. We argue that the observed bump in [$\alpha$/Fe] is indicative of a large starburst some 2--3 Gyr ago \citep[e.g.,][]{Harris&Zaritsky2009},
which is then responsible for enriching the galaxy from [Fe/H] = $-$1.0 to [Fe/H]$\approx$ $-$0.2. Without the recent starburst, likely created by a close interaction with the SMC \citep[e.g.,][]{Harris&Zaritsky2009,Besla2012}, the LMC chemical-abundance patterns and MDF would be substantially different than we see today.
While we do not observe increasing [$\alpha$/Fe] with increasing [Fe/H] beyond the knee in any of the other galaxies, we do observe a flat [$\alpha$/Fe] with increasing [Fe/H] for the SMC, Sgr, and Fornax \citep[from][]{Hendricks2014}. A flat [$\alpha$/Fe] abundance pattern can be produced with a weak starburst (or starbursts), that enrich the ISM with metals and prevent the further dilution of [$\alpha$/Fe] by Type Ia SNe, but are too weak to produce substantial [$\alpha$/Fe] enhancement.

\citet{Hendricks2014} found that they were able to recreate the Fornax $\alpha$-element abundance patterns by using a chemical evolution model with three distinct starbursts that varied in SFE, but were all more efficient than the initial starburst. There is a thread of work in the literature that finds evidence for a recent merger in Fornax from spatial and kinematical substructure \citep[e.g.,][]{Coleman2004,Yozin2012,delPino2015}, which could be responsible for causing these starbursts.  In the case of the MCs, the starbursts were likely triggered by a close interaction between the MCs some 2--3 Gyr ago. Evidence for such an interaction is motivated by cotemporal starbursts, dynamical simulations, and studies of Magellanic stellar substructures
\citep[e.g.,][]{Harris&Zaritsky2009,Besla2012,Besla2016,Choi2018a,Choi2018b,Belokurov2018}.
% REFERENCES!!! (reference list, magellanic stream, etc.)
%SH: In the last sentence above, I added what motivated that starbursts...how does
%it sound? Also, need the starburst references I mentioned above.
%SH: Just checked and Harris&Zaritsky do indeed suggest the star formation burst might have been triggered by LMC-SMC interaction
\section{Summary}
\label{sec:summary}

We have obtained 3800 high-resolution $H$-band spectra of stars in the Large and Small Magellanic Clouds (MCs) using the new Southern APOGEE instrument on the du Pont telescope at Las Campanas Observatory.  This sizable stellar sample covers a large radial and azimuthal range of the MCs.  The stars cover a large metallicity range ($-$2.5$<$[Fe/H]$<$ $-$0.2), and the $\alpha$-element distributions reveal important insights into the chemical evolution of the MCs. The main conclusions from our analysis of our red giant branch sample are:

\begin{enumerate}
\item The [$\alpha$/Fe]--[Fe/H] distributions of the MCs are quite flat over a large range in metallicity, $-$1.2$<$[Fe/H]$<$ $-$0.2.
\item There is an increase of $\sim$0.1 dex in [$\alpha$/Fe] from [Fe/H] = $-$1.0 to [Fe/H] = $-$0.5, with a small decrease for the youngest and most metal-rich LMC stars at [Fe/H] $>$ $-$0.5. This behavior can be explained by a recent increase in sta-formation activity in the MCs.  Our one-zone chemical evolution models are able to reproduce this $\alpha$-element abundance ``bump'' with such a recent starburst.  
%This feature in the star formation history is vital to reproducing the bump and metal-rich distribution of the MCs.
\item We constrain the position of the ``$\alpha$-knee'' to be at [Fe/H]$\lesssim$ $-$2.2 for both the LMC and the SMC. 
We define a new metric, [Fe/H]$_{\alpha0.15}$, the metallicity where the $\alpha$-element trendline crosses [$\alpha$/Fe]=$+$0.15, to reliably inter-compare $\alpha$-element abundance trends between dwarf galaxies.
 Chemical evolution models fitted to the abundances of the metal-poor stars in the LMC find a low SFE of $\approx$0.01 Gyr$^{-1}$ with a gas consumption timescale of $\approx$100 Gyr.
%[[SH to DLN: We should rephrase this bullet point to introduce the shin?]] DLN: done
\item The LMC and SMC $\alpha$-element abundance trendlines are more metal poor than those for less massive MW satellites such as Fornax, Sculptor, or Sagittarius and the MCs are large outliers in [Fe/H]$_{\alpha0.15}$--$M_{\rm V}$ for MW satellites.  This counter-intuitive result suggests that the MCs formed in a lower-density environment, which is consistent with the paradigm that the MCs fell into the MW potential only recently.
\end{enumerate}

\acknowledgments
DAGH and FDA acknowledge support from the State Research Agency (AEI) of the Spanish Ministry of Science, Innovation and Universities (MCIU) and the European Regional Development Fund (FEDER) under grant AYA2017-88254-P.
S.H. is supported by an NSF Astronomy and Astrophysics Postdoctoral Fellowship under award AST-1801940.
H.J. acknowledges support from the Crafoord Foundation, Stiftelsen Olle Engkvist Byggm\"astare, and Ruth och Nils-Erik Stenb\"acks stiftelse.
T.C.B. acknowledges partial support for this work from grant PHY 14-30152; Physics Frontier Center/JINA Center for the Evolution
of the Elements (JINA-CEE), awarded by the US National Science Foundation.
R.R.M. acknowledges partial support from project BASAL AFB-$170002$ as well as FONDECYT project N$^{\circ}1170364$.
M.R. acknowledges the UNAM-DGAPA-PAPIIT grant IN109919.
We thank the anonymous referee for useful comments that improved the manuscript.

Funding for the Sloan Digital Sky Survey IV has been provided by the Alfred P. Sloan Foundation, the U.S. Department of Energy Office of Science, and the Participating Institutions. SDSS-IV acknowledges
support and resources from the Center for High-Performance Computing at
the University of Utah. The SDSS web site is www.sdss.org.

SDSS-IV is managed by the Astrophysical Research Consortium for the 
Participating Institutions of the SDSS Collaboration including the 
Brazilian Participation Group, the Carnegie Institution for Science, 
Carnegie Mellon University, the Chilean Participation Group, the French Participation Group, Harvard-Smithsonian Center for Astrophysics, 
Instituto de Astrof\'isica de Canarias, The Johns Hopkins University, 
Kavli Institute for the Physics and Mathematics of the Universe (IPMU) / 
University of Tokyo, Lawrence Berkeley National Laboratory, 
Leibniz Institut f\"ur Astrophysik Potsdam (AIP),  
Max-Planck-Institut f\"ur Astronomie (MPIA Heidelberg), 
Max-Planck-Institut f\"ur Astrophysik (MPA Garching), 
Max-Planck-Institut f\"ur Extraterrestrische Physik (MPE), 
National Astronomical Observatories of China, New Mexico State University, 
New York University, University of Notre Dame, 
Observat\'ario Nacional / MCTI, The Ohio State University, 
Pennsylvania State University, Shanghai Astronomical Observatory, 
United Kingdom Participation Group,
Universidad Nacional Aut\'onoma de M\'exico, University of Arizona, 
University of Colorado Boulder, University of Oxford, University of Portsmouth, 
University of Utah, University of Virginia, University of Washington, University of Wisconsin, 
Vanderbilt University, and Yale University.

\bibliographystyle{apj}
\bibliography{ref_og.bib}

% Bibtex will create a .bbs file in the directory and before sending to the editor, I should replace the bibliography call by this file.

%SH: What's the deal with this table?
\begin{deluxetable*}{lcccrrcccccccccc}
\tablecaption{APOGEE Magellanic Clouds Fields}
\tablecolumns{14}
\tablewidth{0pt}
\tablehead{
\colhead{Name} & \colhead{RA} & \colhead{DEC} & 
\colhead{$L_{\rm MS}$} & \colhead{$B_{\rm MS}$} & 
\colhead{$R$} & \colhead{PA} & 
\colhead{N$_{\rm Visits}$} & \colhead{N$_{\rm Visits}$} &
\colhead{N$_{\rm MC}$} & \colhead{N$_{\rm MC,RGB}$} &
\colhead{N$_{\rm MC,RGB}$} & \colhead{N$_{\rm MC,RGB,S/N>40}$} \\
\colhead{} & \colhead{(J2000)} & \colhead{(J2000)} & \colhead{(deg)} & \colhead{(deg)} &
\colhead{(deg)} & \colhead{(deg)} &
\colhead{Planned} & \colhead{ } & \colhead{Targets} & \colhead{Targets} & \colhead{Members} & \colhead{Members}
}
\startdata
SMC1 & 00:20:16.9 & $-$77:13:21.3 & $-$11.96 & $-$14.98 & 4.9 & 202 & 12 & 11 & 332 & 82 & 48 & 29 \\
47Tuc & 00:24:39.5 & $-$72:09:26.8 & $-$16.91 & $-$13.34 & 2.2 & 284 & 12 & 10 & 411 & 98 & 87 & 86 \\
SMC2 & 00:41:58.6 & $-$67:45:25.4 & $-$20.66 & $-$10.50 & 5.2 & 349 & 12 & 8 & 365 & 132 & 109 & 77 \\
SMC3 & 00:45:00.1 & $-$73:13:45.1 & $-$15.37 & $-$12.26 & 0.7 & 234 & 12 & 8 & 442 & 289 & 273 & 259 \\
NGC362 & 00:57:32.9 & $-$71:05:40.3 & $-$16.99 & $-$10.54 & 1.8 & 13 & 12 & 10 & 309 & 146 & 139 & 138 \\
SMC4 & 01:07:56.2 & $-$75:35:35.4 & $-$12.52 & $-$11.80 & 3.0 & 161 & 12 & 8 & 275 & 188 & 180 & 175 \\
SMC5 & 01:20:41.2 & $-$73:04:48.0 & $-$14.37 & $-$9.86 & 2.1 & 100 & 12 & 6 & 307 & 174 & 169 & 158 \\
SMC6 & 01:38:29.9 & $-$71:09:11.1 & $-$15.30 & $-$7.69 & 3.9 & 70 & 12 & 6 & 279 & 189 & 182 & 142 \\
SMC7 & 02:07:23.7 & $-$73:24:29.7 & $-$12.15 & $-$7.30 & 5.4 & 105 & 12 & 9 & 351 & 88 & 55 & 40 \\
LMC1 & 04:10:18.1 & $-$71:56:58.5 & $-$5.54 & $-$0.93 & 6.6 & 243 & 9 & 4 & 323 & 173 & 147 & 85 \\
LMC2 & 04:10:22.6 & $-$68:26:59.8 & $-$6.77 & 2.34 & 7.0 & 273 & 9 & 6 & 329 & 138 & 108 & 71 \\
LMC3 & 04:46:05.2 & $-$75:19:58.6 & $-$2.17 & $-$3.46 & 6.3 & 205 & 9 & 10 & 282 & 190 & 186 & 184 \\
LMC4 & 04:52:44.1 & $-$68:48:13.5 & $-$2.96 & 3.04 & 3.3 & 285 & 9 & 4 & 400 & 263 & 255 & 252 \\
LMC5 & 04:55:38.7 & $-$71:18:17.7 & $-$2.27 & 0.62 & 3.0 & 238 & 9 & 7 & 366 & 239 & 234 & 202 \\
LMC6 & 05:11:02.4 & $-$65:40:47.7 & $-$1.66 & 6.39 & 4.5 & 338 & 9 & 6 & 359 & 206 & 202 & 200 \\
LMC7 & 05:14:06.1 & $-$62:33:57.9 & $-$1.65 & 9.52 & 7.4 & 348 & 9 & 3 & 278 & 177 & 175 & 171 \\
LMC8 & 05:19:43.7 & $-$72:49:04.1 & $-$0.25 & $-$0.65 & 3.0 & 191 & 9 & 12 & 349 & 228 & 225 & 225 \\
LMC9 & 05:22:23.8 & $-$69:42:36.7 & $-$0.24 & 2.47 & 0.5 & 289 & 9 & 5 & 251 & 152 & 142 & 129 \\
LMC10 & 05:31:22.4 & $-$76:10:04.7 & 0.69 & $-$3.96 & 6.3 & 178 & 9 & 6 & 275 & 176 & 168 & 162 \\
30Dor & 05:36:13.6 & $-$69:07:40.0 & 0.96 & 3.09 & 1.1 & 47 & 1 & 4 & 173 & 53 & 41 & 4 \\
LMC11 & 05:41:31.0 & $-$63:34:43.1 & 1.53 & 8.63 & 6.4 & 14 & 9 & 2 & 264 & 186 & 182 & 179 \\
LMC12 & 05:44:47.3 & $-$60:23:34.1 & 2.00 & 11.81 & 9.6 & 13 & 9 & 2 & 223 & 128 & 90 & 36 \\
LMC13 & 05:46:05.6 & $-$67:42:34.9 & 1.89 & 4.49 & 2.7 & 40 & 9 & 2 & 258 & 177 & 172 & 168 \\
LMC14 & 05:53:21.1 & $-$70:58:13.7 & 2.36 & 1.20 & 2.4 & 120 & 9 & 4 & 262 & 157 & 154 & 152 \\
LMC15 & 06:08:30.6 & $-$63:35:53.7 & 4.55 & 8.39 & 7.4 & 38 & 9 & 0 & 260 & 215 & 0 & 0 \\
LMC16 & 06:32:17.2 & $-$75:11:03.3 & 4.54 & $-$3.37 & 7.2 & 145 & 9 & 6 & 252 & 149 & 122 & 97 \\
LMC17 & 06:32:47.1 & $-$70:18:34.9 & 5.68 & 1.37 & 5.6 & 102 & 9 & 10 & 269 & 178 & 172 & 170 
\enddata
%\tablenotetext{a}{Fields with more than 300 targets }
\label{table_fields}
\end{deluxetable*}

\end{document}